\documentclass[preprint,amsmath,amssymb,nofootinbib]{revtex4}
\usepackage[utf8]{inputenc}
\usepackage{rotating}
\usepackage{hyperref}
\usepackage{array}
\usepackage{subfig}
\usepackage{float}
\usepackage{graphicx}
\usepackage{slashed}
\usepackage{color}
\usepackage{ctable}
\definecolor{orange}{rgb}{1,0.5,0}
\usepackage[T1]{fontenc}

\begin{document}

\title{Angular and polarization observables for Majorana-mediated B decays with effective interactions.}

\author{Luc\'{\i}a Duarte}
\email{lucia@fisica.edu.uy}
 \affiliation{Instituto de F\'{\i}sica, Facultad de Ciencias,
 Universidad de la Rep\'ublica \\ Igu\'a 4225,(11400) 
Montevideo, Uruguay.}

\author{Gabriel Zapata}
\author{Oscar A. Sampayo}
\email{sampayo@mdp.edu.ar}
\affiliation{Instituto de F\'{\i}sica de Mar del Plata (IFIMAR)\\ CONICET, UNMDP\\ Departamento de F\'{\i}sica,
Universidad Nacional de Mar del Plata \\
Funes 3350, (7600) Mar del Plata, Argentina}

\begin{abstract}

We probe the effective field theory extending the Standard Model with a sterile neutrino in B meson decays at B factories and lepton colliders, using angular and polarization observables. We put bounds on different effective operators characterized by their distinct Dirac-Lorentz structure, and probe the $N$-mediated B decays sensitivity to these interactions. 
We define a Forward-Backward asymmetry $A_{FB}^{\ell \gamma}$ between the muon and photon directions for the $B \to \mu \nu \gamma$ decay, which allows us to separate the SM contribution from the effective lepton number conserving and violating processes, mediated by a near on-shell $N$. Using the most stringent constraints on the effective parameter space from Belle and BaBar we find that a measurement of the final polarization $P_{\tau}$ in the rare $B^-\rightarrow  \ell^-_{1}\ell^-_{2} \pi^+$ decays can help us infer the scalar or vector interaction content in the $N$ production or decay vertices. We find that the B meson decays are more sensitive to scalar operators. 

\end{abstract}

\maketitle

\section{Introduction.}{\label{intro}}

Besides the remarkable performance of the standard model (SM) of particle physics in describing nature, neutrino oscillations are currently the most compelling experimental evidence of the need to extend the SM in order to include mechanisms for neutrino mass generation. In the recent years, the LHC experiments have put stringent constraints on the existence of new physics involving colored states, but still the possible extensions of the electroweak (EW) sector are far less restricted. A variety of new physics scenarios may be hidden at energies well above the EW scale, and their study is being consistently tackled by the use of the standard model effective field theory (SMEFT) \cite{Buchmuller:1985jz, Grzadkowski:2010es}. 

But also new physics might involve weakly coupled fields at the EW scale. Many models leading to neutrino masses predict the existence of sterile right-handed neutrinos with Majorana masses, as the Type I \cite{Minkowski:1977sc, Mohapatra:1979ia, Yanagida:1980xy, GellMann:1980vs, Schechter:1980gr} and also the linear and inverse seesaw mechanisms. This possibly not-that-heavy degrees of freedom can be described by an EFT including them (SMNEFT) \cite{delAguila:2008ir, Aparici:2009fh, Liao:2016qyd, Bhattacharya:2015vja}, and here we concentrate on a simplified scenario with only one right-handed neutrino added \cite{delAguila:2008ir}. The phenomenology of models extending the Type I seesaw renormalizable Lagrangian with effective interactions of higher dimension for the right-handed neutrinos are being studied \cite{Caputo:2017pit, Jones-Perez:2019plk}, and recently complemented with the implications of the Minimal Flavor Violation ansatz on the new heavy neutrino interactions \cite{Graesser:2007yj, Barducci:2020ncz}. 

The simplified scenario considering only one sterile state $N$ has started to get attention in connection to the novel dimension five Higgs-neutrino interactions \cite{Butterworth:2019iff} and the heavy neutrino magnetic moment dipole portal \cite{Magill:2018jla}. In particular, some studies have set constraints on the effective operators in this simplified scenario using existing LHC searches \cite{Alcaide:2019pnf} and recent work explores the matching between the off-shell EW-scale operator basis and a low-energy on-shell basis \cite{Chala:2020vqp, Dekens:2020ttz}. 

On the other hand, works on approaches as the so called neutrino non standard interactions (NSI) and general neutrino interactions (GNI) are incorporating right-handed neutrinos to the SMEFT \cite{Bischer:2019ttk, Han:2020pff} considering their Majorana and/or Dirac nature. Also effective neutrino long-range interactions are being studied \cite{Bolton:2020xsm} in an EFT approach. 

The presence of a Majorana mass term for the right-handed neutrinos is the source of lepton number violation (LNV) in these models. While the phenomenology of LNV has been thoroughly studied in the past in the context of seesaw scenarios, for recent reviews see \cite{Drewes:2019byd, Cai:2017mow}, the study of lepton number violating and conserving (LNC) interference effects in final states with light neutrinos is usually discarded. Here we assess the chances of disentangling possible contributions from effective operators with distinct Dirac-Lorentz structure to B leptonic decays where the final neutrinos do not allow for the identification of LNV or LNC interactions. 

In this article we focus on the simplified scenario with only one heavy Majorana neutrino $N$, and neglect the effect of the renormalizable Yukawa term $N L \phi$ giving the heavy-active neutrino mixings $U_{l N}$, given that it is strongly constrained not only by the naive seesaw relation $U_{lN}^2 \sim m_{\nu}/M_N \sim 10^{-14}-10^{-10}$ required to account for the light $\nu$ masses \cite{Cai:2017mow, Atre:2009rg}, but also by the experimental constraints on a toy-like model in which the SM is extended with a massive Majorana neutral fermion, assumed to have non-negligible mixings with the active states, without making any hypothesis on the neutrino mass generation mechanism \cite{Abada:2017jjx, Pascoli:2018heg}. Such a minimal SM extension leads to contributions to LNV observables which are already close, or even in conflict, with current data from meson and tau decays, for masses $M_N$ below $10$ GeV (see \cite{Abada:2017jjx, Abada:2018nio} and the references therein). 

Our group has studied the Majorana $N$ decays \cite{Duarte:2015iba,Duarte:2016miz} and phenomenology mostly in colliders for masses $m_N$ above the EW-scale  \cite{Peressutti:2011kx,Peressutti:2014lka,Duarte:2014zea,Duarte:2018xst,Duarte:2018kiv} focusing on LNV processes, and in the order $1-10 ~GeV$ scale, where its decay is dominated by the $N\to \nu \gamma$ channel \cite{Duarte:2016caz}, which was also considered in \cite{Yue:2017mmi}.  

The pure leptonic and radiative decays of pseudoscalar mesons mediated by a sterile $N$ in this effective scenario were initially studied in \cite{Yue:2018hci}, where the authors only took into account vectorial operators contributions. The study of $N$-mediated lepton number violation in rare B meson decays has been pursued, for example, in \cite{Chun:2019nwi, Abada:2017jjx, Asaka:2016rwd, Cvetic:2016fbv, Cvetic:2015naa, Wang:2014lda, Cvetic:2010rw, Zhang:2010um, Helo:2010cw, Atre:2009rg, Ali:2001gsa} and the references therein. Also, the tension with the SM values of the ratios of branching fractions $R(D)$ and $R(D^*)$ in semileptonic B decays \cite{Bifani:2018zmi} measured at Belle, BaBar and LHCb, has led to proposals involving sterile neutrinos as solutions, in seesaw scenarios \cite{Cvetic:2017gkt} and also including EFTs with right-handed neutrinos \cite{Mandal:2020htr}, with sterile $N$ interactions mediated by leptoquarks \cite{Azatov:2018kzb} or a $W^{\prime}$ $SU(2)_L$ singlet vector boson \cite{Greljo:2018ogz, Robinson:2018gza}.   

In this work we go on studying the effect on B meson decays of the presence of a Majorana $N$ with effective interactions. In a previous paper \cite{Duarte:2019rzs} we studied the constraints imposed by the bound on the $B^{-}\to \mu^{\pm}\mu^{\mp}\pi^{+}$ decay by LHCb \cite{Aaij:2014aba} and in the radiative leptonic $B\to \mu \nu \gamma$ decay by Belle \cite{Gelb:2018end} on the effective operators. Here we propose to use final tau polarization and angular observables to disentangle the possible contributions of a Majorana neutrino $N$ with effective interactions to the $B^{-} \to  \tau^{-} \nu$, the radiative $B\to \mu \nu \gamma$ and the LNV $B^-\rightarrow  \ell^-_{1}\ell^-_{2} \pi^+$ decays in future experiments. 

The paper is organized as follows. In Sec. \ref{sec:eff_form} we introduce the effective Lagrangian formalism for the Majorana $N$, discuss the existing bounds on the couplings weighting the different effective operators contributions \ref{sec:bounds_alpha}, and review the experimental prospects for B meson leptonic decays \ref{sec:B_lept_dec_exp}. In Sec.\ref{sec:BtotauN} we study the B tauonic decay. The radiative $B\to \ell \nu \gamma$ decay is assessed in Sec.\ref{sec:bdec_rad} and the LNV $B^-\rightarrow  \ell^-_{1}\ell^-_{2} \pi^+$ in Sec.\ref{sec:NmedBdec_taus}. We summarize our results in Sec.\ref{sec:summary}.

\subsection{Effective interactions formalism.}\label{sec:eff_form}

We extend the SM Lagrangian including only \emph{one} right-handed neutrino $N_R$ with a Majorana mass term, giving a relatively light massive state $N$ as an observable degree of freedom. The new physics effects are parameterized by a set of effective operators $\mathcal{O}_\mathcal{J}$ constructed with the SM and the Majorana neutrino fields and satisfying the $SU(2)_L \otimes U(1)_Y$ gauge symmetry \cite{delAguila:2008ir, Liao:2016qyd, Wudka:1999ax}. 

The effect of these operators is suppressed by inverse powers of the new physics scale $\Lambda$. The total Lagrangian is organized as follows:
\begin{eqnarray}\label{eq:lagrangian}
\mathcal{L}=\mathcal{L}_{SM}+\sum_{n=5}^{\infty}\frac1{\Lambda^{n-4}}\sum_{\mathcal{J}} \alpha_{\mathcal{J}} \mathcal{O}_{\mathcal{J}}^{(n)}
\end{eqnarray}
where $n$ is the mass dimension of the operator $\mathcal{O}_{\mathcal{J}}^{(n)}$ \footnote{Note that we do not include the Type-I seesaw Lagrangian terms giving the Majorana and Yukawa terms for the sterile neutrinos.}.

\begin{table}[t]
 \centering
 \begin{tabular}{c l c r c l c r}
 \firsthline \specialrule{.1em}{.05em}{.05em} 
 Operator &  Notation & Type & Coupling  &  Operator &  Notation & Type & Coupling   \\
 $\;\; (\phi^{\dag}\phi)(\bar L_i N \tilde{\phi}) \;\;$ & $\;\;  \mathcal{O}^{(i)}_{LN\phi} \;\;$   &~S~~& $\alpha^{(i)}_{\phi}$ &  &  &  \\
 \specialrule{.03em}{.03em}{.03em}
 $\;\; i(\phi^T \epsilon D_{\mu}\phi)(\bar N \gamma^{\mu} l_i)\;\;$ & $ \;\;\mathcal{O}^{(i)}_{Nl\phi}\;\;$&~V~~& $\alpha^{(i)}_W$ & $\;\;  i(\phi^{\dag}\overleftrightarrow{D_{\mu}}\phi)(\bar N \gamma^{\mu} N) \;\;$& $\;\; \mathcal{O}_{NN\phi} \;\;$&~V~~& $\alpha_Z$\\ 
\specialrule{.06em}{.05em}{.05em} 
 $\;\;(\bar N \gamma_{\mu} l_i) (\bar d'_i \gamma^{\mu} u'_i)\;\;$& $\;\; \mathcal{O}^{(i)}_{duNl} \;\;$&~V~~& $\alpha^{(i)}_{V_0}$ &  $\;\;(\bar N \gamma_{\mu}N) (\bar f_i \gamma^{\mu}f_i) \;\;$& $\;\; \mathcal{O}^{(i)}_{fNN} \;\;$ &~V~~& $\alpha^{(i)}_{V_{f}}$ \\
 \specialrule{.03em}{.03em}{.03em}
 $\;\; (\bar Q'_i u'_i)(\bar N L_i) \;\;$& $\;\;\mathcal{O}^{(i)}_{QuNL}\;\;$ &~S~~& $\alpha^{(i)}_{S_1}$ & $ \;\;(\bar L_i N)\epsilon (\bar L_i l_i)\;\; $ & $\;\;\mathcal{O}^{(i)}_{LNLl}\;\;$&~S~~& $\alpha^{(i)}_{S_0}$\\
 $\;\;(\bar L_i N) \epsilon (\bar Q'_i d'_i)\;\;$& $\;\;\mathcal{O}^{(i)}_{LNQd}\;\;$&~S~~& $\alpha^{(i)}_{S_2}$ & $\;\;|\bar N L_i|^2\;\; $& $\;\;\mathcal{O}^{(i)}_{LN}\;\;$&~S~~& $\alpha^{(i)}_{S_4}$\\
 $\;\;(\bar Q'_i N)\epsilon (\bar L_i d'_i)\;\;$ & $\;\;\mathcal{O}^{(i)}_{QNLd}\;\;$ &~S~~&  $\alpha^{(i)}_{S_3}$ & & &\\
  \specialrule{.06em}{.05em}{.05em} 
 $\;\; (\bar L_i \sigma^{\mu\nu} \tau^I N) \tilde \phi W_{\mu\nu}^I \;\;$& $\;\;\mathcal{O}^{(i)}_{NW}\;\;$ & T & $\alpha^{(i)}_{NW}$ &  $\;\;(\bar L_i \sigma^{\mu\nu} N) \tilde \phi B_{\mu\nu}\;\;$& $\;\;\mathcal{O}^{(i)}_{NB}\;\;$ & T &$\alpha^{(i)}_{NB}$ \\
\specialrule{.1em}{.05em}{.05em}
\lasthline 
 \end{tabular}
\caption{\small{Basis of dimension 6 baryon (and lepton) number conserving operators with a right-handed neutrino $N$ \cite{delAguila:2008ir, Liao:2016qyd}. Here $l_i$, $u'_i$, $d'_i$ and $L_i$, $Q'_i$ denote, for the family labeled $i$, the right handed $SU(2)$ singlet and the left-handed $SU(2)$ doublets, respectively (collectively $f_i$). The field $\phi$ is the scalar doublet, $B_{\mu\nu}$ and $W_{\mu\nu}^I$ represent the $U(1)_{Y}$ and $SU(2)_{L}$ field strengths respectively. Also $\sigma^{\mu \nu}$ is the Dirac tensor, $\gamma^{\mu}$ are the Dirac matrices, and $\epsilon=i\sigma^{2}$ is the antisymmetric symbol in two dimensions. Types V, S and T stand for scalar, vectorial and tensorial (one-loop level generated) structures.}}\label{tab:Operators}
\end{table}

The dimension 5 operators were studied in detail in \cite{Aparici:2009fh}. These include the Weinberg operator $\mathcal{O}_{W}\sim (\bar{L}\tilde{\phi})(\phi^{\dagger}L^{c})$ \cite{Weinberg:1979sa} which contributes to the light neutrino masses, $\mathcal{O}_{N\phi}\sim (\bar{N}N^{c})(\phi^{\dagger} \phi)$ which gives Majorana masses and couplings of the heavy neutrinos to the Higgs (its LHC phenomenology has been studied in \cite{Caputo:2017pit, Graesser:2007yj, Jones-Perez:2019plk}), and the operator $\mathcal{O}^{(5)}_{NB}\sim (\bar{N}\sigma_{\mu \nu}N^{c}) B^{\mu \nu}$ inducing magnetic moments for the heavy neutrinos, which is identically zero if we include just one sterile neutrino $N$ in the theory. 

In the following, as the dimension 5 operators do not contribute to the studied processes -discarding the heavy-light neutrino mixings- we will only consider the contributions of the dimension 6 operators, following the treatment presented in \cite{delAguila:2008ir, Liao:2016qyd}, and shown in Tab.\ref{tab:Operators}. 

The effective operators above can be classified by their Dirac-Lorentz structure into \emph{scalar}, \emph{vectorial} and \emph{tensorial}. The couplings of the tensorial operators are naturally suppressed by a loop factor $1/(16\pi^2)$, as they are generated at one-loop level in the UV complete theory \cite{delAguila:2008ir, Arzt:1994gp}.

In this paper we will consider the B decays $B\to \tau N$ in Sec.\ref{sec:BtotauN}, $B^- \to \mu^- \nu \gamma$ in Sec.\ref{sec:bdec_rad} and $B^-\to \ell^-_{1}\ell^-_{2} \pi^+$ in Sec.\ref{sec:NmedBdec_taus}, mediated by a Majorana neutrino $N$. We can thus take into account the following effective Lagrangian terms involved in those processes:
\begin{eqnarray}\label{eq:lag_tree}
  \mathcal{L}^{tree}& = &\mathcal{L}_{SM} + \frac{1}{\Lambda^2} \sum_{i,j}\Big\{ -\alpha^{(i)}_W \frac{ ~v ~m_W}{\sqrt{2}}\, \overline{l_i} \gamma^{\nu} P_R N  \, W^{-}_{\mu}    
		+  \alpha^{(i)}_{V_0} \, \overline{u'_j} \gamma^{\nu} P_R d'_j \,  \, \overline{l_i} \gamma_{\nu} P_R N   \nonumber \\
		&& +  \alpha^{(i)}_{S_1}\, \, \overline{u'_j} P_L d'_j \, \, \overline{l_i} P_R N -  \alpha^{(i)}_{S_2}\, \overline{u'_j} P_R d'_j \, \, \overline{l_i} P_R N  + \alpha^{(i)}_{S_3}\, \overline{u'_j} P_R N \, \,  \overline{l_i} P_R d'_j + \mbox{h.c.} \Big\}  
\end{eqnarray}
and
\begin{eqnarray}\label{eq:lag_1loop}
 \mathcal{L}^{1-loop}= -i \frac{\sqrt{2} v}{\Lambda^2} {(\alpha_{NB}^{(i)}c_W +  \alpha_{NW}^{(i)}s_W)  (P^{(A)}_{\mu} ~\bar \nu_{L,i} \sigma^{\mu\nu}N_R~ A_{\nu})}.
\end{eqnarray}
The one-loop generated effective Lagrangian contributes to the $N\to \nu \gamma$ decay. Here $-P^{(A)}$ is the 4-momentum of the outgoing photon and $s_W$ and $c_W$ are the sine and cosine of the weak mixing angle. 

In the effective four-fermion terms in \eqref{eq:lag_tree} the quark fields are flavor eigenstates with family $j=1,2,3$. In order to find the contribution of the effective Lagrangian to the $B^{-}$ decays we are studying, we must write it in terms of the massive quark fields. Thus, we consider that the contributions of the dimension 6 effective operators to the Yukawa Lagrangian are suppressed by the new physics scale with a factor $\frac{1}{\Lambda^2}$, and neglect them, so that the matrices that diagonalize the quark mass matrices are the same as in the pure SM. 

Writing with a prime symbol the flavor fields, we take the matrices $U_{R}, ~U_{L}, D_{R}$ and $D_{L}$ to diagonalize the SM quark mass matrix in the Yukawa Lagrangian. Thus the left- and right- handed quark flavor fields (subscript $j$) are written in terms of the massive fields (subscript $\beta$) as:
\begin{eqnarray}\label{eq:mass_basis}
&& u^{'}_{(R,L) j} = U_{(R,L)}^{j,\beta} u_{(R,L) \beta}  \;,  \; \; \; \; \; \overline{u^{'}_{(R,L)j}}=\overline{u_{(R,L)\beta}} (U_{(R,L)}^{j, \beta})^{\dagger}    \nonumber \\
&& d^{'}_{(R,L) j} = D_{(R,L)}^{j,\beta} d_{(R,L) \beta} \;,  \; \; \; \; \; \overline{d^{'}_{(R,L)j}}=\overline{d_{(R,L)\beta}} (D_{(R,L)}^{j, \beta})^{\dagger} .
\end{eqnarray}
With this notation, the SM $V_{CKM}$ mixing matrix corresponds to the term $V^{\beta \beta'} = \sum_{j=1}^{3} (U_{L}^{j \beta})^{\dagger}D_L^{j \beta'} $ appearing in the charged SM current $J^{\mu}_{CC}= \overline{u_{\beta}} ~V^{\beta \beta'} \gamma^{\mu}  P_L   d_{\beta'}$. 

Thus in the mass basis the tree-level generated Lagrangian in \eqref{eq:lag_tree} can be written in terms of the massive quarks as
\begin{eqnarray}\label{eq:lag_massiveq}
  \mathcal{L}^{tree}&&= \mathcal{L}_{SM} + \frac{1}{\Lambda^2}  \Big\{ -\sum_{i} \alpha^{(i)}_W \frac{ ~v ~m_W}{\sqrt{2}}\,  \overline{l_i} \gamma^{\nu} P_R N  \, W^{-}_{\nu} \nonumber \\   
 		&& + \sum_{i,j} \left( \alpha^{(i)}_{V_0}~ U_{R}^{\beta j~*} ~D_{R}^{j \beta'} \, \overline{u}_{\beta} \gamma^{\nu} P_R d_{\beta'} \,  \, \overline{l_i} \gamma_{\nu} P_R N  +  \alpha^{(i)}_{S_1}~ U_{R}^{\beta j~*} ~D_{L}^{j \beta'} \, \, \overline{u}_{\beta} P_L d_{\beta'} \, \, \overline{l_i} P_R N \right. \nonumber \\ 
		&& \left. -  \alpha^{(i)}_{S_2}~ U_{L}^{\beta j~*} D_{R}^{j \beta'} \, \overline{u}_{\beta} P_R d_{\beta'} \, \, \overline{l_i} P_R N  + \alpha^{(i)}_{S_3}~  U_{L}^{\beta j~*} D_{R}^{j \beta'} \,\,  \overline{u}_{\beta} P_R N \, \,  \overline{l_i} P_R d_{\beta'} \right) +\mbox{h.c.} \Big\}.
\end{eqnarray}
For the sake of simplicity, we will rename the new quark mixing matrix element products as
\begin{eqnarray}\label{eq:mix_beta_betap} 
Y^{\beta \beta'}_{RR} \equiv  \sum_{j} U_{R}^{\beta j*} ~D_{R}^{j \beta'}, \qquad 
Y^{\beta \beta'}_{RL} \equiv  \sum_{j} U_{R}^{\beta j*} ~D_{L}^{j \beta'},  \qquad
Y^{\beta \beta'}_{LR} \equiv  \sum_{j} U_{L}^{\beta j*} D_{R}^{j \beta'} \qquad 
\end{eqnarray}
These new quark flavor-mixing matrices combinations $Y^{\beta \beta'}$ are unknown, and in principle their entries may be found by independent measurements, as is done in the case of the SM $V_{CKM}$ matrix. In this occasion we will make an ansatz and consider that all the $Y^{\beta \beta'}$ values in \eqref{eq:mix_beta_betap} shall be of the order of the SM $V^{\beta \beta'}_{CKM}$ value, taking it as a measure of the strength of the coupling between the respective quarks: $Y^{\beta \beta'}_{RR} = Y^{\beta \beta'}_{RL} = Y^{\beta \beta'}_{LR} \approx V_{CKM}^{\beta \beta'}$. This will be done in the numerical calculations, while we leave the explicit expressions in our analytical equations\footnote{The matrices in \eqref{eq:mass_basis} could also be reabsorbed into the definition of the effective couplings $\alpha_{\mathcal{J}}$, as is done for instance in \cite{Bolton:2020xsm}.}.

\subsection{Numerical bounds on the effective couplings}\label{sec:bounds_alpha}

Besides depending on the quark mixing matrices $Y^{\beta \beta'}$ values, all the quantities we calculate of course depend on the numerical values of the effective couplings $\alpha_{\mathcal{J}}^{(i)}$ which weight the contribution of each operator $\mathcal{O}_{\mathcal{J}}^{(i)}$ in Tab. \ref{tab:Operators} to the Lagrangian \eqref{eq:lagrangian}. Their numerical value can be constrained considering the current experimental bounds on the active-heavy neutrino mixing parameters $U_{\ell N}$ in low scale minimal seesaw models appearing in the charged $(V-A)$ interactions $N L W$ when neutrino mixing is taken into account \cite{Atre:2009rg}. Inspired in this interaction we consider the combination
\begin{equation} \label{eq:u2}
U^2=(\alpha v^2 /(2 \Lambda^2))^2
\end{equation}
which is derived from the Lagrangian term coming from the operator $\mathcal{O}^{(i)}_{LN\phi}$ in Tab.\ref{tab:Operators} and allows a direct comparison with the mixing angles in the Type I seesaw scenarios \cite{Atre:2009rg}. Updated reviews of the existing bounds on the Type I seesaw mixings $U_{\ell N}$ can be found in refs. \cite{Bolton:2019pcu,Chun:2019nwi}. 

Some of the operators involving the first fermion family (with indices $i=1$) are strongly constrained by the neutrinoless double beta decay bounds, currently obtained by the KamLAND-Zen collaboration \cite{KamLAND-Zen:2016pfg}. Following the treatment already made in \cite{Duarte:2016miz}, the values of the $0\nu \beta \beta$-decay constrained couplings $\alpha^{(1)}_W,\,\alpha^{(1)}_{V_0},\,\alpha^{(1)}_{S_{1,2,3}}$ are taken as equal to the bound $\alpha^b_{0\nu\beta\beta}=3.2\times 10^{-2}\left(\frac{m_N}{100 GeV} \right)^{1/2}$ for $\Lambda=1~ TeV$.
These operators, although not directly involved in the processes we consider in this work, appear as contributions to the $\Gamma_N$ total width. The numerical treatment of the second and third fermion family effective couplings $\alpha_{\mathcal{J}}^{(i)}$ for $i=2,3$ (involved in the processes studied in this work) is explained below.

As we would like to disentangle the kind of new physics contributing to the Majorana neutrino interactions, for the numerical analysis we will consider different benchmark scenarios for the effective couplings, where we switch on/off the operators with distinct Dirac-Lorentz structure: vectorial, scalar and the tensorial one-loop generated operators. If we call ($V, S, T$) the factors multiplying the vectorial, scalar and tensorial one-loop generated operators respectively, we can define six sets, presented in the table \ref{tab:alpha-sets}. 
\begin{table}[t]
 \centering
 \begin{tabular}{|l l c |c c c c c c|}
\firsthline
Operators & Couplings  & $\;$ Type $\;$ & \bf{Set1} &  \bf{Set2}  &  \bf{Set3} & \bf{Set4} &  \bf{Set5} & \bf{ Set6}\\
\hline
$\mathcal{O}_{L N \phi}$, $\mathcal{O}_{duNL}$  & $\alpha^{(i)}_{W}$ $\alpha^{(i)}_{V_0}$  & V & 1 & 1 &0 &1 & 1& 0 \\  
$\mathcal{O}_{QuNL}$, $\mathcal{O}_{LNQd}$, $\mathcal{O}_{QNLd}\qquad$ & $\alpha^{(i)}_{S_1} $ $\alpha^{(i)}_{S_2}$ $\alpha^{(i)}_{S_3}$ & S & 1  & 0 & 1& 1 & 0 & 1   \\
$\mathcal{O}_{NB}$, $\mathcal{O}_{NW}$ & $\alpha^{(i)}_{NB}$ $\alpha^{(i)}_{NW}$  & T & 1 & 1 & 1& 0 & 0  & 0 \\ 
\lasthline
 \end{tabular}
\caption{Effective operators benchmark sets. }\label{tab:alpha-sets}
\end{table}
Throughout the paper we will take all the vectorial couplings numerical values equal to a generic $\alpha_{V}^{(i)}$, the scalar couplings as $\alpha_{S}^{(i)}$ and the tensorial ones as $\alpha_{T}^{(i)}$ for $i=2,3$. In the different sections of the paper, we will take complementary approaches to assign these values, that will be explained in a timely manner. However, we will use the sets defined in Tab. \ref{tab:alpha-sets} to see which operators are taken into account in each case. 

For the calculation of the total decay width of the Majorana neutrino $\Gamma_{N}$ we include all the kinematically allowed channels for $m_N$ in the range $1 ~GeV< m_N < 5 ~GeV$. The details of the calculation are described in our ref. \cite{Duarte:2016miz}. The sets 1, 2 and 3 in Tab. \ref{tab:alpha-sets} take into account the contributions of the tensorial one-loop-generated effective couplings in \eqref{eq:lag_1loop} to the $N$ decay width. In particular these sets allow for the existence of the $N\to \nu \gamma$ decay channel. As we found in \cite{Duarte:2016miz}, this channel gives the dominant contribution to the $N$ decay width for the low mass $m_N$ range considered in this work. The sets 4, 5 and 6 discard this contribution. The total $\Gamma_N$ width is around three orders of magnitude higher in the sets 1, 2 and 3 than in the sets 4, 5 and 6 \cite{Duarte:2019rzs}.

\subsection{B meson leptonic and semileptonic decays}\label{sec:B_lept_dec_exp}

Many ongoing and future experiments will copiously produce B mesons, enabling the study of its leptonic and semileptonic decays. The number of B mesons is expected to be over $10^{10}$ at Belle II, over $10^{13}$ at the $300 fb^{-1}$ luminosity LHCb upgrade, and a similar number is expected at SHiP. The estimated number of B mesons to decay in the MATHUSLA detector volume is over $10^{14}$, and the expected number of B meson pairs to be produced at the Z peak at the FCC-ee is above $10^{11}$ \cite{Chun:2019nwi}.    

Measuring pure leptonic B decays is challenging. The $B \to \tau \nu$ predicted branching ratio is of order $10^{-5}$ in the SM. For muonic and electronic B decays it is much smaller due to the $m^2_{\ell}$ factor coming from helicity suppression. The spinless B meson -like the pion- prefers to decay to the heaviest possible charged lepton because balancing the spins of the outgoing leptons requires them to have the same handedness, and the neutrino forces its charged partner into the unfavored helicity. The SM decay $B^{-} \to  \tau^{-} \bar{\nu}_{\tau}$ (or its conjugate) has not yet been measured by one single experiment with $5\sigma$ significance, and in Sec. \ref{sec:BtotauN} we use the value of the combined measurement \cite{Tanabashi:2018oca} to put bounds on the effective couplings $\alpha_{\mathcal{J}}$ involved in the $B \to \tau N$ decay. However, the prospects are for Belle II to achieve a measurement with approximately $2 ~ab^{-1}$ (assuming the SM branching ratio)\cite{Kou:2018nap}. This would be a first step towards disentangling the interference with possible new physics, as $B \to \tau N$ decays, for instance with the aid of the final tau lepton polarization information.  

Final taus are the only fermions whose polarization is accessible by means of the energy and angular distribution of its decay products. These measurements rely on the dependence of kinematic distributions of the observed tau decay products on the helicity of the parent tau. Tau lepton polarization measurements at Belle have focused on the $R_{(D^*)}$ anomaly \cite{Hirose:2016wfn, Hirose:2017dxl}. In Sec. \ref{sec:NmedBdec_taus} we find that measuring the final tau polarization in eventual LNV $B^-\rightarrow  \ell^-_{1}\ell^-_{2} \pi^+$ decays with $\ell^-_{1  ~\text{and/or}~ 2}=\tau^{-}$ could help to discern between the contributions of different types of effective operators. The Belle II prospects for measuring SM rare semi-tauonic modes as $B^{+}\to \tau^{+} \tau^{-} \pi^{+}$ are discouraging, because it is very difficult to reconstruct both final taus, and the lepton flavor violating $B^{+}\to \tau^{+} \mu^{-} \pi^{+}$ decays with final taus and muons are expected to be easier to measure \cite{Kou:2018nap}. One can hope that the LNV semi-tauonic decays could be discovered in experiments with more B meson events, as MATHUSLA and SHiP \cite{Chun:2019nwi}. However, these would not allow for the reconstruction of the taus polarization, leaving us with the eventual measurements at the FCC-ee \cite{Kamenik:2017ghi, Abada:2019zxq} to test our proposal. 

B factories as Belle II produce B mesons at the center of mass (CM) energy of the $\Upsilon(4S)$ resonance, which decays in 50\% to $B^+ B^-$ pairs. The decay of the $\Upsilon(4S)$ produces B mesons in pairs: if one (the tag $B_{tag}$) is reconstructed, the rest of the event must be a B meson (the signal $B_{sig}$). In hadronic B tags all the charged and neutral particles are identified, and used to reconstruct the tag-side B meson. The efficiency of this method is low -of order $10^{-3}$- but a very pure sample of B mesons is obtained. In semileptonic B tags a charmed meson is reconstructed together with a high momentum lepton. The efficiency of this method is higher, of order $10^{-2}$, though the obtained sample is not so pure. Inclusive tagging combines the four-momenta of all particles in the rest of the event of the signal-side B candidate. The achieved tagging efficiency is usually one order of magnitude above the hadronic and semileptonic tagging approaches \cite{Kou:2018nap}.

The SM radiative leptonic B decays have been extensively studied in the literature \cite{Beneke:2011nf,Wang:2016qii,DescotesGenon:2002mw,Korchemsky:1999qb,Wang:2018wfj,Beneke:2018wjp}, as they are a means of probing the strong and weak SM interactions in a heavy meson system. While the measurement of pure leptonic B decays is very difficult due to helicity suppression and the fact of having only one detected final state particle, the radiative modes can be larger as they escape helicity suppression, and are easier to reconstruct because of the extra real final photon. In Sec. \ref{sec:NmedBdec_taus} we use the bounds imposed on the effective couplings by the last Belle measurements on the $B \to \mu \nu \gamma$ partial branching fraction \cite{Gelb:2018end}. Now, as the signal yields for the $B \to \mu \nu \gamma$ decay are expected to be three times higher in Belle II compared to Belle \cite{Kou:2018nap}, in Sec. \ref{sec:bdec_rad} we explore the possibility to, if the signal is found, use angular information from the hard photon and the muon flight directions to study the interference of $N$ mediation in this decay. We are particularly interested in the inclusive tagging technique mentioned above, which allows for the reconstruction of the $B_{sig}$ meson flight direction, and the use of its rest frame, without a high loss in efficiency, as has been recently done in \cite{Prim:2019gtj}. This technique could be used for implementing the angular Forward-Backward asymmetry $A_{FB}^{\ell \gamma}$ between the final charged lepton and photon flight directions we propose in Sec. \ref{sec:subsec_AFB_lg}.

\section{B tauonic decay: final tau polarization}\label{sec:BtotauN}

In this section we aim to study the possibility to measure the effects of the presence of a Majorana neutrino $N$, considering the final tau polarization in the SM decay $B^{-} \to  \tau^{-} \bar{\nu}_{\tau}$ and the effective $B \to \tau N$ decay. The latter mode can leave the same final tau plus missing energy $\slashed{E_{T}}$ signal as the SM process when the $N$ escapes the detector before decaying into observable particles.    

The current value of the branching fraction $Br(B \to \tau \nu)= (1.09 \pm 0.24)\times 10^{-4}$ \cite{Tanabashi:2018oca}, which averages the latest measurements by Belle \cite{Adachi:2012mm, Kronenbitter:2015kls} and BaBar \cite{Lees:2012ju, Aubert:2009wt} allows us to impose bounds on the effective couplings $\alpha_{\mathcal{J}}$ involved in the $B \to \tau N$ decay.

The expression for the effective decay width $\Gamma_{B \to \tau N}$ is given below. The details of the calculation\footnote{Changing $\mu \to \tau$ in the given formulas.} can be found in our ref. \cite{Duarte:2019rzs}. The result is
\begin{eqnarray}\label{eq:BtotauN_width}
\Gamma_{B \to \tau N}&=&\frac{1}{16\pi m_B}\left(\frac{f_B m_B^2}{2 \Lambda^2} \right)^2 \left\{ |A_V|^2 \left[
(1+x_{\tau}^{2}-x_N^{2})(1-x_{\tau}^{2}+x_N^{2})-(1-x_{\tau}^{2}-x_N^{2})\right] 
\right. 
\nonumber \\
&+& \left.
|A_S|^2\frac{(1-x_{\tau}^{2}-x_N^{2})}{(x_u+x_b)^2} + (A^*_S A_V+A^*_V A_S) \frac{x_{\tau}}{(x_u+x_b)}(1-x_{\tau}^{2}+x_N^{2}) \right\}
\nonumber \\
& \times & ((1-x_{\tau}^{2}+x_N^{2})^2-4 x_N^{2})^{1/2},
\end{eqnarray}
where $x_{\tau}= m_{\tau}/m_B , \;\; x_N=m_{N}/m_B,  \;\; x_{u} = m_{u}/m_B, \;\; x_{b}=m_{b}/m_B$,
\begin{displaymath}
 A_V =  \left( \alpha^{(3)}_{V_0} Y^{ub}_{RR} + \alpha_W^{(3)} V^{ub} \right) \;\text{and}\;\;  
 A_S =  \left( \alpha^{(3)}_{S_1} Y^{ub}_{RL} + (\alpha^{(3)}_{S_2}  +   \frac12 \alpha^{(3)}_{S_3}) Y^{ub}_{LR} \right). 
\end{displaymath}
Here $f_B$ is the B meson decay constant. 

The effective couplings in $A_{V,S}$ -as the subscript indicates- correspond to \emph{vectorial} and \emph{scalar} interactions. We find that, due to the pseudo-scalar nature of the B meson the scalar operators contribution to the $B\to \tau N$ decay width is enhanced with respect to the vectorial contribution. This leads to the presence of the quark masses $m_u$ and $m_b$ in the denominator of the scalar term in \eqref{eq:BtotauN_width}. In turn, this will enable us to put more stringent bounds on the scalar operators contributions to this process \cite{Duarte:2019rzs}. 

In order to find the allowed values for the effective couplings in this context, we compare the experimental value $Br^{exp}(B \to \tau \nu)= (1.09 \pm 0.24)\times 10^{-4}$ to the theoretical prediction
 $Br^{th}= (\Gamma^{SM}_{B \to  \tau \nu}+ \Gamma_{B \to  \tau N})/\Gamma^{total}_{B} $ subject to the constraint $\tau_{N}> 10^{3} ~ps$ on the $N$ lifetime, enforcing it to escape undetected \cite{Aaij:2014aba}. 

The SM result for the branching ratio is 
\begin{equation}
 Br^{SM}_{B \to  \tau \bar{\nu}_{\tau}} = \frac{\Gamma^{SM}}{\Gamma_B}= \frac{G^2_F ~ m_B ~ m^{2}_{\tau} }{8 \pi \Gamma_B} \left(1- \frac{m^{2}_{\tau}}{m^{2}_{B}}\right)^{2} f^2_B ~|V^{ub}|^{2}.  
\end{equation}
The uncertainty in this calculation comes from the B meson decay constant $f_B$, the CKM mixing $V^{ub}$, the total width $\Gamma_B$ and the values of the masses. Summing in quadrature the uncertainties in the SM ($\Delta Br^{SM}$) and experimental ($\Delta Br^{exp}=0.24\times 10^{-4}$) values, we find the region in the $(m_N, \alpha)$ plane consistent with the conditions
\begin{equation}
 Br^{exp}-\sqrt{(\Delta Br^{SM})^2+ (\Delta Br^{exp})^2} \leq  Br^{th} \leq Br^{exp}+ \sqrt{(\Delta Br^{SM})^2+ (\Delta Br^{exp})^2}
\end{equation}
and $\tau_{N}> 10^{3} ~ps$.  

In the numerical calculation we consider all the values of the effective couplings to be equal: $\alpha_V^{(i)}=\alpha_S^{(i)}=\alpha_T^{(i)}= \alpha$, and let them be on/off depending on the benchmark set defined in Tab.\ref{tab:alpha-sets} considered. As we need the Majorana $N$ to decay outside the detector, the sets 1, 2 and 3 do not give a contribution, as the $N$ decays very fast if the radiative $N\to \nu \gamma$ channel is available. Thus we are left with sets 4, 5 and 6.

\begin{figure*}
\centering
\subfloat[Set 4: vector and scalar operators.]{\label{fig:Bnutau_pol_s4}\includegraphics[totalheight=6.cm]{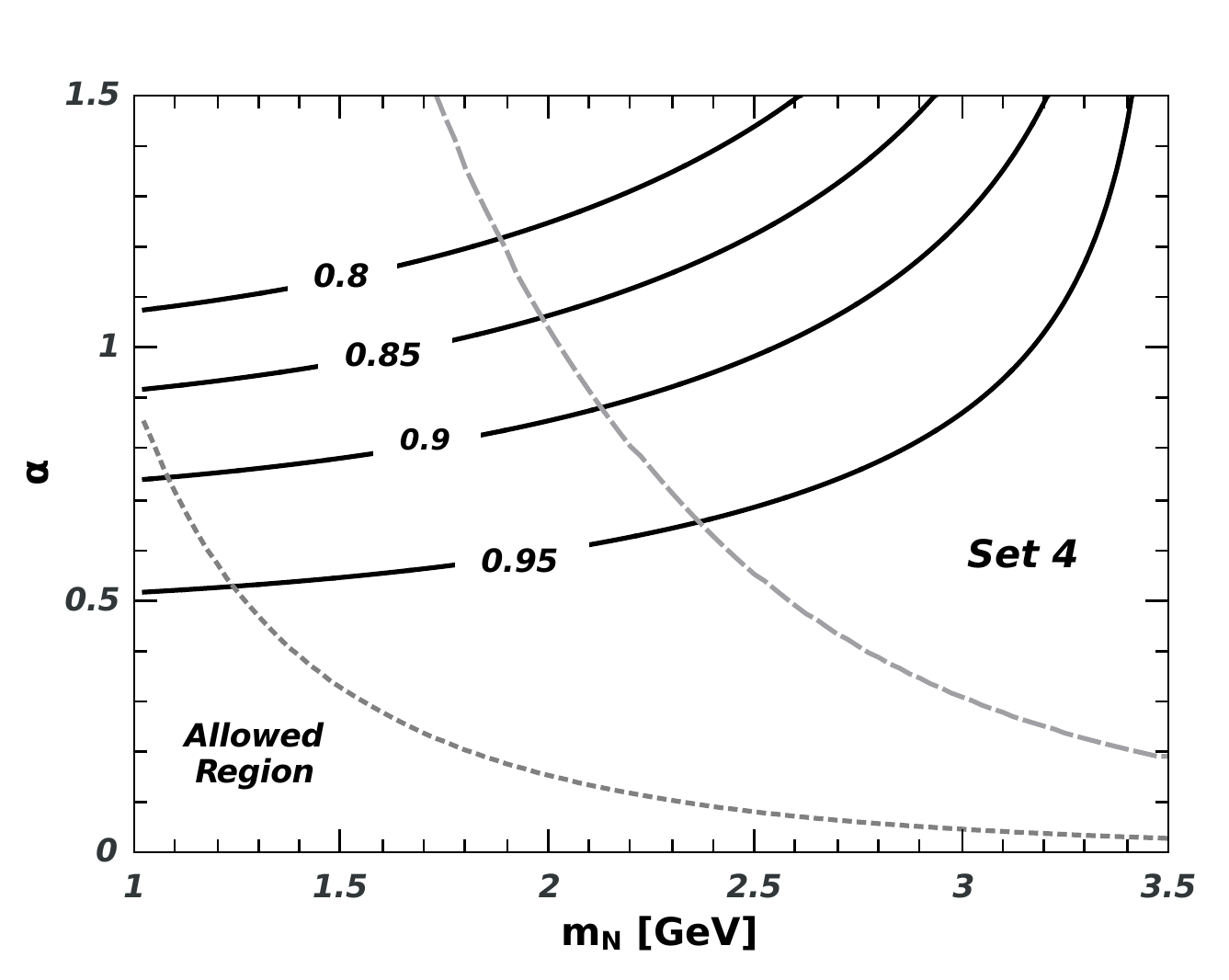}}~
\subfloat[Set 6: only scalar operators.]{\label{fig:Bnutau_pol_s6}\includegraphics[totalheight=6.cm]{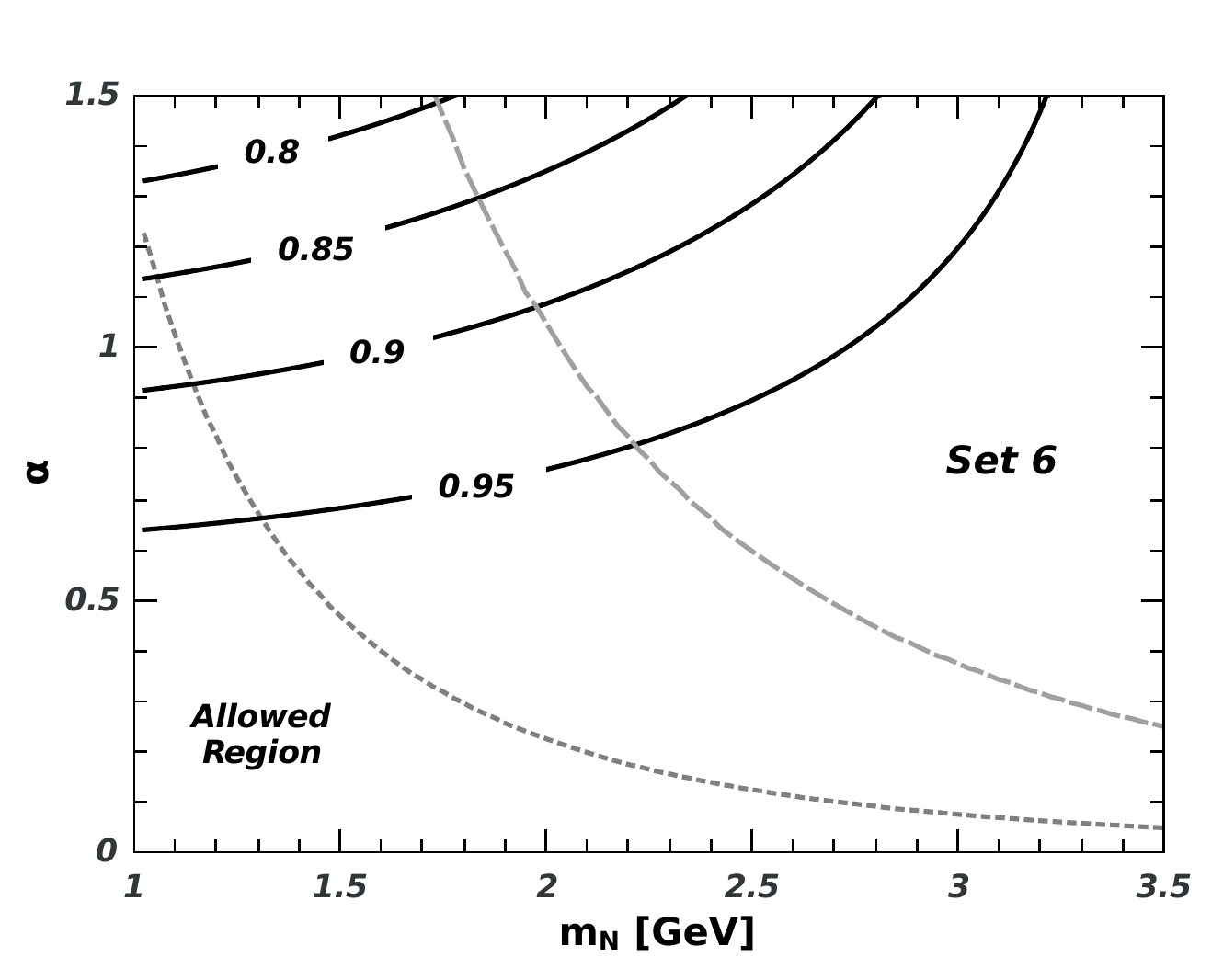}}
\caption{Tau polarization $P_{\tau}$ contours (solid black), and allowed regions in the $(m_N,\alpha)$ plane (dotted and dashed) for $B\to \tau \nu$ decay.}\label{fig:Bnutaupol}
\end{figure*}

The allowed regions are shown in Fig.\ref{fig:Bnutaupol}, for sets 4 and 6. The left panel shows the results for set 4, which includes all the scalar and vector operators, while the right panel shows the result for set 6, where only scalar interactions are active. As we mentioned above, the constraints for the vectorial operators in set 5 are loose as compared to the scalar operators, and we do not show them here. The dashed border (light gray) is obtained when we consider that only the third lepton family (tau) can have effective interactions: this is implemented numerically taking $\alpha^{(1,2)}_{S}=\alpha^{(1,2)}_{V}=0$, with only $\alpha^{(3)}_{S}=\alpha^{(3)}_{V}= \alpha \neq 0$. The dotted border (dark gray), includes the three fermion families in both sets. As can be seen in the plot, this condition reduces the allowed region, because the $N$ lifetime is shortened when it can decay to the light charged leptons, which is not allowed if only the third family interactions are included.

We now turn to calculate the polarized $B \to \tau \nu$ and $B \to \tau N$ decays. We will consider the final tau polarization observable $P_{\tau}$ in terms of the decay widths $\Gamma(h)$, with $h=\pm 1$ for the positive and negative tau helicity states. Thus we write
\begin{equation}
 P_{\tau}=\frac{\Gamma(+1)-\Gamma(-1)}{\Gamma(+1)+\Gamma(-1)}.
\end{equation}
We will also assume that tau decays preserve $CP$ invariance and therefore the distributions for $h=\pm$ anti-taus $\tau^{+}$ follow those of $h=\mp$ taus $\tau^{-}$. 

The SM calculation result, summarized in the Appendix \ref{sec:apx_polarSM}, is $P_{\tau}^{SM}=1$. The expression for the polarized decay width $\Gamma(h)_{B\to \tau N}$ is 
\begin{eqnarray}\label{eq:BtotauN_h}
&&\Gamma(h)_{B\to \tau N}=\frac{1}{32\pi m_B}\left(\frac{f_B m_B^2}{2 \Lambda^2} \right)^2 \left\{ |A_V|^2 \left[
x_{\tau}^{2}+x_N^{2}-(x_{\tau}^{2}-x_N^{2})(x_{\tau}^{2}-x_N^{2}+ h R)\right] 
\right. 
\nonumber \\
&+& \left.
 |A_S|^2\frac{1}{(x_u+x_b)^2} (1-x_{\tau}^{2}-x_N^{2}-h R) + (A^*_S A_V+A^*_V A_S) \frac{x_{\tau}}{(x_u+x_b)}(1-x_{\tau}^{2}+x_N^{2}-h R) \right\}
\nonumber \\
& \times & ((1-x_{\tau}^{2}+x_N^{2})^2-4 x_N^{2})^{1/2},
\end{eqnarray}
where $R=[(1-x_N^2)^2-2(1+x_N^2)x_{\tau}^2+x_{\tau}^4]^{1/2}$ and we use the notation in \eqref{eq:BtotauN_width}. We find that the pure scalar contribution in the second term is divided by the squared sum of the quark mass quotients $x_u$ and $x_b$. This enhances its contribution (and also the contribution of the second mixed term) compared to the pure vectorial contribution of the first term.

The contour curves in the $(m_N, \alpha)$ plane that give values of $P_{\tau}=0.8, ~0.85, ~0.9, ~0.95$ are shown in Fig.\ref{fig:Bnutaupol} for the sets 4 and 6 in solid black lines. While set 4 takes into account both the vectorial and scalar operators contribution to the to the polarized $B\to \tau N$ decay, the set 5 only considers the vectorial contributions, and set 6 the scalars. As we saw from eq. \eqref{eq:BtotauN_h}, the departure from the SM value of the polarization ($P^{SM}_{\tau}=1$) is expected to be dominated by the scalar contribution present in sets 4 and 6. 

\section{B radiative decay: lepton-photon angular asymmetry} \label{sec:bdec_rad}

In this section we investigate the sensitivity of a Forward-Backward asymmetry between the flight directions of the final lepton and photon in $B\to \ell \nu \gamma$ to the different kinds of -vectorial or scalar- effective interactions.  

For the case of the B leptonic radiative decay, we consider the lepton number conserving (LNC) and violating (LNV) contributions to $B^{-}\to \ell^{-} \bar{\nu} \gamma $ and $ B^{-}\to \ell^{-} \nu \gamma$, as in the experiment we cannot distinguish between neutrinos and anti-neutrinos in the final state. 

 \begin{figure*}[h]
 \centering
  \includegraphics[width=0.8\textwidth]{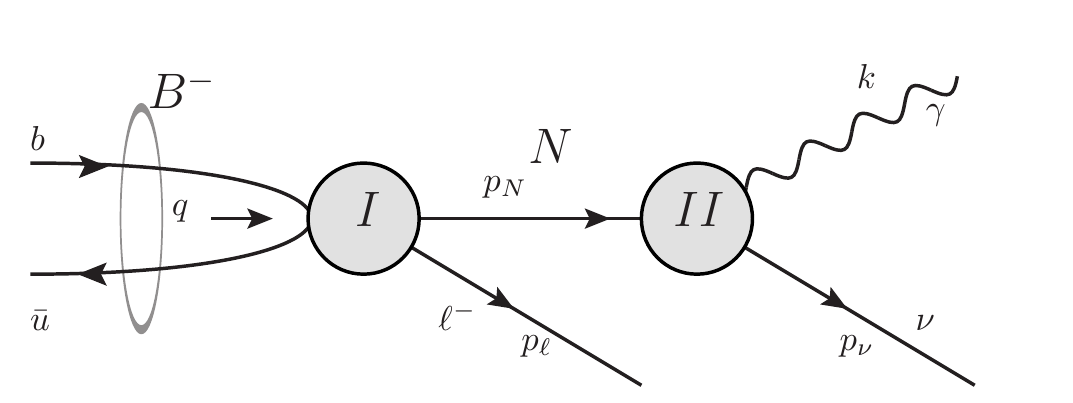}
 \caption{\label{fig:Blnugamma} Schematic representation for the effective contribution to the decay $B^-\to \ell^-\nu \gamma$.}
 \end{figure*}
%

In the case of a specific final lepton flavor, take $\ell=\mu$, we consider the amplitude in the SM, which conserves lepton number: $B^{-} \to \mu^{-} \bar{\nu}_{\mu} \gamma$ ($\mathcal{M}^{SM}_{\bar{\nu}_{\mu}}$) and the one mediated by the Majorana neutrino represented in Fig.\ref{fig:Blnugamma}. This effective process gives a LNC contribution to the same final state ($\mathcal{M}^{N}_{\bar{\nu}_{\mu}}$) together with LNC contributions to final states with the other two anti-neutrino flavors ($\mathcal{M}^{N}_{\bar{\nu}_{(e,\tau)}}$) and the LNV $B^{-} \to \mu^{-} \nu_{j} \gamma$ ($\mathcal{M}^{N}_{LNV}$) with final neutrinos. The first two must be added coherently\footnote{In the Majorana case only the $ \mu \overline{\nu}_{\mu} \gamma $ final state adds coherently to the SM amplitude.} in order to calculate the squared matrix element, while the last must be added incoherently. Thus we write:
\begin{equation}
 |\mathcal{M}_{B^{-}\to \ell^{-} \nu(\bar{\nu}) \gamma}|^2= | \mathcal{M}^{SM}_{\bar{\nu}_{\mu}} + \mathcal{M}^{N}_{\bar{\nu}_{\mu}}|^2 + | \mathcal{M}^{N}_{\bar{\nu}_{(e,\tau)}}|^2 + | \mathcal{M}^{N}_{{LNV}}|^2. 
\end{equation}
The SM amplitude is written following the treatment in \cite{Beneke:2011nf}
\begin{equation}\label{eq:Mlnug_SM}
\mathcal{M}^{SM}= e G_F V_{ub} \sqrt{2} \left(\overline{u_{\ell}}(p_{\ell}\right) \gamma^{\alpha} v_{\nu}(p_{\nu}))~ \varepsilon^{* \delta}(k) \left[f_V ~\epsilon_{\alpha \delta \rho \omega}~ v^{\rho} k^{\omega} -i f_A ~(g_{\alpha \delta}~ v.k - v_{\alpha} k_{\delta})\right],
\end{equation}
with the B meson momentum $q^{\rho}= m_B v^{\rho}$, $p_{\ell}$ and $p_{\nu}$ the final lepton and anti-neutrino momenta, and $k$ the photon momentum. The constants $f_V$ and $f_A$ are the vector and axial-vector form factors in \cite{Beneke:2011nf, Wang:2018wfj}, $e$ is the electron charge and $G_F$ is the Fermi constant.  

The LNC Majorana contribution to the amplitude is
\begin{equation}\label{eq:munugamma_LNC}
 \mathcal{M}^{N}_{{LNC}}= \frac{i}{2} \mathbf{C} f_B P_{N}(p_{N}^{2}) k^{\alpha} \varepsilon^{* \delta}(k) \left( \overline{u_{\ell}}(p_{\ell}) (C_{V} \slashed{q} + C^B_{S}) \slashed{p}_{N} \sigma^{\alpha \delta} P_{L} v_{\nu}(p_{\nu})\right)
\end{equation}
and the LNV part is 
\begin{equation}\label{eq:munugamma_LNV}
\mathcal{M}^{N}_{LNV}= -\frac{i}{2} m_N \mathbf{C} f_B P_{N}(p_{N}^{2}) k^{\alpha} \varepsilon^{* \delta}(k) \left( \overline{u_{\ell}}(p_{\ell}) (C_{V} \slashed{q} + C^B_{S}) \sigma^{\alpha \delta} P_{R} v_{\nu}(p_{\nu})\right). 
\end{equation}
Here $p_{N}$ is the intermediate Majorana momentum and $P_{N}(p_{N}^{2})= 1/(p_{N}^{2}-m_N^2)$. The coefficients are $~\mathbf{C}=\frac{v \sqrt{2}}{\Lambda^4}(\alpha_{NB}^{(j)} c_W +  \alpha_{NW}^{(j)} s_W)$, 
\begin{displaymath}
 C_{V}= \left(\alpha_{V_0}^{(i)} Y^{ub}_{RR} + \alpha_{W}^{(i)} V^{ub}\right) \;\text{and}\;\;  C^{B}_{S}=\left(\alpha_{S_1}^{(i)} Y^{ub}_{RL}+ (\alpha_{S_2}^{(i)}+ \frac12 \alpha_{S_3}^{(i)}) Y^{ub}_{LR} \right)\frac{m_B^2}{m_{u}+m_{b}}.  
\end{displaymath}
The indices $(j)$ in $\mathbf{C}$ correspond to the final (anti-)neutrino flavor, and $(i)$ in $C_V$ and $C^{B}_{S}$ to the final charged lepton flavor. A sum over the corresponding indices $j=1,2,3$ in \eqref{eq:munugamma_LNV} is understood. The details of the calculation can be found in the Appendix C in our ref. \cite{Duarte:2019rzs}. 

With the amplitudes above, we perform a phase space Monte Carlo integration, and use it to calculate a Forward-Backward asymmetry between the final charged lepton and photon flight directions in the B rest frame, which can be used as an observable to disentangle the effective Majorana from the SM contribution to the leptonic radiative B decay. 

\subsubsection{Forward-Backward lepton-photon asymmetry}\label{sec:subsec_AFB_lg}

The Forward-Backward asymmetry $A_{FB}^{\ell \gamma}$ between the final charged lepton and photon flight directions is defined as 
\begin{equation}\label{eq:AFB_lg}
 A_{FB}^{\ell \gamma}= \frac{N_{+}-N_{-}}{N_{+}+N_{-}}, 
\end{equation}
where $N_{\pm}$ is the number of events with a positive (negative) value of $\cos(\theta)$, the angle between the final lepton and the photon flight directions in the B meson rest frame.  

\begin{figure*}[h]
 \centering
  \subfloat[]{\label{fig:AFBs1v0}\includegraphics[totalheight=6.cm]{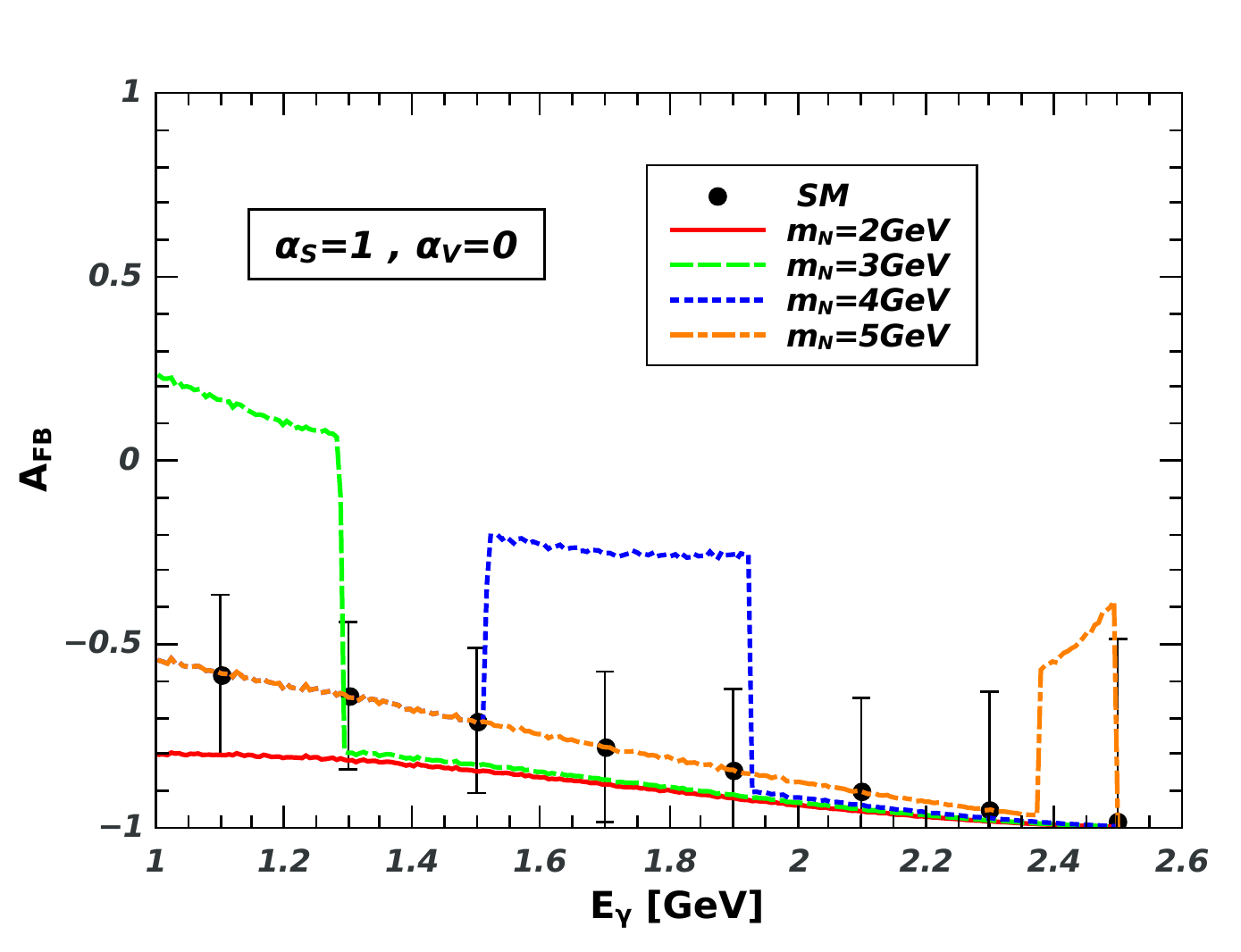}}~
  \subfloat[]{\label{fig:AFBs0v1}\includegraphics[totalheight=6.cm]{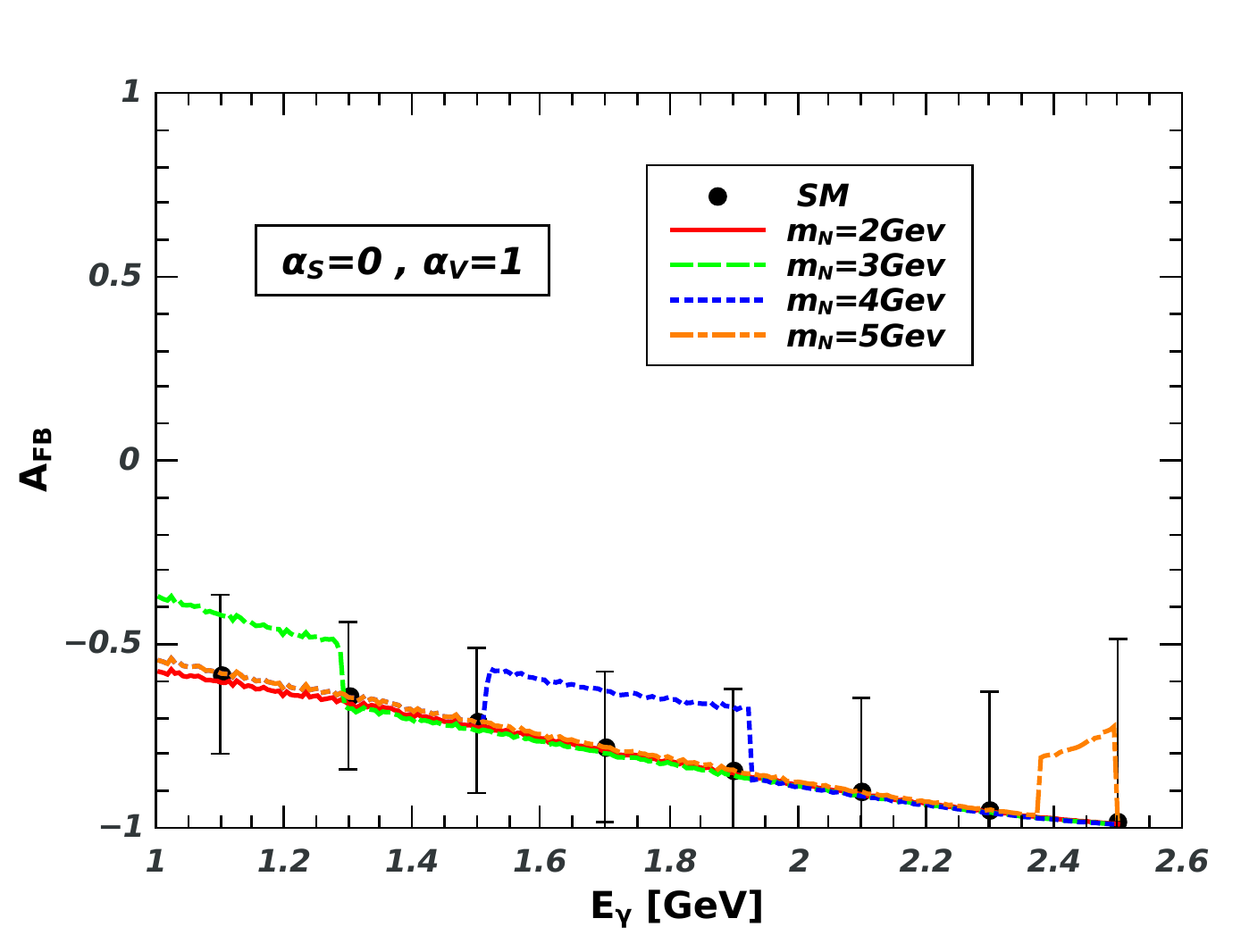}}
 \caption{\label{fig:AFB_E} $A_{FB}^{\ell \gamma}$ asymmetry for the SM process $B^{-}\to \mu^{-} \bar{\nu}_{\mu} \gamma$, for each photon energy bin $E_{i}$.}
 \end{figure*}

In Fig.\ref{fig:AFB_E} we show our results for the value of the $A_{FB}^{\ell \gamma}$ asymmetry for the SM process $B^{-}\to \mu^{-} \bar{\nu}_{\mu} \gamma$ considering a final muon, for each photon energy bin $E_{\gamma}$ (black dots with error bars). The SM asymmetry value is negative in all the allowed photon energy range. The photon has maximum energy when the muon and neutrino have parallel momenta, opposed to $\vec{p_{\gamma}}$. In this case, the asymmetry value is strictly $A_{FB}^{\ell \gamma}=-1$, as can be seen in the plot. In the massless muon limit, the energy allowed for the photon in the B rest frame is
\begin{equation}\label{eq:E_gamma_max}
 E_{\gamma}= \frac{m_B}{2}+ \frac{E_{\mu}E_{\nu}}{m_B}\left(\cos(\varphi)-1\right) 
\end{equation}
where $\varphi$ is the angle between the muon and neutrino momenta $\vec{p_{\mu}}$ and $\vec{p_{\nu}}$. Thus the photon has maximum energy when the muon and neutrino have the same flight directions in the B rest frame, meaning the photon and muon momenta are back-to-back.   

The error bars $\delta_i$ for the SM calculation in each photon energy bin are obtained by adding in quadrature the statistical and theoretical uncertainties in the numbers of events $N_{\pm}$. The leading theoretical error comes from the uncertainty in the values of the form factors $f_A$ and $f_V$ in \eqref{eq:Mlnug_SM}, which are taken from \cite{Wang:2018wfj}.

 \begin{figure*}[h]
 \centering
  \subfloat[$E_{\gamma}$ allowed region, as a function of $m_N$.]{\label{fig:reg_ene}\includegraphics[totalheight=6.cm]{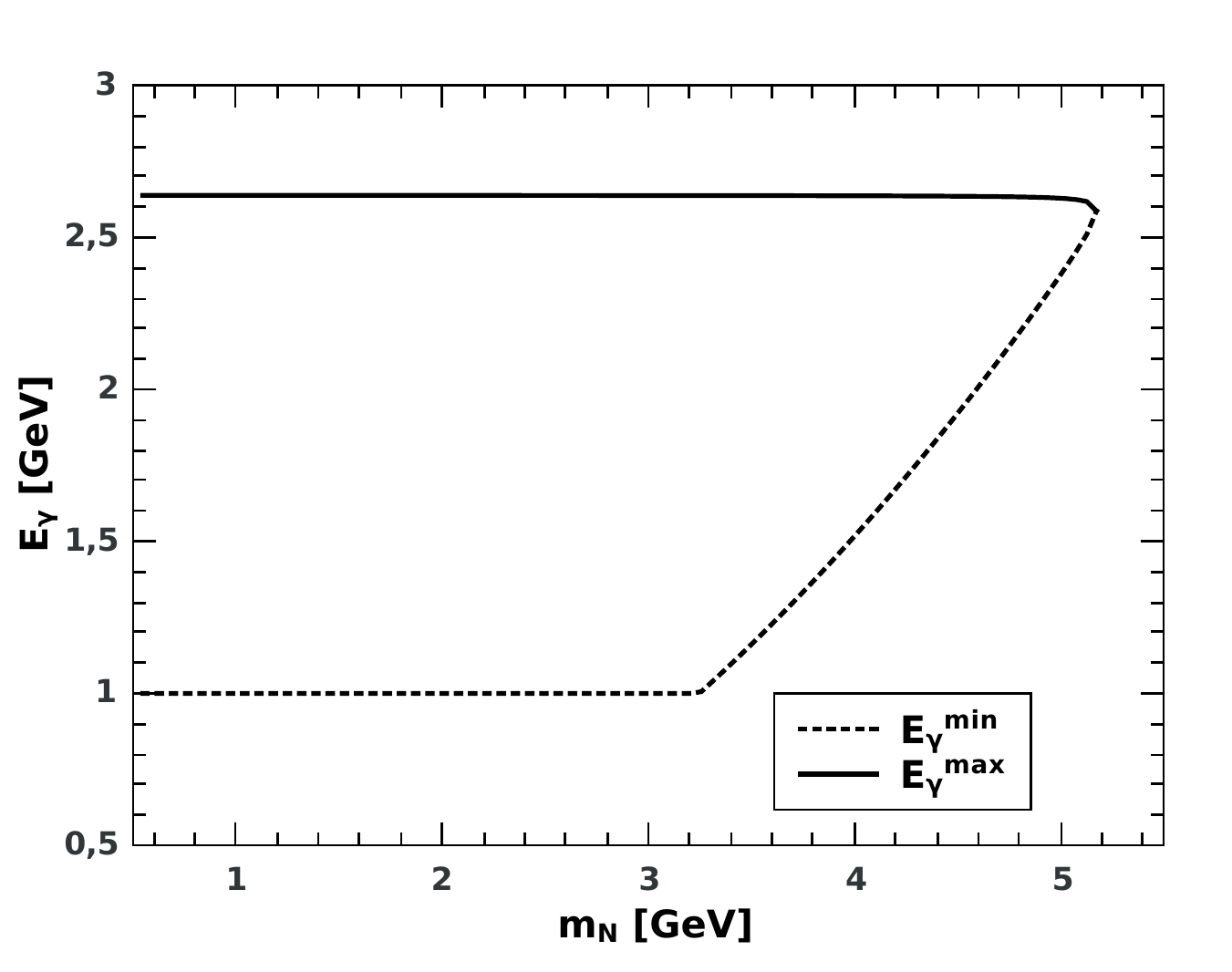}}~
  \subfloat[Angle between $\vec{p_{\mu}}$ and $\vec{p_{\gamma}}$ as a function of $E_{\gamma}$. ]{\label{fig:cos_theta}\includegraphics[totalheight=6.cm]{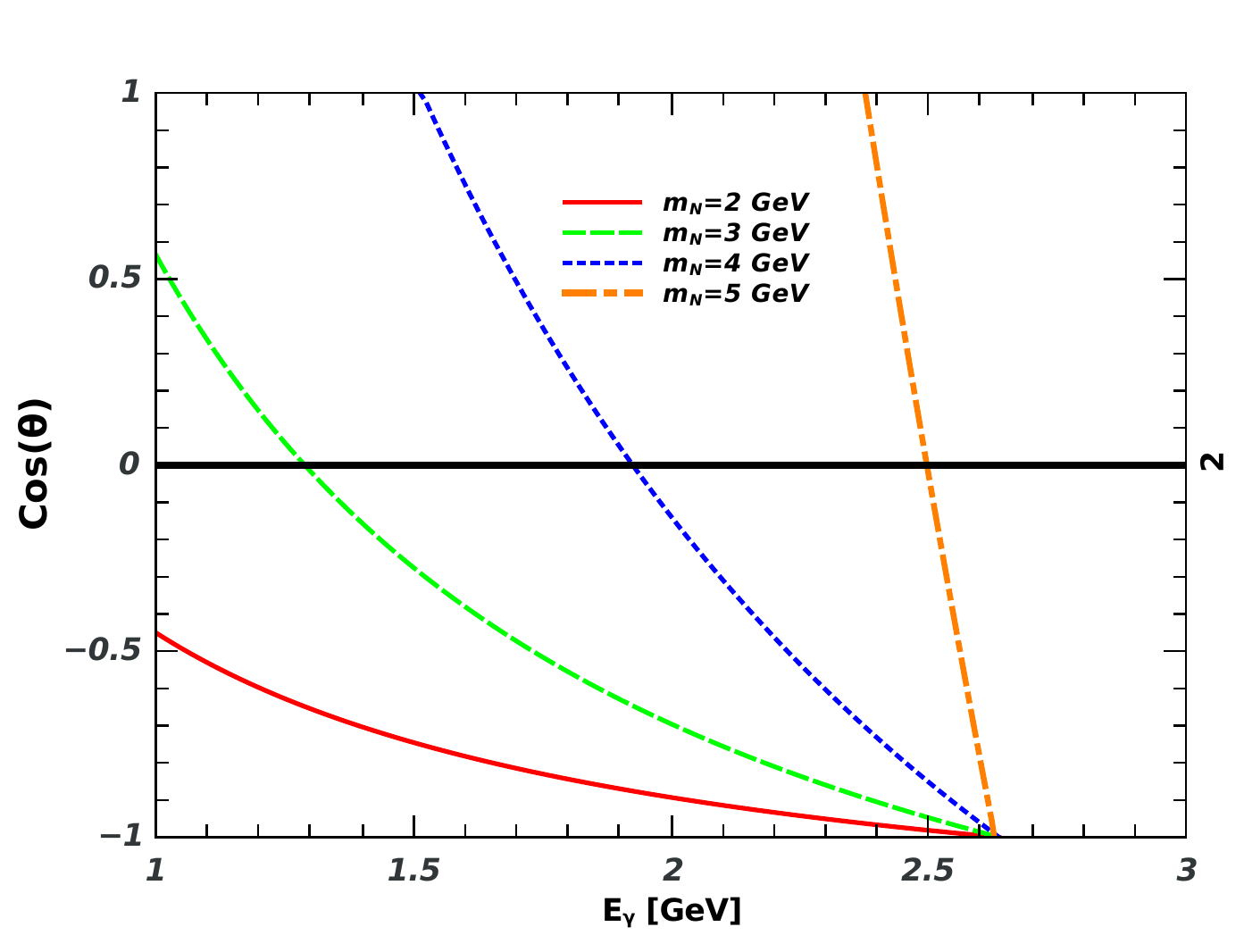}}
  \caption{\label{fig:Eg_limit} $B^-\rightarrow  \mu^- \nu \gamma$ kinematics in the B rest frame.}
 \end{figure*}

\subsubsection{Sensitivity to effective couplings}

The numerical value of the effective contribution to the Forward-Backward asymmetry $A_{FB}^{\ell \gamma}$ depends on the Majorana neutrino mass $m_N$ and the effective couplings $\alpha_{\mathcal{J}}$.

If the number of events in Belle II allows us to detect the $B^{-}\to \mu^{-} \nu (\bar\nu) \gamma$ process, and measure the Forward-Backward lepton-photon asymmetry in \eqref{eq:AFB_lg}, we would like to see which values of the effective couplings $\alpha_{\mathcal{J}}$ would be compatible (or not) with the SM prediction. 

As we saw previously, the effective radiative decay can only occur when the tensorial couplings $\alpha_{T}$ are included in the calculation. However, the vector and scalar contributions to the numerical value of the asymmetry can change, predicting different results. For the sake of concreteness, we will consider a scenario where the tensorial couplings take a numerical value as $\alpha_{T}= \frac{|\alpha_{V}|+|\alpha_S|}{2}$ to avoid unnatural cancellations, and let the numerical values of $\alpha_V$ and $\alpha_S$ change freely. This means that when both $\alpha_V=\alpha_S=0$ there is no effective contribution to $A_{FB}^{\ell \gamma}$.

In Fig.\ref{fig:AFBs1v0} we show the result for the sum of the effective and SM contributions $A^{Tot}_{FB}$ to the asymmetry for four different $m_N$ values taking into account only scalar contributions, fixing $\alpha_S=1$ and $\alpha_V=0$. Fig.\ref{fig:AFBs0v1} shows the result for pure vectorial contributions, fixing $\alpha_S=0$ and $\alpha_V=1$. 

The different curves in both panels in Fig.\ref{fig:AFB_E} show that the effective contribution to the asymmetry is moderate, except for an interval in the photon energy $E_{\gamma}$ range, where it steps towards less negative values. This means that for this energy interval, some of the effective events contribute with a positive value of $\cos(\theta)$ to the asymmetry. This interval is different for different $m_N$ values. 

The effect can be explained by kinematical reasons. In the effective $B \to \mu N \to \mu \nu \gamma$ decay, for each intermediate $m_N$ mass, the photon energy $E_{\gamma}$ in the B meson rest frame can take the values shown in Fig.\ref{fig:reg_ene}. This is due in part to the cut value $E_{cut}=1~GeV$ introduced to ensure the validity of the QCD treatment in the calculation \cite{Wang:2018wfj}, and to the minimal allowed energy $E_{\gamma}^{min}=E_N (1+ \beta_N)/2$ the photon can have in the B rest frame, according to the energy $E_N$ and boost velocity $\beta_N$ of the intermediate Majorana neutrino. The maximal energy is given by the formula corresponding to \eqref{eq:E_gamma_max} for $m_{\mu} \neq 0$. Thus, the effective contribution to the asymmetry only exists in the region between the curves in Fig.\ref{fig:reg_ene}. The minimum value of $E_{\gamma}$ in Fig.\ref{fig:reg_ene} for each $m_N$ is the value where the effective contribution starts \cite{Duarte:2019rzs}. This contribution stops to rise the asymmetry value when the scalar product $ \vec{p_{\mu}}. \vec{p_{\gamma}}= |\vec{p_{\mu}}| |\vec{p_{\gamma}}| \cos(\theta)$ starts to be negative again:
\begin{equation}\label{eq:costheta}
 \cos(\theta)= \frac{1}{\beta_{N}}\left(\frac{m_N}{2 \gamma_N E_{\gamma}}-1\right).
\end{equation}
In Fig.\ref{fig:cos_theta}, we plot eq. \eqref{eq:costheta} for the $m_N$ values in Fig.\ref{fig:AFB_E} as an aid to visualize the $E_{\gamma}$ value where the sign of $\cos(\theta)$ starts to be negative. 

\begin{figure*}
\centering
\subfloat[]{\label{fig:chi_m2}\includegraphics[totalheight=6.cm]{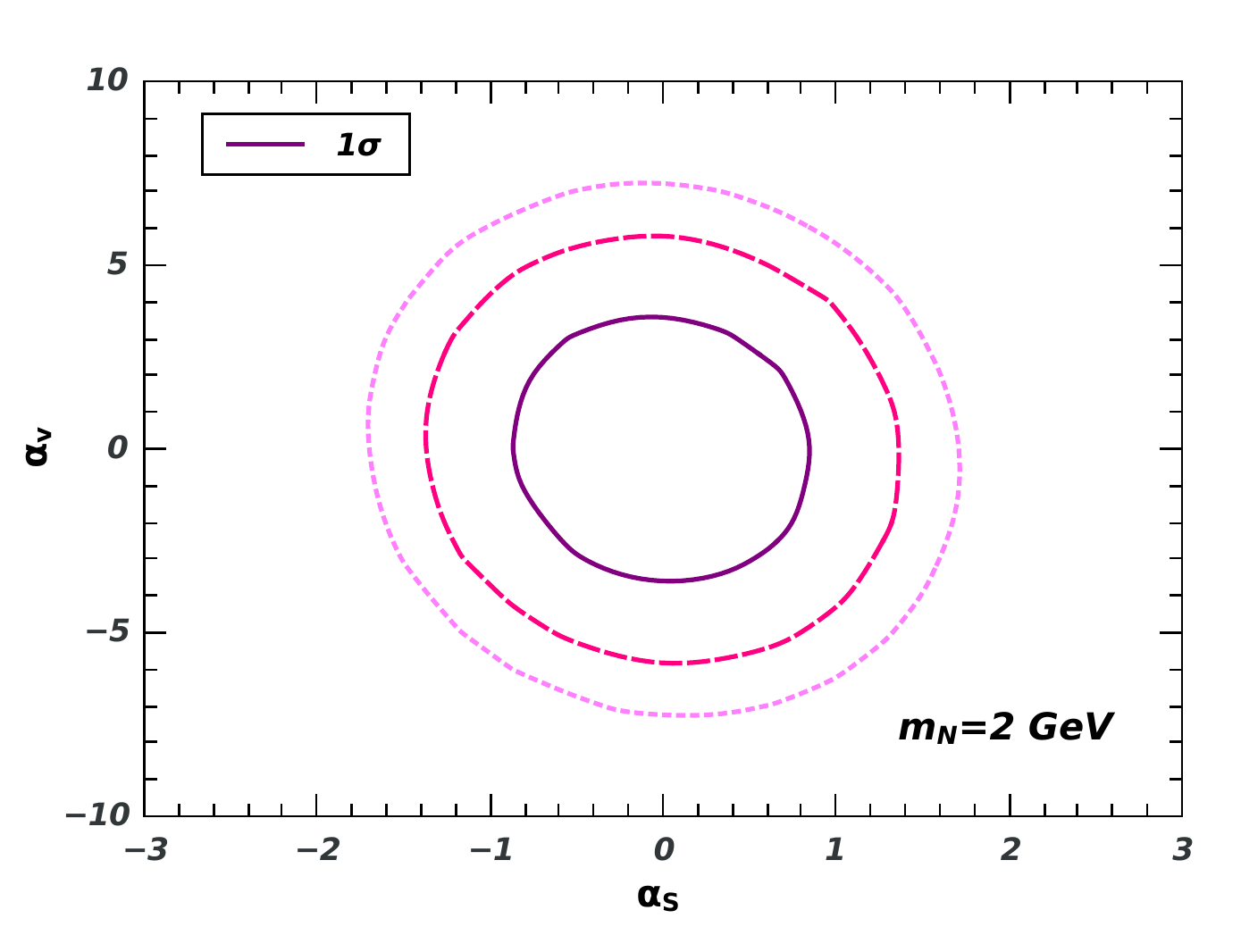}}~
\subfloat[]{\label{fig:chi_m3}\includegraphics[totalheight=6.cm]{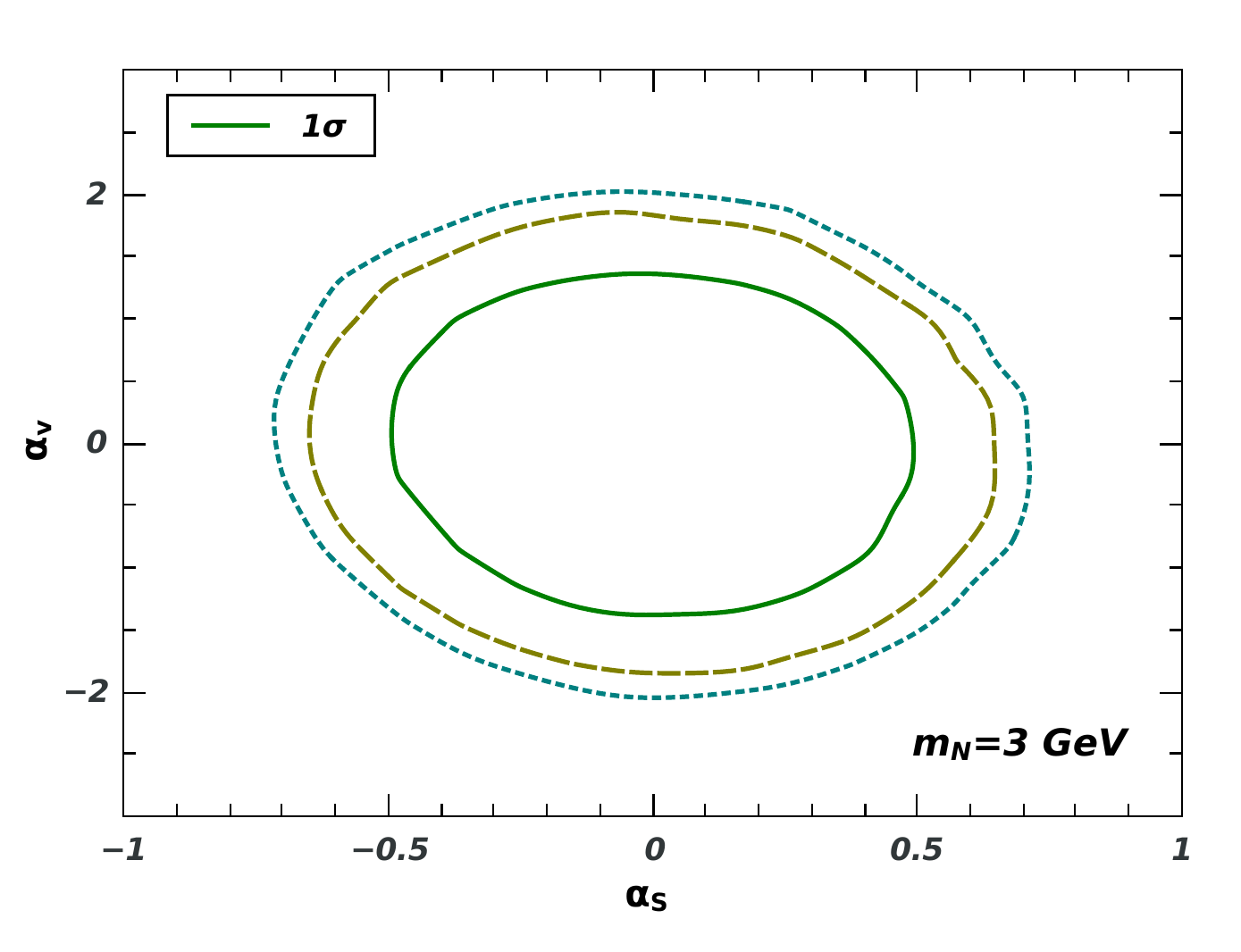}}

\subfloat[]{\label{fig:chi_m4}\includegraphics[totalheight=6.cm]{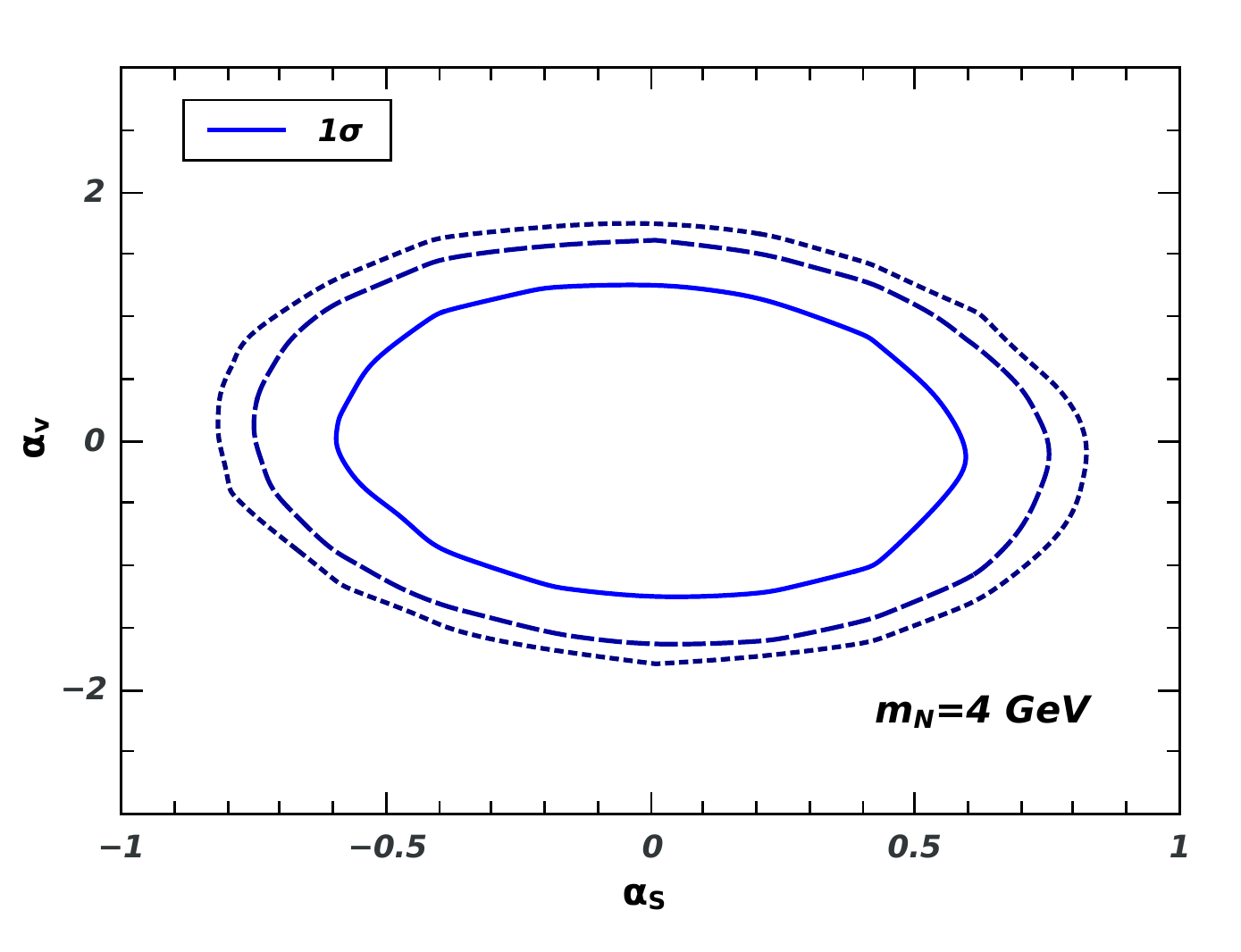}}~
\subfloat[]{\label{fig:chi_m5}\includegraphics[totalheight=6.cm]{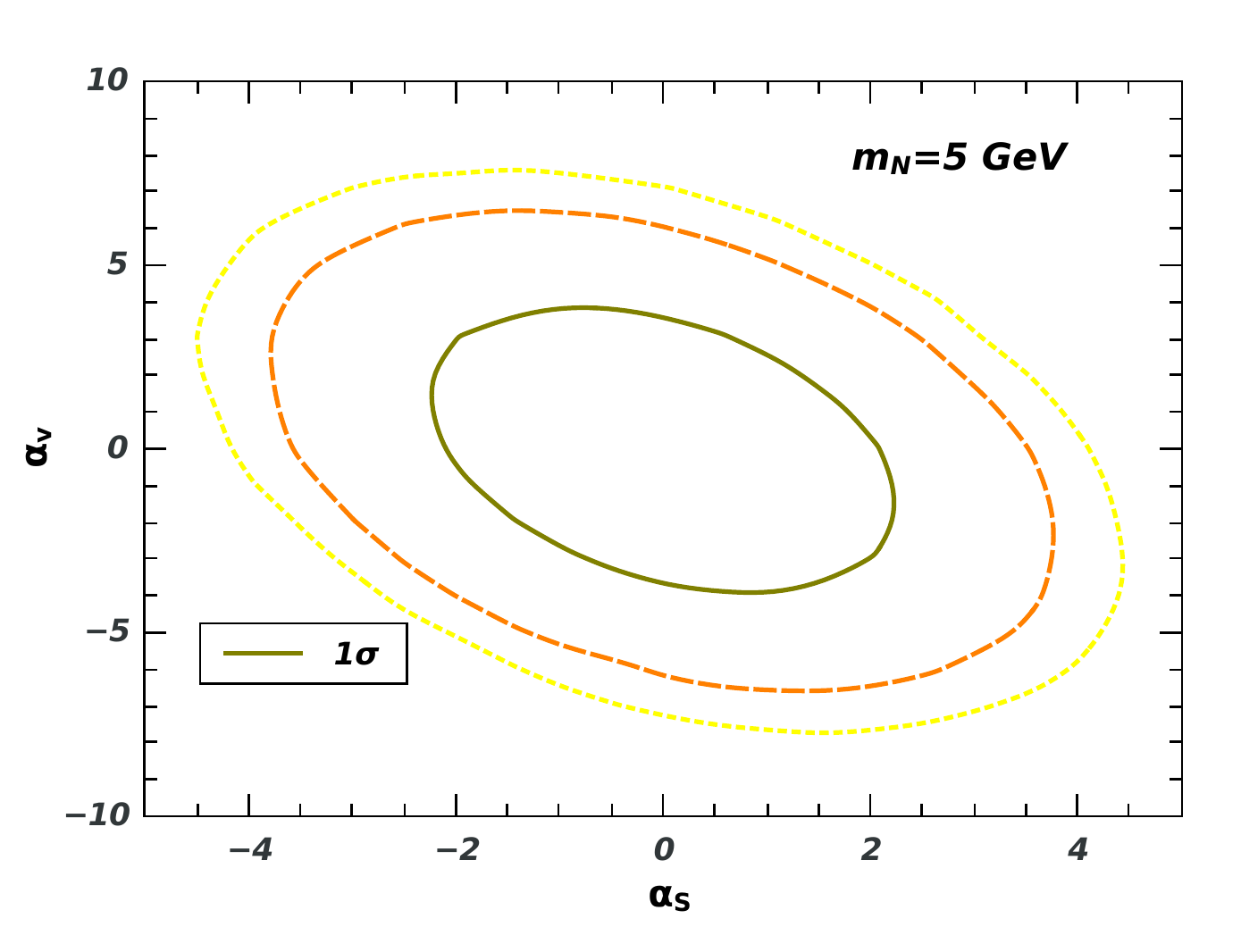}}
\caption{Sensitivity of $A_{FB}^{\ell \gamma}$ to the effective couplings.}\label{fig:chi2}
\end{figure*}

Let us consider the curves for $m_N=4~GeV$ in Fig.\ref{fig:AFB_E}. The step in both panels starts at $E_{\gamma}= 1.5 ~GeV$, which is the minimum value we obtain from the curve in Fig.\ref{fig:reg_ene}. At this energy $\cos(\theta)$ is positive (it is entering the allowed range $|\cos(\theta)|\leq 1$), and changes sign when $E_{\gamma}=1.95 ~GeV$, as can be seen in Fig.\ref{fig:cos_theta}. Here the step in the curves in Fig.\ref{fig:AFB_E} comes to an end. We also note that when we consider only vectorial operators, in Fig.\ref{fig:AFBs0v1}, the curve for $m_N=4 ~GeV$ stands between the SM value error bars, while the scalar contribution in Fig.\ref{fig:AFBs1v0} increases above them. This can be studied by considering a $\Delta \chi^2$ distribution to see which values of the effective couplings $\alpha_{\mathcal{J}}$ would be compatible (or not) with the SM prediction. We construct a $\Delta \chi^2$ function as 
\begin{equation}
 \Delta \chi^2= \sum_{E_i}\frac{ \left(A^{SM}_{FB}(E_i)- A^{Tot}_{FB}(E_i, m_N,\alpha_V, \alpha_S)\right)^2}{\delta_i^2},
\end{equation}
where $E_i$ are the photon energy bins shown in Fig.\ref{fig:AFB_E}, and $\delta_i$ are the errors for the SM prediction, and $A^{Tot}_{FB}$ is the sum of the SM and the effective value for $A_{FB}^{\ell \gamma}$ in each photon energy bin. In Fig.\ref{fig:chi2} we show the contours in the $(\alpha_S,\alpha_V)$ plane corresponding to 
o a coverage probability  $(1-\alpha)$ of 68.27\% ($1 \sigma$), 90\% and 95\% from the SM prediction for $A_{FB}^{\ell \gamma}$, for different $m_N$ values.

Let us consider the panel for $m_N=4~GeV$ in Fig.\ref{fig:chi_m4}. The point $\alpha_S=0, \alpha_V=1$ lies in the region inside the $1\sigma$ contour, and thus the effective contribution to $A^{Tot}_{FB}$ lies between the error bars in Fig.\ref{fig:AFBs0v1}, while the point $\alpha_S=1, \alpha_V=0$ is far outside the outer contour, and the effective contribution can be well distinguished from the pure SM curve in Fig.\ref{fig:AFBs1v0}. The same effect occurs for other $m_N$ values: the asymmetry is more sensitive to the scalar interactions contribution. This can be seen easily by inspection of eqs. \eqref{eq:munugamma_LNV} and \eqref{eq:munugamma_LNC}. The scalar contribution term $C^B_S$ is weighted by a factor of order $m_B/(m_u+m_b)$ with respect to the vectorial contribution, and this is again due to the pseudoscalar nature of the B meson.

\section{N mediated LNV B decays with final taus}\label{sec:NmedBdec_taus}

\begin{figure*}[]
 \centering
  \includegraphics[width=0.8\textwidth]{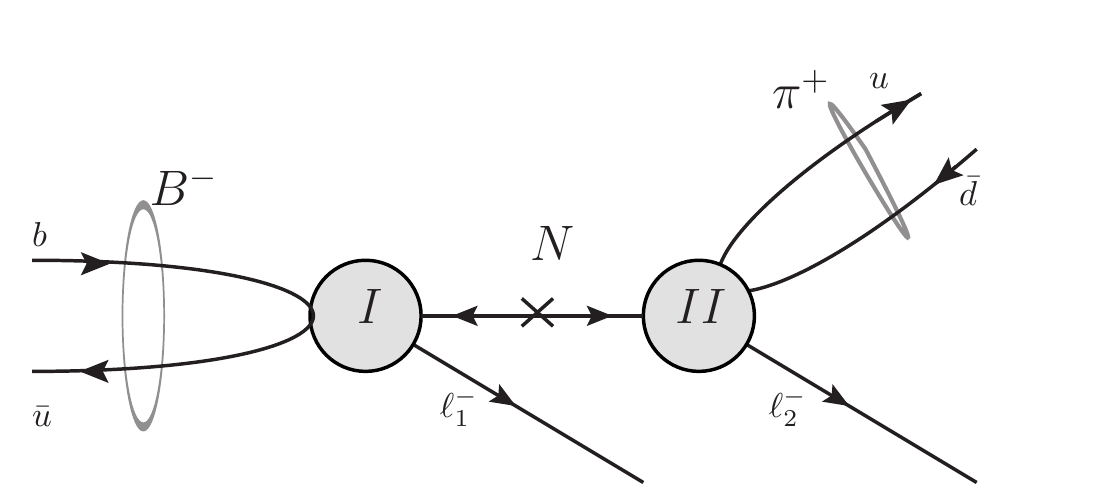}
 \caption{\label{fig:Bllpi} Schematic representation for the effective contribution to the decay $B^-\rightarrow  \ell^-_{1}\ell^-_{2} \pi^+$.}
 \end{figure*}

In this section we study the LNV decay $B^-\rightarrow  \ell^-_{1}\ell^-_{2} \pi^+$, mediated by a Majorana neutrino, as depicted in Fig.\ref{fig:Bllpi}. In the case where one or two of the final leptons are tau leptons, we would like to see if a measurement of their polarization could give a hint on the kind of new physics involved in the production or decay of the intermediate $N$. As we will see, the polarization of the final taus in these decays could be used to distinguish the vectorial and scalar operators contributions to the $N$ production and decay vertices, complementing those proposed for a higher $m_N$ scale in \cite{Duarte:2018kiv}.

As we want to keep track of the final taus polarization, we perform a calculation of the decay amplitude considering the intermediate $N$ to be near the mass shell, but without recurring to the narrow width approximation. The details of the calculation are summarized in the Appendix \ref{sec:apx_polar_LNV}.   

The amplitude for the LNV decay $B^-\rightarrow  \ell^-_{1}\ell^-_{2} \pi^+$ can be written as
\begin{equation}\label{eq:M_B-llpi}
 \mathcal{M}^{N}_{B^-\rightarrow  \ell^-_{1}\ell^-_{2} \pi^+}= -i \frac{f_{\pi} f_{B}}{4 \Lambda^4} m_N P_{N}(p_{N}^2)~ 
 \overline{u}(p_2) \left(C_V^{(II)} \slashed{k} - C_S^{\pi} \right) P_R \left(C_V^{(I)} \slashed{q} - C_S^{B} \right) v(p_1),
\end{equation}
with
\begin{displaymath}
 C_{V}^{(I)}= \left(\alpha_{V_0}^{(i)} Y^{ub}_{RR} + \alpha_{W}^{(i)} V^{ub}\right) \;\;\;\;  C^{B}_{S}=\left(\alpha_{S_1}^{(i)} Y^{ub}_{RL}+ (\alpha_{S_2}^{(i)}+ \frac12 \alpha_{S_3}^{(i)}) Y^{ub}_{LR} \right)\frac{m_B^2}{m_{u}+m_{b}}  
\end{displaymath}
and
\begin{displaymath}
 C_{V}^{(II)}= \left(\alpha_{V_0}^{(j)} Y^{ud}_{RR} + \alpha_{W}^{(j)} V^{ud}\right) \;\;\;\;  C^{\pi}_{S}=\left(\alpha_{S_1}^{(j)} Y^{ud}_{RL}+ (\alpha_{S_2}^{(j)}+ \frac12 \alpha_{S_3}^{(j)}) Y^{ud}_{LR} \right)\frac{m_{\pi}^2}{m_{u}+m_{d}}.  
\end{displaymath}
The indices $(i)$ and $(j)$ for the effective couplings correspond to the $\ell_1$ and $\ell_2$ flavors, respectively. In order to keep track of the final leptons polarizations, we use the relations in \eqref{eq:ApX_polar}, which allow us to perform a Monte Carlo simulation and obtain the number of decay events with positively or negatively polarized taus in the final state.  

We study the polarization of final taus in the processes depicted in Fig.\ref{fig:Bllpi}. The different possible processes  interchanging the final leptons flavor ($\ell_1=\tau, \mu$ and $\ell_2=\tau, \mu$) depend on the value of the mediator mass $m_N$. The possibilities for the different regions are summarized in Tab. \ref{tab:kinematic_regions}.

\begin{table}[h]
 \centering
 \begin{tabular}{| c | c | c|}
\firsthline
 \bf{I}  & $\;$ \bf{II} $\;$ & \bf{III} \\
\hline
 $m_{\pi}+m_{\mu} < m_{N} < m_{\pi}+m_{\tau}$  & $m_{\pi}+m_{\tau} < m_{N} < m_{B}-m_{\tau}$ &   $m_{B}-m_{\tau} < m_{N} < m_{B}-m_{\mu}$  \\  
 $\ell_{1}= \tau, $\; \;$ \ell_{2}= \mu $ & $\ell_{1}= \tau, \mu $\; \;$ \ell_{2}= \mu, \tau $   &  $\ell_{1}= \mu,  $\; \;$ \ell_{2}= \tau $     \\
\lasthline
 \end{tabular}
\caption{Kinematic regions}\label{tab:kinematic_regions}
\end{table}

In the case of only one final tau, we define the final-state polarization as
 \begin{equation}\label{eq:pol_tau_}
P_{\tau}=\frac{N_{+}-N_{-}}{N_{+}+N_{-}}=\frac{N_{+}-N_{-}}{N},
 \end{equation}
and the two-taus final state polarization in $B^{-} \to \tau^{-} \tau^{-} \pi ^{+}$ as
 \begin{equation}\label{eq:pol_taus}
 P_{\tau}=\frac{N_{++}+N_{+-}-N_{-+}-N_{--}}{N_{++}+N_{+-}+N_{-+}+N_{--}}.
 \end{equation}
Here the number of events ($N_{\pm}$, $N_{{\pm}{\pm} }$) with subscripts $+$ and $-$ correspond respectively to the number of $h=+$ and $h=-$ polarization states of each final tau $\ell_1$ or $\ell_2$ in Fig.\ref{fig:Bllpi} measured in the experiment, as considered in \eqref{eq:Apx_h}. The defined final state polarizations, being a quotient, are independent of the total number of B decay events considered.  


In our previous paper \cite{Duarte:2019rzs}, we studied the bounds that can be set on the different couplings $\alpha_{\mathcal{J}}$ in the effective dimension 6 Lagrangian \eqref{eq:lagrangian} involved in $N$ mediated B decays by exploiting the experimental results existing on the $B^{-} \to \mu^{-} \mu^{-} \pi^{+}$ \cite{Aaij:2014aba} and $B^{-} \to \mu^{-} \nu \gamma$ \cite{Gelb:2018end} processes. The Fig.6 in \cite{Duarte:2019rzs} shows different $U^2 (m_N)$ \eqref{eq:u2} curves for the upper bounds obtained for the $U^2$ combination varying with the $m_N$ mass. The most restrictive one corresponds to a benchmark scenario (set 1) where all the types of operators involved in the effective $B^{-} \to \mu^{-} \mu^{-} \pi^{+}$ decay process (vectorial, scalar and tensorial) are considered to be present in the Lagrangian. In this numerical calculation we use this curve $U^2 (m_N)$ to fix the value of the remaining couplings (generically $\alpha_{\mathcal{J}}$, not involved in $0\nu \beta \beta$-decay) inverting for the $\alpha(m_N)$ value in \eqref{eq:u2}. While this constraint is obtained from interactions involving the $\ell= \mu, ~i=2$ family, we will use it here to fix also the $\alpha^{(3)}_{\mathcal{J}}$ couplings in \eqref{eq:M_B-llpi}, for the sake of simplicity\footnote{This treatment yields more stringent bounds than those obtained from previous Belle bounds \cite{Liventsev:2013zz} using the relation in \eqref{eq:u2}.}. 

To appreciate the ability of the final taus polarization to determine the kind of effective operators involved in the studied interaction, we define a parameter $\lambda \in [0,1]$ to measure the proportion of vectorial and scalar operators contributing to the processes. Thus we multiply the vector operators by $\lambda$ and the scalars by $(1-\lambda)$, so that $\lambda=1$ means pure vectorial and $\lambda=0$ means pure scalar interactions. 

In our numerical code, given a value of the scalar - vector interaction content $\lambda$, we let the value of the intermediate $m_N$ mass vary in the allowed interval for each region in Tab. \ref{tab:kinematic_regions}. After fixing the numerical values of the effective couplings ($\alpha$) and mixing matrices ($Y$) in \eqref{eq:M_B-llpi}, this gives us as output an interval of values for the final-state polarization $P_{\tau}$ as defined in eqs. \eqref{eq:pol_tau_} and \eqref{eq:pol_taus} for one or two tau processes, consisting of a band in the ($\lambda$, $P_{\tau}$) plane. 

\begin{figure*}
\centering
\subfloat[Reg.{\bf{I}}. $\ell_{1}=\tau,\ell_{2}=\mu $]{\label{fig:polmutauI}\includegraphics[totalheight=6.cm]{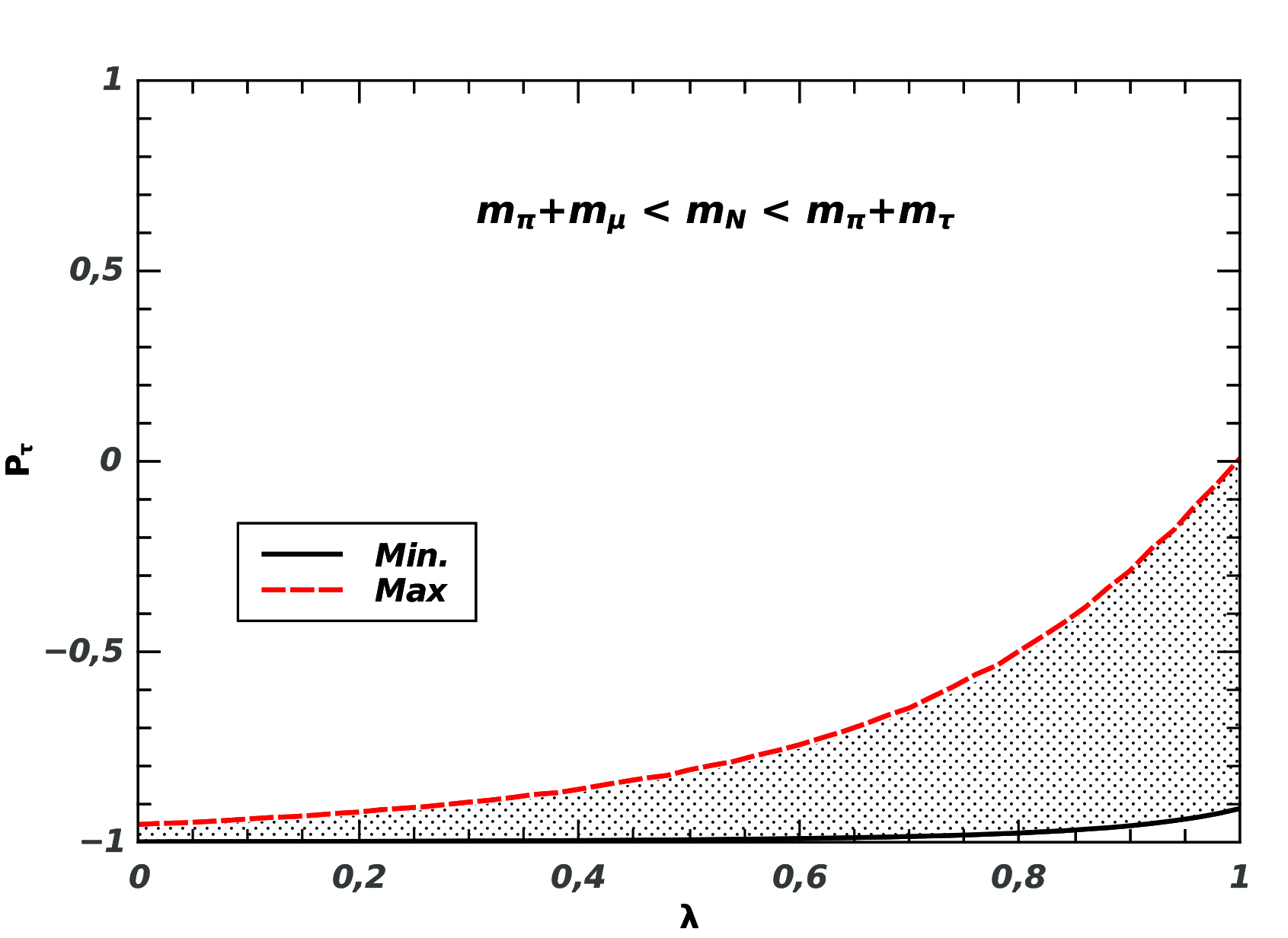}}~
\subfloat[Reg.{\bf{II}}. $\ell_{1}=\tau, \mu, ~ \ell_{2}=\mu, \tau$]{\label{fig:polmutauII}\includegraphics[totalheight=6.cm]{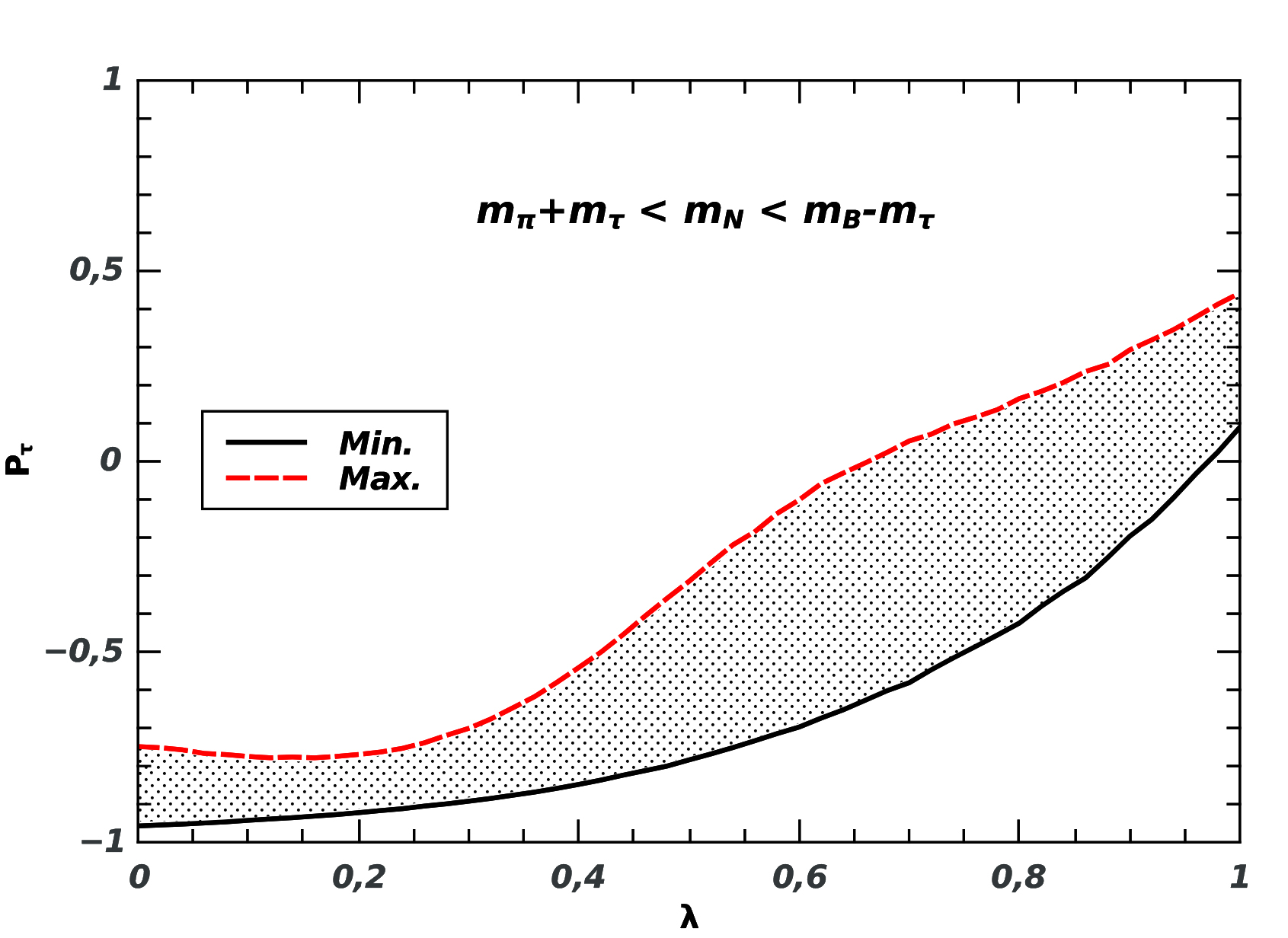}}

\subfloat[Reg.{\bf{III}}. $\ell_{1}=\mu,\ell_{2}=\tau $]{\label{fig:polmutauIII}\includegraphics[totalheight=6.cm]{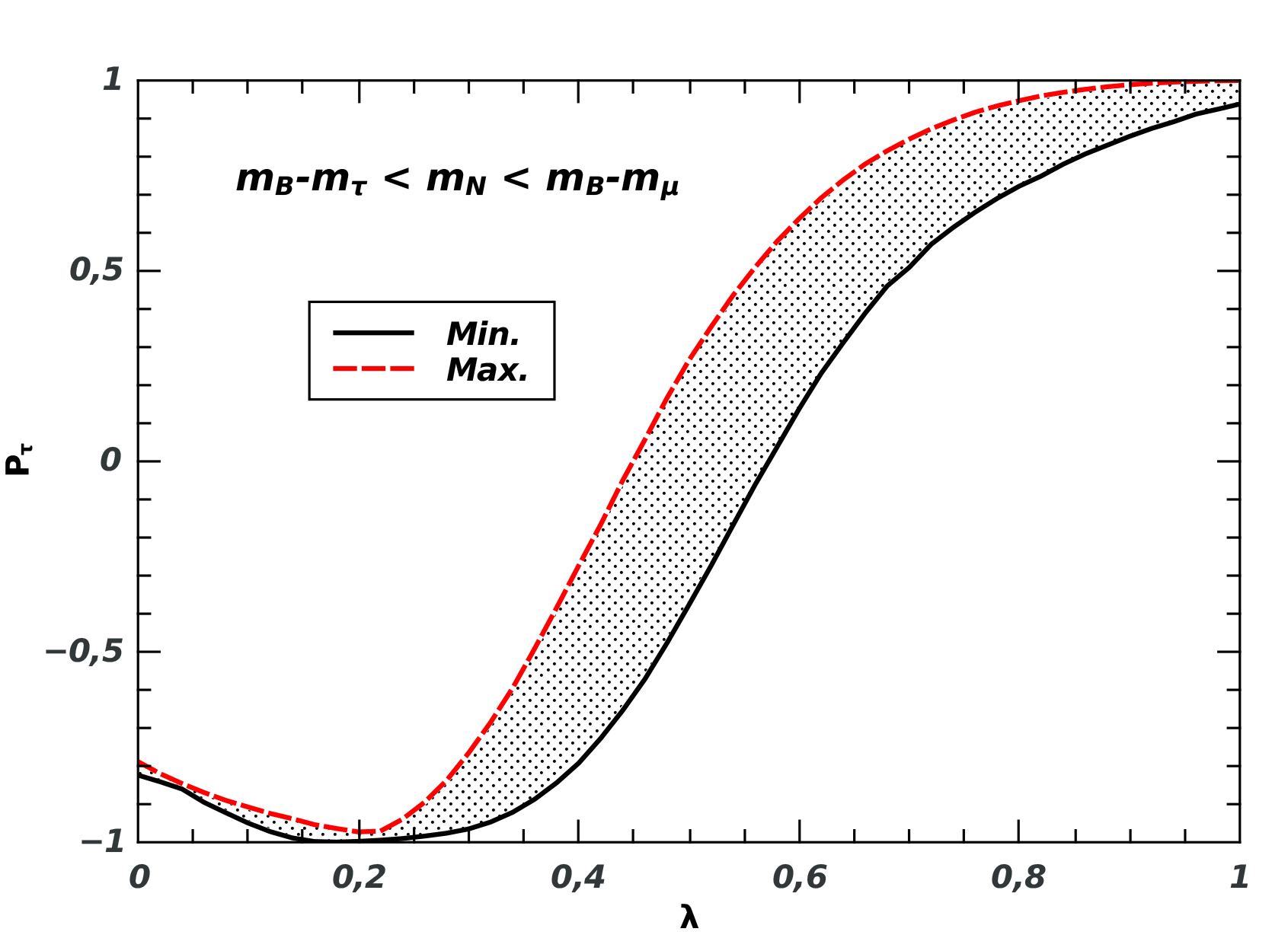}}~
\subfloat[Reg.{\bf{II}. $\ell_{1}=\tau,\ell_{2}=\tau $}]{\label{fig:poltautau}\includegraphics[totalheight=6.cm]{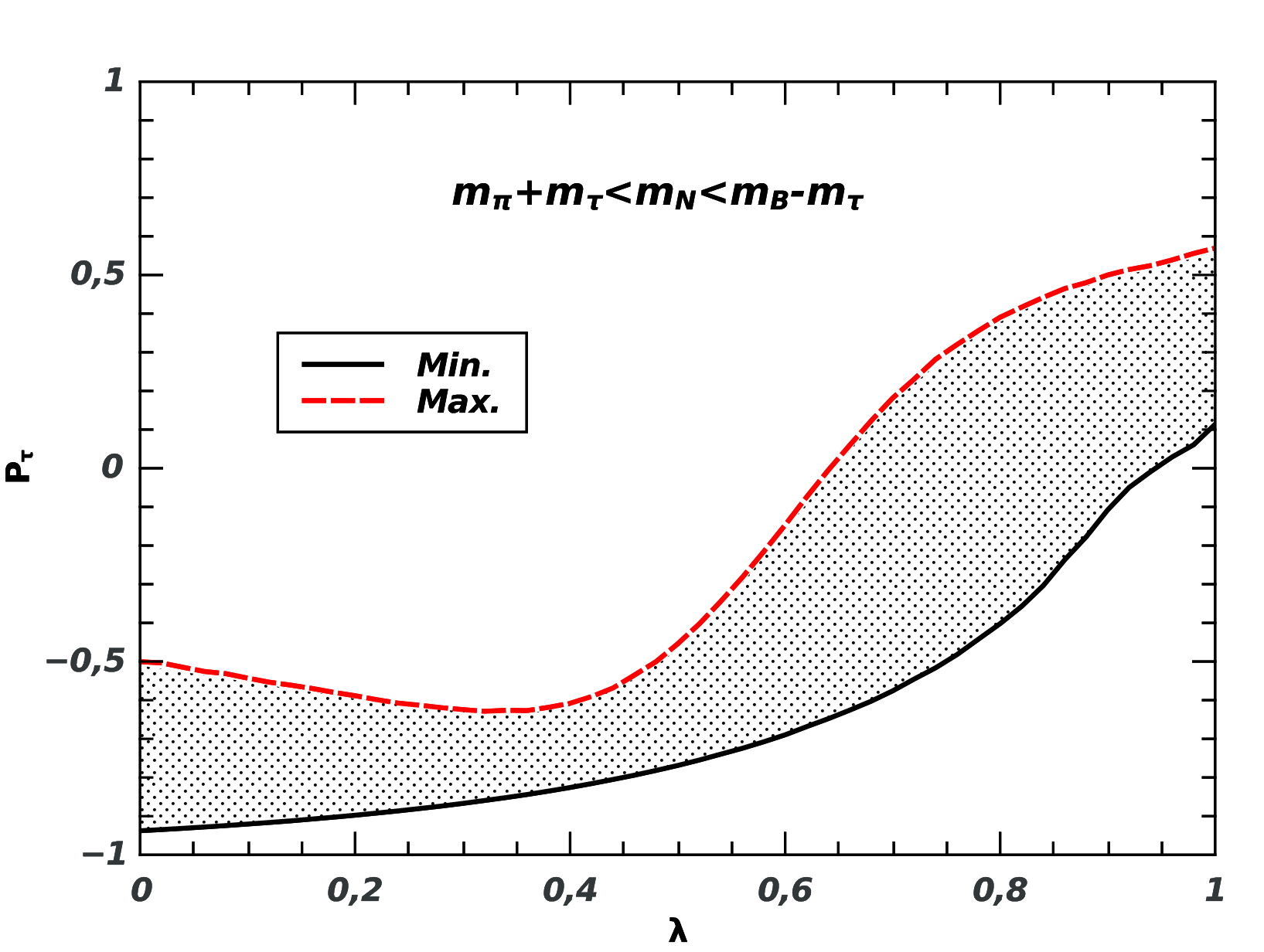}}
\caption{Final state polarization $P_{\tau}$ for the different kinematic regions I, II and III defined in table \ref{tab:kinematic_regions}. For $\lambda=1$ only vectorial interactions are considered, and for $\lambda=0$, only scalar. }\label{fig:Pol_maxmin_lam}
\end{figure*}

In Fig.\ref{fig:Pol_maxmin_lam} we show the curves of maximum and minimum values giving the allowed values for the final tau polarization $P_{\tau}$ in \eqref{eq:pol_tau_} depending on the $\lambda$ parameter for Majorana neutrino masses $m_N$ in the kinematic regions {\bf{I}} (Fig.\ref{fig:polmutauI}), {\bf{II}} (Fig.\ref{fig:polmutauII}) and {\bf{III}} (Fig.\ref{fig:polmutauIII}) defined in Table \ref{tab:kinematic_regions}. For the case of two final taus \eqref{eq:pol_taus}, the curves are shown in Fig.\ref{fig:poltautau}.  

In the experiment, such as FCC-ee as we discussed in Sec.\ref{sec:B_lept_dec_exp}, the first and second leptons can in principle be distinguished, as $\ell_1$ should have the higher momentum. Then, the $m_N$ mass range can be reconstructed with a measurement of the invariant mass $M(\ell_2, \pi)$. This would allow us to distinguish the three regions {\bf{I}}, {\bf{II}} and {\bf{III}}.   

If only one tau lepton is found in the final state, we can distinguish between regions {\bf{I}} and {\bf{III}} by comparing its momentum with the muon momentum. If $p_{\tau} > p_{\mu}$, we should be in region {\bf{I}}, and the invariant mass $M(\mu, \pi)$ value should confirm this. In this case, the final tau is produced together with the $N$ in the primary vertex (see Fig.\ref{fig:Bllpi}). This means in this case we are probing the $C_V^{(I)}$ and $C_S^{B}$ coefficients in \eqref{eq:M_B-llpi}. As we find in Fig.\ref{fig:polmutauI}, this tau leptons are expected to be mostly negatively-polarized if we only consider scalar interactions ($\lambda=0$), and an unpolarized final tau state could be reached in the case of purely vector interactions ($\lambda=1$). If $p_{\mu} > p_{\tau}$, we should be in region {\bf{III}}, which can be checked with the invariant mass $M(\tau, \pi)$. Here, by measuring the tau polarization we access the $N$ decay vertex, and test the  $C_V^{(II)}$ and $C_S^{\pi}$ coefficients in \eqref{eq:M_B-llpi}. Fig.\ref{fig:polmutauIII} shows that finding a negative $P_{\tau}$ would point towards a dominant scalar interaction, while a positive value would signal a vector-dominated $N$ decay.

The case of region {\bf{II}} shows that we expect mostly a negative final state tau polarization, indicating the dominance of the scalar interactions in both the $N$ production and decay vertices. This is consistent with the fact that the scalar interactions are weighted in both cases by the quark masses in the denominators in \eqref{eq:M_B-llpi}, which as we mentioned in the discussion of \eqref{eq:BtotauN_width}, enhances the scalar operators contribution, due to the pseudoscalar nature of both the B and $\pi$ mesons.       


\section{Summary}\label{sec:summary}

The effective field theory extending the standard model with right-handed neutrinos (SMNEFT) parameterizes new high-scale weakly coupled physics in a model independent manner. We consider massive Majorana neutrinos coupled to ordinary matter by dimension 6 effective operators, focusing on a simplified scenario with only one right-handed neutrino added, which provides us with a manageable parameter space to probe. These massive neutrinos would be produced in leptonic and semileptonic decays of B mesons, a theoretically clean system to probe new physics effects in lepton colliders and B factories. We exploit the known bounds on B meson decays to constrain the effective parameter space, and propose to use final taus polarization and angular observables to disentangle vectorial or scalar type effective contributions.

The B meson couples preferably to the scalar operators in the left column of Tab.\ref{tab:Operators} in comparison to the vectorial ones. Thus, as we already found in \cite{Duarte:2019rzs}, probing its decay to a Majorana neutrino $N$ we can either put more stringent bounds or give more precise predictions on the scalar operators effective couplings. 

In Sec.\ref{sec:BtotauN} we considered the bounds on the tauonic B decay  \cite{Adachi:2012mm,Kronenbitter:2015kls,Lees:2012ju,Aubert:2009wt} to constrain the allowed regions in the $(m_N, \alpha)$ plane (Fig.\ref{fig:Bnutaupol}), and find that a measure of the final tau polarization could help to separate the pure SM decay, which gives $P_{\tau}=1$, from the effective $B\to \tau N$, where the $N$ escapes undetected. This contribution can lower the $P_{\tau}$ value to $0.8$ in the allowed region (with $m_N< 3.5 ~GeV$) for heavy neutrinos coupling only to the third lepton family. In Tab.\ref{tab:alpha-sets} we define six different numerical coupling sets which turn on/off the scalar, vector or tensorial interactions. We find that the sets 4 and 6, which involve scalar $N$ interactions, are the ones giving the more interesting results.

In Sec.\ref{sec:bdec_rad} we define a Forward-Backward asymmetry $A_{FB}^{\ell \gamma}$ between the muon and photon directions for the $B \to \mu \nu \gamma$ decay, which allows us to separate the SM lepton number conserving contribution from the effective lepton number conserving and violating processes, mediated by a near on-shell $N$. We find that the effective contributions tend to deviate $A_{FB}^{\ell \gamma}$ from the SM value in photon energy intervals depending on the mass $m_N$, according to the boost of the $N$ in the B meson rest frame (Fig.\ref{fig:AFB_E}). This measure could be done in Belle II with the aid of inclusive tagging techniques, which allow for the reconstruction of the signal B frame \cite{Prim:2019gtj}. Again, the asymmetry is more sensitive to the scalar interactions, as is found in Fig.\ref{fig:chi2}. 

In Sec.\ref{sec:NmedBdec_taus} we study the final taus polarization in the lepton number violating $B^-\rightarrow  \ell^-_{1}\ell^-_{2} \pi^+$ decay when $\ell^-_{1  ~\text{and/or}~ 2}=\tau^{-}$. Using the bounds on the effective couplings obtained in \cite{Duarte:2019rzs} we probe the chances to disentangle different effective operators contributions to these decays. Weighting the vectorial and scalar operators by a factor $\lambda \in [0,1]$ with $\lambda = 1$ (purely vectorial) and $\lambda = 0$ (purely scalar) contributions. In Fig.\ref{fig:Pol_maxmin_lam} we find that a measurement of the final polarization $P_{\tau}$ can help us infer the scalar or vector content in the $N$ production or decay vertices, provided that the invariant mass $M(\ell_2, \pi)$ can be reconstructed. 

The B meson provides a rich environment to search for and constrain possible high-energy physics effects in the case a GeV-mass heavy $N$ is the only new accessible state. This study complements probes of lepton number violating same-sign dilepton signals, Higgs-neutrino interactions and other collider searches to characterize the Standard Model Neutrino effective field theory parameter space.

\appendix

\section{Polarized final taus in B decays}\label{sec:apx_polar}

\subsection{Tau polarization in the SM B tauonic decay}\label{sec:apx_polarSM}

In section \ref{sec:BtotauN} we showed the $B \to \tau \bar\nu_{\tau}$ polarized decay result in the SM. For reference purposes, we summarize the calculation here. The SM amplitude can be written as
\begin{equation*}
 \mathcal{M}_{B\to \tau \bar\nu_{\tau}}= i\frac{g^2}{2}\frac{V^{ub}}{m_W^2} \langle 0|\bar{u}\gamma^{\beta} P_L b|B^-\rangle \langle \tau^- \bar\nu_{\tau} |\bar{\tau} \gamma_{\beta} P_L \nu|0 \rangle = i\frac{g^2}{2}\frac{V^{ub}}{m_W^2}  i f_B q^{\beta} (\bar{u}_{\tau}(p) \gamma_{\beta} P_L v_{\nu}(k)), 
\end{equation*}
with $q\ll m_W$ the B meson momentum and $q= p+k$, the tau and anti-neutrino momenta, respectively.

We take the squared amplitude $|\mathcal{M}|^{2}$, considering the spin projectors for final leptons with mass $m$, spin $s$ and momentum $p$, with $p.s=0$:
\begin{eqnarray}\label{eq:ApX_polar}
 u(p,s)_{\rho}~ \overline{u}(p,s)_{\omega} &=&\left[(\slashed{p}+m)\frac{(1+ \gamma^{5} \slashed{s})}{2} \right]_{\rho \omega} \nonumber \\
 v(p,s)_{\rho}~ \overline{v}(p,s)_{\omega} &=& \left[(\slashed{p}-m)\frac{(1+ \gamma^{5} \slashed{s})}{2} \right]_{\rho \omega}.
\end{eqnarray}
The leptons spin vectors can be written in terms of their helicity $h=\pm$ as
\begin{equation}\label{eq:Apx_h}
 s^{\beta}=h \left( \frac{|\vec{p}|}{m},\frac{E}{m} \frac{\vec{p}}{|\vec{p}|} \right).
\end{equation}
Thus we have, in the B meson rest frame
\begin{equation*}
 |\mathcal{M}_{B\to \tau \bar\nu_{\tau}}|^2= G_F^2 f_B^2 {V^{ub}}^2 m_{\tau}^2 (m_B^2-m_{\tau}^2 +2 m_{\tau} ~q.s), 
\end{equation*}
which gives us a polarized decay width
\begin{equation*}
 \Gamma_{B\to \tau \bar\nu_{\tau}}(h)= G_F^2 f_B^2 ~|V^{ub}|^2 m_B^3 x_{\tau}(1-x_{\tau}^2)(1+h),  
 \end{equation*}
with $x_{\tau}=m_{\tau}/m_B$, which shows that we only have taus with $h=1$ in the final state, giving a polarization $P_{\tau}=1$ in the SM.

\subsection{Final taus polarization in $B\to \ell \ell \pi$ decays}\label{sec:apx_polar_LNV}

In section \ref{sec:NmedBdec_taus} we presented the decay $B^{-} \to \ell_1^{-} \ell_2^{-} \pi^{+}$ mediated by the Majorana neutrino $N$, where we considered final polarized taus $\ell^-_{1  ~\text{and/or}~ 2}=\tau^{-}$. 

In the calculation, the amplitude $\mathcal{M}^{N}({B^-\rightarrow  \ell^-_{1}\ell^-_{2} \pi^+})$ is initially written as $\mathcal{M}=\langle \ell_1^{-} \ell_2^{-} \pi^{+}| \hat{\mathcal{O}}|B^{-} \rangle$. The interaction $\hat{\mathcal{O}}= \hat{\mathcal{O}}_{I} \cdot \hat{\mathcal{O}}_{II}$ includes the terms coming from the $N$ production vertex $I$ in Fig.\ref{fig:Bllpi}, considering the Lagrangian terms in \eqref{eq:lag_massiveq} with the up-type quark $\bar{u}_\beta=u$, the down-type quark $d_\beta'=b$, the charged lepton $l_i=\ell_1$ and the $N$ decay vertex $II$, with $\bar{u}_\beta=u$, $d_\beta'=d$ and $l_i=\ell_2$. 

The leptonic and hadronic parts of the final state can be separated and one considers the intermediate $N$ in $\langle 0|T N \overline{N^{c}}|0 \rangle= i \frac{\slashed{p}_N+ m_N}{(p_N^{2}-m_N^{2})}$, taking the charge conjugate of the $N$ decay terms. 

The only non-zero hadronic currents for a pseudoscalar meson M (this includes both the B and the pion) are the the pseudo-vector $\langle 0| \bar{u}\gamma^{\mu}\gamma^{5} d| M \rangle = i f_M q^{\mu}$ and the pseudo-scalar $\langle 0| \bar{u}\gamma^{5} d| M \rangle= -i  f_M \frac{m_{M}^2}{m_u+m_d}$ currents, where $u$ and $d$ are up-type and down-type quarks, $f_M$ is the meson constant, and $q$ its momentum. 

The amplitude is finally written as in \eqref{eq:M_B-llpi}
\begin{equation}\label{eq:M_B-llpi_apx}
 \mathcal{M}^{N}_{B^-\rightarrow  \ell^-_{1}\ell^-_{2} \pi^+}= -i \frac{f_{\pi} f_{B}}{4 \Lambda^4} m_N P_{N}(p_{N}^2)~ 
 \overline{u}(p_2) \left(C_V^{(II)} \slashed{k} - C_S^{\pi} \right) P_R \left(C_V^{(I)} \slashed{q} - C_S^{B} \right) v(p_1).
\end{equation}
Here $p_{1,2}$ are the $\ell_{1,2}$ momenta, $f_B$ and $f_{\pi}$ are the $B$ and $\pi$ mesons constants and
\begin{displaymath}
 C_{V}^{(I)}= \left(\alpha_{V_0}^{(i)} Y^{ub}_{RR} + \alpha_{W}^{(i)} V^{ub}\right) \;\;\;\;  C^{B}_{S}=\left(\alpha_{S_1}^{(i)} Y^{ub}_{RL}+ (\alpha_{S_2}^{(i)}+ \frac12 \alpha_{S_3}^{(i)}) Y^{ub}_{LR} \right)\frac{m_B^2}{m_{u}+m_{b}}  
\end{displaymath}
are the terms corresponding to the hadronic $B$ current in the $N$ production vertex. On the $N$ decay side we have the terms corresponding to the hadronic $\pi$ current,
\begin{displaymath}
 C_{V}^{(II)}= \left(\alpha_{V_0}^{(j)} Y^{ud}_{RR} + \alpha_{W}^{(j)} V^{ud}\right) \;\;\;\;  C^{\pi}_{S}=\left(\alpha_{S_1}^{(j)} Y^{ud}_{RL}+ (\alpha_{S_2}^{(j)}+ \frac12 \alpha_{S_3}^{(j)}) Y^{ud}_{LR} \right)\frac{m_{\pi}^2}{m_{u}+m_{d}}.  
\end{displaymath}
We then take the squared amplitude, and as we did in \ref{sec:apx_polarSM}, we use the expressions in eq. \eqref{eq:ApX_polar} and \eqref{eq:Apx_h} to calculate the polarized decay width.


\bibliographystyle{bibstyle.bst}
\bibliography{Bib_N_5_2020}

\begin{thebibliography}{10}%
\makeatletter
\providecommand \@ifxundefined [1]{%
 \ifx #1\undefined \expandafter \@firstoftwo
 \else \expandafter \@secondoftwo
\fi
}%
\providecommand \@ifnum [1]{%
 \ifnum #1\expandafter \@firstoftwo
 \else \expandafter \@secondoftwo
\fi
}%
\providecommand \enquote [1]{``#1''}%
\providecommand \bibnamefont  [1]{#1}%
\providecommand \bibfnamefont [1]{#1}%
\providecommand \citenamefont [1]{#1}%
\providecommand\href[0]{\@sanitize\@href}%
\providecommand\@href[1]{\endgroup\@@startlink{#1}\endgroup\@@href}%
\providecommand\@@href[1]{#1\@@endlink}%
\providecommand \@sanitize [0]{\begingroup\catcode`\&12\catcode`\#12\relax}%
\@ifxundefined \pdfoutput {\@firstoftwo}{%
 \@ifnum{\z@=\pdfoutput}{\@firstoftwo}{\@secondoftwo}%
}{%
 \providecommand\@@startlink[1]{\leavevmode\special{html:<a href="#1">}}%
 \providecommand\@@endlink[0]{\special{html:</a>}}%
}{%
 \providecommand\@@startlink[1]{%
  \leavevmode
  \pdfstartlink
   attr{/Border[0 0 1 ]/H/I/C[0 1 1]}%
   user{/Subtype/Link/A<</Type/Action/S/URI/URI(#1)>>}%
  \relax
 }%
 \providecommand\@@endlink[0]{\pdfendlink}%
}%
\providecommand \url  [0]{\begingroup\@sanitize \@url }%
\providecommand \@url [1]{\endgroup\@href {#1}{\urlprefix}}%
\providecommand \urlprefix [0]{URL }%
\providecommand \Eprint[0]{\href }%
\@ifxundefined \urlstyle {%
  \providecommand \doi [1]{doi:\discretionary{}{}{}#1}%
}{%
  \providecommand \doi [0]{doi:\discretionary{}{}{}\begingroup
  \urlstyle{rm}\Url }%
}%
\providecommand \doibase [0]{http://dx.doi.org/}%
\providecommand \Doi[1]{\href{\doibase#1}}%
\providecommand \bibAnnote [3]{%
  \BibitemShut{#1}%
  \begin{quotation}\noindent
    \textsc{Key:}\ #2\\\textsc{Annotation:}\ #3%
  \end{quotation}%
}%
\providecommand \bibAnnoteFile [2]{%
  \IfFileExists{#2}{\bibAnnote {#1} {#2} {\input{#2}}}{}%
}%
\providecommand \typeout [0]{\immediate \write \m@ne }%
\providecommand \selectlanguage [0]{\@gobble}%
\providecommand \bibinfo [0]{\@secondoftwo}%
\providecommand \bibfield [0]{\@secondoftwo}%
\providecommand \translation [1]{[#1]}%
\providecommand \BibitemOpen[0]{}%
\providecommand \bibitemStop [0]{}%
\providecommand \bibitemNoStop [0]{.\EOS\space}%
\providecommand \EOS [0]{\spacefactor3000\relax}%
\providecommand \BibitemShut [1]{\csname bibitem#1\endcsname}%
\bibitem{Buchmuller:1985jz}%
  \BibitemOpen
  \bibfield{author}{%
  \bibinfo {author} {\bibfnamefont{W.}~\bibnamefont{Buchmuller}}\ and\ \bibinfo
  {author} {\bibfnamefont{D.}~\bibnamefont{Wyler}},\ }%
  \emph{\bibinfo {title} {{Effective Lagrangian Analysis of New Interactions
  and Flavor Conservation}}},\ \bibfield{journal}{%
  \Doi{10.1016/0550-3213(86)90262-2}{\bibinfo {journal} {Nucl. Phys. B}}\ }%
  \textbf{\bibinfo {volume} {268}},\ \bibinfo {pages} {621} (\bibinfo {year}
  {1986}).%
  \bibAnnoteFile{Stop}{Buchmuller:1985jz}%
\bibitem{Grzadkowski:2010es}%
  \BibitemOpen
  \bibfield{author}{%
  \bibinfo {author} {\bibfnamefont{B.}~\bibnamefont{Grzadkowski}}, \bibinfo
  {author} {\bibfnamefont{M.}~\bibnamefont{Iskrzynski}}, \bibinfo {author}
  {\bibfnamefont{M.}~\bibnamefont{Misiak}}\ and\ \bibinfo {author}
  {\bibfnamefont{J.}~\bibnamefont{Rosiek}},\ }%
  \emph{\bibinfo {title} {{Dimension-Six Terms in the Standard Model
  Lagrangian}}},\ \bibfield{journal}{%
  \Doi{10.1007/JHEP10(2010)085}{\bibinfo {journal} {JHEP}}\ }%
  \textbf{\bibinfo {volume} {1010}},\ \bibinfo {pages} {085} (\bibinfo {year}
  {2010}),\ \Eprint{http://arxiv.org/abs/1008.4884}{arXiv:1008.4884 [hep-ph]}.%
  \bibAnnoteFile{Stop}{Grzadkowski:2010es}%
\bibitem{Minkowski:1977sc}%
  \BibitemOpen
  \bibfield{author}{%
  \bibinfo {author} {\bibfnamefont{P.}~\bibnamefont{Minkowski}},\ }%
  \emph{\bibinfo {title} {{$\mu \rightarrow e \gamma$ at a rate of one out of
  1-billion muon decays?}}},\ \bibfield{journal}{%
  \Doi{10.1016/0370-2693(77)90435-X}{\bibinfo {journal} {Phys.Lett.}}\ }%
  \textbf{\bibinfo {volume} {B67}},\ \bibinfo {pages} {421} (\bibinfo {year}
  {1977}).%
  \bibAnnoteFile{Stop}{Minkowski:1977sc}%
\bibitem{Mohapatra:1979ia}%
  \BibitemOpen
  \bibfield{author}{%
  \bibinfo {author} {\bibfnamefont{R.~N.}\ \bibnamefont{Mohapatra}}\ and\
  \bibinfo {author} {\bibfnamefont{G.}~\bibnamefont{Senjanovic}},\ }%
  \emph{\bibinfo {title} {{Neutrino Mass and Spontaneous Parity Violation}}},\
  \bibfield{journal}{%
  \Doi{10.1103/PhysRevLett.44.912}{\bibinfo {journal} {Phys.Rev.Lett.}}\ }%
  \textbf{\bibinfo {volume} {44}},\ \bibinfo {pages} {912} (\bibinfo {year}
  {1980}).%
  \bibAnnoteFile{Stop}{Mohapatra:1979ia}%
\bibitem{Yanagida:1980xy}%
  \BibitemOpen
  \bibfield{author}{%
  \bibinfo {author} {\bibfnamefont{T.}~\bibnamefont{Yanagida}},\ }%
  \emph{\bibinfo {title} {{Horizontal Symmetry and Masses of Neutrinos}}},\
  \bibfield{journal}{%
  \Doi{10.1143/PTP.64.1103}{\bibinfo {journal} {Prog.Theor.Phys.}}\ }%
  \textbf{\bibinfo {volume} {64}},\ \bibinfo {pages} {1103} (\bibinfo {year}
  {1980}).%
  \bibAnnoteFile{Stop}{Yanagida:1980xy}%
\bibitem{GellMann:1980vs}%
  \BibitemOpen
  \bibfield{author}{%
  \bibinfo {author} {\bibfnamefont{M.}~\bibnamefont{Gell-Mann}}, \bibinfo
  {author} {\bibfnamefont{P.}~\bibnamefont{Ramond}}\ and\ \bibinfo {author}
  {\bibfnamefont{R.}~\bibnamefont{Slansky}},\ }%
  \emph{\bibinfo {title} {{Complex Spinors and Unified Theories}}},\
  \bibfield{journal}{%
  \bibinfo {journal} {Conf.Proc.}\ }%
  \textbf{\bibinfo {volume} {C790927}},\ \bibinfo {pages} {315} (\bibinfo
  {year} {1979}),\ \Eprint{http://arxiv.org/abs/1306.4669}{arXiv:1306.4669
  [hep-th]}.%
  \bibAnnoteFile{Stop}{GellMann:1980vs}%
\bibitem{Schechter:1980gr}%
  \BibitemOpen
  \bibfield{author}{%
  \bibinfo {author} {\bibfnamefont{J.}~\bibnamefont{Schechter}}\ and\ \bibinfo
  {author} {\bibfnamefont{J.~W.~F.}\ \bibnamefont{Valle}},\ }%
  \emph{\bibinfo {title} {{Neutrino Masses in SU(2) x U(1) Theories}}},\
  \bibfield{journal}{%
  \Doi{10.1103/PhysRevD.22.2227}{\bibinfo {journal} {Phys. Rev.}}\ }%
  \textbf{\bibinfo {volume} {D22}},\ \bibinfo {pages} {2227} (\bibinfo {year}
  {1980}).%
  \bibAnnoteFile{Stop}{Schechter:1980gr}%
\bibitem{delAguila:2008ir}%
  \BibitemOpen
  \bibfield{author}{%
  \bibinfo {author} {\bibfnamefont{F.}~\bibnamefont{del Aguila}}, \bibinfo
  {author} {\bibfnamefont{S.}~\bibnamefont{Bar-Shalom}}, \bibinfo {author}
  {\bibfnamefont{A.}~\bibnamefont{Soni}}\ and\ \bibinfo {author}
  {\bibfnamefont{J.}~\bibnamefont{Wudka}},\ }%
  \emph{\bibinfo {title} {{Heavy Majorana Neutrinos in the Effective Lagrangian
  Description: Application to Hadron Colliders}}},\ \bibfield{journal}{%
  \Doi{10.1016/j.physletb.2008.11.031}{\bibinfo {journal} {Phys.Lett.}}\ }%
  \textbf{\bibinfo {volume} {B670}},\ \bibinfo {pages} {399} (\bibinfo {year}
  {2009}),\ \Eprint{http://arxiv.org/abs/0806.0876}{arXiv:0806.0876 [hep-ph]}.%
  \bibAnnoteFile{Stop}{delAguila:2008ir}%
\bibitem{Aparici:2009fh}%
  \BibitemOpen
  \bibfield{author}{%
  \bibinfo {author} {\bibfnamefont{A.}~\bibnamefont{Aparici}}, \bibinfo
  {author} {\bibfnamefont{K.}~\bibnamefont{Kim}}, \bibinfo {author}
  {\bibfnamefont{A.}~\bibnamefont{Santamaria}}\ and\ \bibinfo {author}
  {\bibfnamefont{J.}~\bibnamefont{Wudka}},\ }%
  \emph{\bibinfo {title} {{Right-handed neutrino magnetic moments}}},\
  \bibfield{journal}{%
  \Doi{10.1103/PhysRevD.80.013010}{\bibinfo {journal} {Phys. Rev.}}\ }%
  \textbf{\bibinfo {volume} {D80}},\ \bibinfo {pages} {013010} (\bibinfo {year}
  {2009}),\ \Eprint{http://arxiv.org/abs/0904.3244}{arXiv:0904.3244 [hep-ph]}.%
  \bibAnnoteFile{Stop}{Aparici:2009fh}%
\bibitem{Liao:2016qyd}%
  \BibitemOpen
  \bibfield{author}{%
  \bibinfo {author} {\bibfnamefont{Y.}~\bibnamefont{Liao}}\ and\ \bibinfo
  {author} {\bibfnamefont{X.-D.}\ \bibnamefont{Ma}},\ }%
  \emph{\bibinfo {title} {{Operators up to Dimension Seven in Standard Model
  Effective Field Theory Extended with Sterile Neutrinos}}},\
  \bibfield{journal}{%
  \Doi{10.1103/PhysRevD.96.015012}{\bibinfo {journal} {Phys. Rev.}}\ }%
  \textbf{\bibinfo {volume} {D96}},\ \bibinfo {pages} {015012} (\bibinfo {year}
  {2017}),\ \Eprint{http://arxiv.org/abs/1612.04527}{arXiv:1612.04527
  [hep-ph]}.%
  \bibAnnoteFile{Stop}{Liao:2016qyd}%
\bibitem{Bhattacharya:2015vja}%
  \BibitemOpen
  \bibfield{author}{%
  \bibinfo {author} {\bibfnamefont{S.}~\bibnamefont{Bhattacharya}}\ and\
  \bibinfo {author} {\bibfnamefont{J.}~\bibnamefont{Wudka}},\ }%
  \emph{\bibinfo {title} {{Dimension-seven operators in the standard model with
  right handed neutrinos}}},\ \bibfield{journal}{%
  \Doi{10.1103/PhysRevD.94.055022, 10.1103/PhysRevD.95.039904}{\bibinfo
  {journal} {Phys. Rev.}}\ }%
  \textbf{\bibinfo {volume} {D94}},\ \bibinfo {pages} {055022} (\bibinfo {year}
  {2016}),\ \bibinfo {note} {[Erratum: Phys. Rev.D95,no.3,039904(2017)]},\
  \Eprint{http://arxiv.org/abs/1505.05264}{arXiv:1505.05264 [hep-ph]}.%
  \bibAnnoteFile{Stop}{Bhattacharya:2015vja}%
\bibitem{Caputo:2017pit}%
  \BibitemOpen
  \bibfield{author}{%
  \bibinfo {author} {\bibfnamefont{A.}~\bibnamefont{Caputo}}, \bibinfo {author}
  {\bibfnamefont{P.}~\bibnamefont{Hernandez}}, \bibinfo {author}
  {\bibfnamefont{J.}~\bibnamefont{Lopez-Pavon}}\ and\ \bibinfo {author}
  {\bibfnamefont{J.}~\bibnamefont{Salvado}},\ }%
  \emph{\bibinfo {title} {{The seesaw portal in testable models of neutrino
  masses}}},\ \bibfield{journal}{%
  \Doi{10.1007/JHEP06(2017)112}{\bibinfo {journal} {JHEP}}\ }%
  \textbf{\bibinfo {volume} {06}},\ \bibinfo {pages} {112} (\bibinfo {year}
  {2017}),\ \Eprint{http://arxiv.org/abs/1704.08721}{arXiv:1704.08721
  [hep-ph]}.%
  \bibAnnoteFile{Stop}{Caputo:2017pit}%
\bibitem{Jones-Perez:2019plk}%
  \BibitemOpen
  \bibfield{author}{%
  \bibinfo {author} {\bibfnamefont{J.}~\bibnamefont{Jones-Pérez}}, \bibinfo
  {author} {\bibfnamefont{J.}~\bibnamefont{Masias}}\ and\ \bibinfo {author}
  {\bibfnamefont{J.}~\bibnamefont{Ruiz-Álvarez}},\ }%
  \emph{\bibinfo {title} {{Search for Long-Lived Heavy Neutrinos at the LHC
  with a VBF Trigger}}} (\bibinfo {month} {12}\ \bibinfo {year} {2019}),\
  \Eprint{http://arxiv.org/abs/1912.08206}{arXiv:1912.08206 [hep-ph]}.%
  \bibAnnoteFile{Stop}{Jones-Perez:2019plk}%
\bibitem{Graesser:2007yj}%
  \BibitemOpen
  \bibfield{author}{%
  \bibinfo {author} {\bibfnamefont{M.~L.}\ \bibnamefont{Graesser}},\ }%
  \emph{\bibinfo {title} {{Broadening the Higgs boson with right-handed
  neutrinos and a higher dimension operator at the electroweak scale}}},\
  \bibfield{journal}{%
  \Doi{10.1103/PhysRevD.76.075006}{\bibinfo {journal} {Phys. Rev.}}\ }%
  \textbf{\bibinfo {volume} {D76}},\ \bibinfo {pages} {075006} (\bibinfo {year}
  {2007}),\ \Eprint{http://arxiv.org/abs/0704.0438}{arXiv:0704.0438 [hep-ph]}.%
  \bibAnnoteFile{Stop}{Graesser:2007yj}%
\bibitem{Barducci:2020ncz}%
  \BibitemOpen
  \bibfield{author}{%
  \bibinfo {author} {\bibfnamefont{D.}~\bibnamefont{Barducci}}, \bibinfo
  {author} {\bibfnamefont{E.}~\bibnamefont{Bertuzzo}}, \bibinfo {author}
  {\bibfnamefont{A.}~\bibnamefont{Caputo}}\ and\ \bibinfo {author}
  {\bibfnamefont{P.}~\bibnamefont{Hernandez}},\ }%
  \emph{\bibinfo {title} {{Minimal flavor violation in the see-saw portal}}}
  (\bibinfo {month} {3}\ \bibinfo {year} {2020}),\
  \Eprint{http://arxiv.org/abs/2003.08391}{arXiv:2003.08391 [hep-ph]}.%
  \bibAnnoteFile{Stop}{Barducci:2020ncz}%
\bibitem{Butterworth:2019iff}%
  \BibitemOpen
  \bibfield{author}{%
  \bibinfo {author} {\bibfnamefont{J.~M.}\ \bibnamefont{Butterworth}}, \bibinfo
  {author} {\bibfnamefont{M.}~\bibnamefont{Chala}}, \bibinfo {author}
  {\bibfnamefont{C.}~\bibnamefont{Englert}}, \bibinfo {author}
  {\bibfnamefont{M.}~\bibnamefont{Spannowsky}}\ and\ \bibinfo {author}
  {\bibfnamefont{A.}~\bibnamefont{Titov}},\ }%
  \emph{\bibinfo {title} {{Higgs phenomenology as a probe of sterile
  neutrinos}}},\ \bibfield{journal}{%
  \Doi{10.1103/PhysRevD.100.115019}{\bibinfo {journal} {Phys. Rev. D}}\ }%
  \textbf{\bibinfo {volume} {100}},\ \bibinfo {pages} {115019} (\bibinfo {year}
  {2019}),\ \Eprint{http://arxiv.org/abs/1909.04665}{arXiv:1909.04665
  [hep-ph]}.%
  \bibAnnoteFile{Stop}{Butterworth:2019iff}%
\bibitem{Magill:2018jla}%
  \BibitemOpen
  \bibfield{author}{%
  \bibinfo {author} {\bibfnamefont{G.}~\bibnamefont{Magill}}, \bibinfo {author}
  {\bibfnamefont{R.}~\bibnamefont{Plestid}}, \bibinfo {author}
  {\bibfnamefont{M.}~\bibnamefont{Pospelov}}\ and\ \bibinfo {author}
  {\bibfnamefont{Y.-D.}\ \bibnamefont{Tsai}},\ }%
  \emph{\bibinfo {title} {{Dipole Portal to Heavy Neutral Leptons}}},\
  \bibfield{journal}{%
  \Doi{10.1103/PhysRevD.98.115015}{\bibinfo {journal} {Phys. Rev. D}}\ }%
  \textbf{\bibinfo {volume} {98}},\ \bibinfo {pages} {115015} (\bibinfo {year}
  {2018}),\ \Eprint{http://arxiv.org/abs/1803.03262}{arXiv:1803.03262
  [hep-ph]}.%
  \bibAnnoteFile{Stop}{Magill:2018jla}%
\bibitem{Alcaide:2019pnf}%
  \BibitemOpen
  \bibfield{author}{%
  \bibinfo {author} {\bibfnamefont{J.}~\bibnamefont{Alcaide}}, \bibinfo
  {author} {\bibfnamefont{S.}~\bibnamefont{Banerjee}}, \bibinfo {author}
  {\bibfnamefont{M.}~\bibnamefont{Chala}}\ and\ \bibinfo {author}
  {\bibfnamefont{A.}~\bibnamefont{Titov}},\ }%
  \emph{\bibinfo {title} {{Probes of the Standard Model effective field theory
  extended with a right-handed neutrino}}},\ \bibfield{journal}{%
  \Doi{10.1007/JHEP08(2019)031}{\bibinfo {journal} {JHEP}}\ }%
  \textbf{\bibinfo {volume} {08}},\ \bibinfo {pages} {031} (\bibinfo {year}
  {2019}),\ \Eprint{http://arxiv.org/abs/1905.11375}{arXiv:1905.11375
  [hep-ph]}.%
  \bibAnnoteFile{Stop}{Alcaide:2019pnf}%
\bibitem{Chala:2020vqp}%
  \BibitemOpen
  \bibfield{author}{%
  \bibinfo {author} {\bibfnamefont{M.}~\bibnamefont{Chala}}\ and\ \bibinfo
  {author} {\bibfnamefont{A.}~\bibnamefont{Titov}},\ }%
  \emph{\bibinfo {title} {{One-loop matching in the SMEFT extended with a
  sterile neutrino}}},\ \bibfield{journal}{%
  \Doi{10.1007/JHEP05(2020)139}{\bibinfo {journal} {JHEP}}\ }%
  \textbf{\bibinfo {volume} {05}},\ \bibinfo {pages} {139} (\bibinfo {year}
  {2020}),\ \Eprint{http://arxiv.org/abs/2001.07732}{arXiv:2001.07732
  [hep-ph]}.%
  \bibAnnoteFile{Stop}{Chala:2020vqp}%
\bibitem{Dekens:2020ttz}%
  \BibitemOpen
  \bibfield{author}{%
  \bibinfo {author} {\bibfnamefont{W.}~\bibnamefont{Dekens}}, \bibinfo {author}
  {\bibfnamefont{J.}~\bibnamefont{de~Vries}}, \bibinfo {author}
  {\bibfnamefont{K.}~\bibnamefont{Fuyuto}}, \bibinfo {author}
  {\bibfnamefont{E.}~\bibnamefont{Mereghetti}}\ and\ \bibinfo {author}
  {\bibfnamefont{G.}~\bibnamefont{Zhou}},\ }%
  \emph{\bibinfo {title} {{Sterile neutrinos and neutrinoless double beta decay
  in effective field theory}}} (\bibinfo {month} {2}\ \bibinfo {year} {2020}),\
  \Eprint{http://arxiv.org/abs/2002.07182}{arXiv:2002.07182 [hep-ph]}.%
  \bibAnnoteFile{Stop}{Dekens:2020ttz}%
\bibitem{Bischer:2019ttk}%
  \BibitemOpen
  \bibfield{author}{%
  \bibinfo {author} {\bibfnamefont{I.}~\bibnamefont{Bischer}}\ and\ \bibinfo
  {author} {\bibfnamefont{W.}~\bibnamefont{Rodejohann}},\ }%
  \emph{\bibinfo {title} {{General neutrino interactions from an effective
  field theory perspective}}},\ \bibfield{journal}{%
  \Doi{10.1016/j.nuclphysb.2019.114746}{\bibinfo {journal} {Nucl. Phys. B}}\ }%
  \textbf{\bibinfo {volume} {947}},\ \bibinfo {pages} {114746} (\bibinfo {year}
  {2019}),\ \Eprint{http://arxiv.org/abs/1905.08699}{arXiv:1905.08699
  [hep-ph]}.%
  \bibAnnoteFile{Stop}{Bischer:2019ttk}%
\bibitem{Han:2020pff}%
  \BibitemOpen
  \bibfield{author}{%
  \bibinfo {author} {\bibfnamefont{T.}~\bibnamefont{Han}}, \bibinfo {author}
  {\bibfnamefont{J.}~\bibnamefont{Liao}}, \bibinfo {author}
  {\bibfnamefont{H.}~\bibnamefont{Liu}}\ and\ \bibinfo {author}
  {\bibfnamefont{D.}~\bibnamefont{Marfatia}},\ }%
  \emph{\bibinfo {title} {{Scalar and tensor neutrino interactions}}} (\bibinfo
  {month} {4}\ \bibinfo {year} {2020}),\
  \Eprint{http://arxiv.org/abs/2004.13869}{arXiv:2004.13869 [hep-ph]}.%
  \bibAnnoteFile{Stop}{Han:2020pff}%
\bibitem{Bolton:2020xsm}%
  \BibitemOpen
  \bibfield{author}{%
  \bibinfo {author} {\bibfnamefont{P.~D.}\ \bibnamefont{Bolton}}, \bibinfo
  {author} {\bibfnamefont{F.~F.}\ \bibnamefont{Deppisch}}\ and\ \bibinfo
  {author} {\bibfnamefont{C.}~\bibnamefont{Hati}},\ }%
  \emph{\bibinfo {title} {{Probing New Physics with Long-Range Neutrino
  Interactions: An Effective Field Theory Approach}}} (\bibinfo {month} {4}\
  \bibinfo {year} {2020}),\
  \Eprint{http://arxiv.org/abs/2004.08328}{arXiv:2004.08328 [hep-ph]}.%
  \bibAnnoteFile{Stop}{Bolton:2020xsm}%
\bibitem{Drewes:2019byd}%
  \BibitemOpen
  \bibfield{author}{%
  \bibinfo {author} {\bibfnamefont{M.}~\bibnamefont{Drewes}}, \bibinfo {author}
  {\bibfnamefont{J.}~\bibnamefont{Klari\'c}}\ and\ \bibinfo {author}
  {\bibfnamefont{P.}~\bibnamefont{Klose}},\ }%
  \emph{\bibinfo {title} {{On Lepton Number Violation in Heavy Neutrino Decays
  at Colliders}}},\ \bibfield{journal}{%
  \Doi{10.1007/JHEP11(2019)032}{\bibinfo {journal} {JHEP}}\ }%
  \textbf{\bibinfo {volume} {19}},\ \bibinfo {pages} {032} (\bibinfo {year}
  {2020}),\ \Eprint{http://arxiv.org/abs/1907.13034}{arXiv:1907.13034
  [hep-ph]}.%
  \bibAnnoteFile{Stop}{Drewes:2019byd}%
\bibitem{Cai:2017mow}%
  \BibitemOpen
  \bibfield{author}{%
  \bibinfo {author} {\bibfnamefont{Y.}~\bibnamefont{Cai}}, \bibinfo {author}
  {\bibfnamefont{T.}~\bibnamefont{Han}}, \bibinfo {author}
  {\bibfnamefont{T.}~\bibnamefont{Li}}\ and\ \bibinfo {author}
  {\bibfnamefont{R.}~\bibnamefont{Ruiz}},\ }%
  \emph{\bibinfo {title} {{Lepton Number Violation: Seesaw Models and Their
  Collider Tests}}},\ \bibfield{journal}{%
  \Doi{10.3389/fphy.2018.00040}{\bibinfo {journal} {Front.in Phys.}}\ }%
  \textbf{\bibinfo {volume} {6}},\ \bibinfo {pages} {40} (\bibinfo {year}
  {2018}),\ \Eprint{http://arxiv.org/abs/1711.02180}{arXiv:1711.02180
  [hep-ph]}.%
  \bibAnnoteFile{Stop}{Cai:2017mow}%
\bibitem{Atre:2009rg}%
  \BibitemOpen
  \bibfield{author}{%
  \bibinfo {author} {\bibfnamefont{A.}~\bibnamefont{Atre}}, \bibinfo {author}
  {\bibfnamefont{T.}~\bibnamefont{Han}}, \bibinfo {author}
  {\bibfnamefont{S.}~\bibnamefont{Pascoli}}\ and\ \bibinfo {author}
  {\bibfnamefont{B.}~\bibnamefont{Zhang}},\ }%
  \emph{\bibinfo {title} {{The Search for Heavy Majorana Neutrinos}}},\
  \bibfield{journal}{%
  \Doi{10.1088/1126-6708/2009/05/030}{\bibinfo {journal} {JHEP}}\ }%
  \textbf{\bibinfo {volume} {0905}},\ \bibinfo {pages} {030} (\bibinfo {year}
  {2009}),\ \Eprint{http://arxiv.org/abs/0901.3589}{arXiv:0901.3589 [hep-ph]}.%
  \bibAnnoteFile{Stop}{Atre:2009rg}%
\bibitem{Abada:2017jjx}%
  \BibitemOpen
  \bibfield{author}{%
  \bibinfo {author} {\bibfnamefont{A.}~\bibnamefont{Abada}}, \bibinfo {author}
  {\bibfnamefont{V.}~\bibnamefont{De~Romeri}}, \bibinfo {author}
  {\bibfnamefont{M.}~\bibnamefont{Lucente}}, \bibinfo {author}
  {\bibfnamefont{A.~M.}\ \bibnamefont{Teixeira}}\ and\ \bibinfo {author}
  {\bibfnamefont{T.}~\bibnamefont{Toma}},\ }%
  \emph{\bibinfo {title} {{Effective Majorana mass matrix from tau and
  pseudoscalar meson lepton number violating decays}}},\ \bibfield{journal}{%
  \Doi{10.1007/JHEP02(2018)169}{\bibinfo {journal} {JHEP}}\ }%
  \textbf{\bibinfo {volume} {02}},\ \bibinfo {pages} {169} (\bibinfo {year}
  {2018}),\ \Eprint{http://arxiv.org/abs/1712.03984}{arXiv:1712.03984
  [hep-ph]}.%
  \bibAnnoteFile{Stop}{Abada:2017jjx}%
\bibitem{Pascoli:2018heg}%
  \BibitemOpen
  \bibfield{author}{%
  \bibinfo {author} {\bibfnamefont{S.}~\bibnamefont{Pascoli}}, \bibinfo
  {author} {\bibfnamefont{R.}~\bibnamefont{Ruiz}}\ and\ \bibinfo {author}
  {\bibfnamefont{C.}~\bibnamefont{Weiland}},\ }%
  \emph{\bibinfo {title} {{Heavy Neutrinos with Dynamic Jet Vetoes: Multilepton
  Searches at $\sqrt{s} = 14,~27,$ and $100$ TeV}}} (\bibinfo {year} {2018}),\
  \Eprint{http://arxiv.org/abs/1812.08750}{arXiv:1812.08750 [hep-ph]}.%
  \bibAnnoteFile{Stop}{Pascoli:2018heg}%
\bibitem{Abada:2018nio}%
  \BibitemOpen
  \bibfield{author}{%
  \bibinfo {author} {\bibfnamefont{A.}~\bibnamefont{Abada}}\ and\ \bibinfo
  {author} {\bibfnamefont{A.~M.}\ \bibnamefont{Teixeira}},\ }%
  \emph{\bibinfo {title} {{Heavy neutral leptons and high-intensity
  observables}}},\ \bibfield{journal}{%
  \Doi{10.3389/fphy.2018.00142}{\bibinfo {journal} {Front.in Phys.}}\ }%
  \textbf{\bibinfo {volume} {6}},\ \bibinfo {pages} {142} (\bibinfo {year}
  {2018}),\ \Eprint{http://arxiv.org/abs/1812.08062}{arXiv:1812.08062
  [hep-ph]}.%
  \bibAnnoteFile{Stop}{Abada:2018nio}%
\bibitem{Duarte:2015iba}%
  \BibitemOpen
  \bibfield{author}{%
  \bibinfo {author} {\bibfnamefont{L.}~\bibnamefont{Duarte}}, \bibinfo {author}
  {\bibfnamefont{J.}~\bibnamefont{Peressutti}}\ and\ \bibinfo {author}
  {\bibfnamefont{O.~A.}\ \bibnamefont{Sampayo}},\ }%
  \emph{\bibinfo {title} {{Majorana neutrino decay in an Effective
  Approach}}},\ \bibfield{journal}{%
  \Doi{10.1103/PhysRevD.92.093002}{\bibinfo {journal} {Phys. Rev.}}\ }%
  \textbf{\bibinfo {volume} {D92}},\ \bibinfo {pages} {093002} (\bibinfo {year}
  {2015}),\ \Eprint{http://arxiv.org/abs/1508.01588}{arXiv:1508.01588
  [hep-ph]}.%
  \bibAnnoteFile{Stop}{Duarte:2015iba}%
\bibitem{Duarte:2016miz}%
  \BibitemOpen
  \bibfield{author}{%
  \bibinfo {author} {\bibfnamefont{L.}~\bibnamefont{Duarte}}, \bibinfo {author}
  {\bibfnamefont{I.}~\bibnamefont{Romero}}, \bibinfo {author}
  {\bibfnamefont{J.}~\bibnamefont{Peressutti}}\ and\ \bibinfo {author}
  {\bibfnamefont{O.~A.}\ \bibnamefont{Sampayo}},\ }%
  \emph{\bibinfo {title} {{Effective Majorana neutrino decay}}},\
  \bibfield{journal}{%
  \Doi{10.1140/epjc/s10052-016-4301-8}{\bibinfo {journal} {Eur. Phys. J. C}}\
  }%
  \textbf{\bibinfo {volume} {76}},\ \bibinfo {pages} {453} (\bibinfo {year}
  {2016}),\ \Eprint{http://arxiv.org/abs/1603.08052}{arXiv:1603.08052
  [hep-ph]}.%
  \bibAnnoteFile{Stop}{Duarte:2016miz}%
\bibitem{Peressutti:2011kx}%
  \BibitemOpen
  \bibfield{author}{%
  \bibinfo {author} {\bibfnamefont{J.}~\bibnamefont{Peressutti}}, \bibinfo
  {author} {\bibfnamefont{I.}~\bibnamefont{Romero}}\ and\ \bibinfo {author}
  {\bibfnamefont{O.~A.}\ \bibnamefont{Sampayo}},\ }%
  \emph{\bibinfo {title} {{Majorana Neutrinos Production at NLC in an Effective
  Approach}}},\ \bibfield{journal}{%
  \Doi{10.1103/PhysRevD.84.113002}{\bibinfo {journal} {Phys.Rev.}}\ }%
  \textbf{\bibinfo {volume} {D84}},\ \bibinfo {pages} {113002} (\bibinfo {year}
  {2011}),\ \Eprint{http://arxiv.org/abs/1110.0959}{arXiv:1110.0959 [hep-ph]}.%
  \bibAnnoteFile{Stop}{Peressutti:2011kx}%
\bibitem{Peressutti:2014lka}%
  \BibitemOpen
  \bibfield{author}{%
  \bibinfo {author} {\bibfnamefont{J.}~\bibnamefont{Peressutti}}\ and\ \bibinfo
  {author} {\bibfnamefont{O.~A.}\ \bibnamefont{Sampayo}},\ }%
  \emph{\bibinfo {title} {{Majorana neutrinos in $e$ $\gamma$ colliders from an
  effective Lagrangian approach}}},\ \bibfield{journal}{%
  \Doi{10.1103/PhysRevD.90.013003}{\bibinfo {journal} {Phys. Rev.}}\ }%
  \textbf{\bibinfo {volume} {D90}},\ \bibinfo {pages} {013003} (\bibinfo {year}
  {2014}).%
  \bibAnnoteFile{Stop}{Peressutti:2014lka}%
\bibitem{Duarte:2014zea}%
  \BibitemOpen
  \bibfield{author}{%
  \bibinfo {author} {\bibfnamefont{L.}~\bibnamefont{Duarte}}, \bibinfo {author}
  {\bibfnamefont{G.~A.}\ \bibnamefont{González-Sprinberg}}\ and\ \bibinfo
  {author} {\bibfnamefont{O.~A.}\ \bibnamefont{Sampayo}},\ }%
  \emph{\bibinfo {title} {{Majorana neutrinos production at LHeC in an
  effective approach}}},\ \bibfield{journal}{%
  \Doi{10.1103/PhysRevD.91.053007}{\bibinfo {journal} {Phys. Rev.}}\ }%
  \textbf{\bibinfo {volume} {D91}},\ \bibinfo {pages} {053007} (\bibinfo {year}
  {2015}),\ \Eprint{http://arxiv.org/abs/1412.1433}{arXiv:1412.1433 [hep-ph]}.%
  \bibAnnoteFile{Stop}{Duarte:2014zea}%
\bibitem{Duarte:2018xst}%
  \BibitemOpen
  \bibfield{author}{%
  \bibinfo {author} {\bibfnamefont{L.}~\bibnamefont{Duarte}}, \bibinfo {author}
  {\bibfnamefont{G.}~\bibnamefont{Zapata}}\ and\ \bibinfo {author}
  {\bibfnamefont{O.~A.}\ \bibnamefont{Sampayo}},\ }%
  \emph{\bibinfo {title} {{Angular and polarization trails from effective
  interactions of Majorana neutrinos at the LHeC}}},\ \bibfield{journal}{%
  \Doi{10.1140/epjc/s10052-018-5833-x}{\bibinfo {journal} {Eur. Phys. J.}}\ }%
  \textbf{\bibinfo {volume} {C78}},\ \bibinfo {pages} {352} (\bibinfo {year}
  {2018}),\ \Eprint{http://arxiv.org/abs/1802.07620}{arXiv:1802.07620
  [hep-ph]}.%
  \bibAnnoteFile{Stop}{Duarte:2018xst}%
\bibitem{Duarte:2018kiv}%
  \BibitemOpen
  \bibfield{author}{%
  \bibinfo {author} {\bibfnamefont{L.}~\bibnamefont{Duarte}}, \bibinfo {author}
  {\bibfnamefont{G.}~\bibnamefont{Zapata}}\ and\ \bibinfo {author}
  {\bibfnamefont{O.~A.}\ \bibnamefont{Sampayo}},\ }%
  \emph{\bibinfo {title} {{Final taus and initial state polarization signatures
  from effective interactions of Majorana neutrinos at future $e^{+}e^{-}$
  colliders}}},\ \bibfield{journal}{%
  \Doi{10.1140/epjc/s10052-019-6734-3}{\bibinfo {journal} {Eur. Phys. J.}}\ }%
  \textbf{\bibinfo {volume} {C79}},\ \bibinfo {pages} {240} (\bibinfo {year}
  {2019}),\ \Eprint{http://arxiv.org/abs/1812.01154}{arXiv:1812.01154
  [hep-ph]}.%
  \bibAnnoteFile{Stop}{Duarte:2018kiv}%
\bibitem{Duarte:2016caz}%
  \BibitemOpen
  \bibfield{author}{%
  \bibinfo {author} {\bibfnamefont{L.}~\bibnamefont{Duarte}}, \bibinfo {author}
  {\bibfnamefont{J.}~\bibnamefont{Peressutti}}\ and\ \bibinfo {author}
  {\bibfnamefont{O.~A.}\ \bibnamefont{Sampayo}},\ }%
  \emph{\bibinfo {title} {{Not-that-heavy Majorana neutrino signals at the
  LHC}}},\ \bibfield{journal}{%
  \Doi{10.1088/1361-6471/aa99f5}{\bibinfo {journal} {J. Phys.}}\ }%
  \textbf{\bibinfo {volume} {G45}},\ \bibinfo {pages} {025001} (\bibinfo {year}
  {2018}),\ \Eprint{http://arxiv.org/abs/1610.03894}{arXiv:1610.03894
  [hep-ph]}.%
  \bibAnnoteFile{Stop}{Duarte:2016caz}%
\bibitem{Yue:2017mmi}%
  \BibitemOpen
  \bibfield{author}{%
  \bibinfo {author} {\bibfnamefont{C.-X.}\ \bibnamefont{Yue}}, \bibinfo
  {author} {\bibfnamefont{Y.-C.}\ \bibnamefont{Guo}}\ and\ \bibinfo {author}
  {\bibfnamefont{Z.-H.}\ \bibnamefont{Zhao}},\ }%
  \emph{\bibinfo {title} {{Majorana neutrino signals at Belle-II and ILC}}},\
  \bibfield{journal}{%
  \Doi{10.1016/j.nuclphysb.2017.10.009}{\bibinfo {journal} {Nucl. Phys.}}\ }%
  \textbf{\bibinfo {volume} {B925}},\ \bibinfo {pages} {186} (\bibinfo {year}
  {2017}),\ \Eprint{http://arxiv.org/abs/1710.06144}{arXiv:1710.06144
  [hep-ph]}.%
  \bibAnnoteFile{Stop}{Yue:2017mmi}%
\bibitem{Yue:2018hci}%
  \BibitemOpen
  \bibfield{author}{%
  \bibinfo {author} {\bibfnamefont{C.-X.}\ \bibnamefont{Yue}}\ and\ \bibinfo
  {author} {\bibfnamefont{J.-P.}\ \bibnamefont{Chu}},\ }%
  \emph{\bibinfo {title} {{Sterile neutrino and leptonic decays of the
  pseudoscalar mesons}}},\ \bibfield{journal}{%
  \Doi{10.1103/PhysRevD.98.055012}{\bibinfo {journal} {Phys. Rev.}}\ }%
  \textbf{\bibinfo {volume} {D98}},\ \bibinfo {pages} {055012} (\bibinfo {year}
  {2018}),\ \Eprint{http://arxiv.org/abs/1808.09139}{arXiv:1808.09139
  [hep-ph]}.%
  \bibAnnoteFile{Stop}{Yue:2018hci}%
\bibitem{Chun:2019nwi}%
  \BibitemOpen
  \bibfield{author}{%
  \bibinfo {author} {\bibfnamefont{E.~J.}\ \bibnamefont{Chun}}, \bibinfo
  {author} {\bibfnamefont{A.}~\bibnamefont{Das}}, \bibinfo {author}
  {\bibfnamefont{S.}~\bibnamefont{Mandal}}, \bibinfo {author}
  {\bibfnamefont{M.}~\bibnamefont{Mitra}}\ and\ \bibinfo {author}
  {\bibfnamefont{N.}~\bibnamefont{Sinha}},\ }%
  \emph{\bibinfo {title} {{Sensitivity of Lepton Number Violating Meson Decays
  in Different Experiments}}},\ \bibfield{journal}{%
  \Doi{10.1103/PhysRevD.100.095022}{\bibinfo {journal} {Phys. Rev. D}}\ }%
  \textbf{\bibinfo {volume} {100}},\ \bibinfo {pages} {095022} (\bibinfo {year}
  {2019}),\ \Eprint{http://arxiv.org/abs/1908.09562}{arXiv:1908.09562
  [hep-ph]}.%
  \bibAnnoteFile{Stop}{Chun:2019nwi}%
\bibitem{Asaka:2016rwd}%
  \BibitemOpen
  \bibfield{author}{%
  \bibinfo {author} {\bibfnamefont{T.}~\bibnamefont{Asaka}}\ and\ \bibinfo
  {author} {\bibfnamefont{H.}~\bibnamefont{Ishida}},\ }%
  \emph{\bibinfo {title} {{Lepton number violation by heavy Majorana neutrino
  in $B$ decays}}},\ \bibfield{journal}{%
  \Doi{10.1016/j.physletb.2016.10.070}{\bibinfo {journal} {Phys. Lett.}}\ }%
  \textbf{\bibinfo {volume} {B763}},\ \bibinfo {pages} {393} (\bibinfo {year}
  {2016}),\ \Eprint{http://arxiv.org/abs/1609.06113}{arXiv:1609.06113
  [hep-ph]}.%
  \bibAnnoteFile{Stop}{Asaka:2016rwd}%
\bibitem{Cvetic:2016fbv}%
  \BibitemOpen
  \bibfield{author}{%
  \bibinfo {author} {\bibfnamefont{G.}~\bibnamefont{Cvetic}}\ and\ \bibinfo
  {author} {\bibfnamefont{C.~S.}\ \bibnamefont{Kim}},\ }%
  \emph{\bibinfo {title} {{Rare decays of B mesons via on-shell sterile
  neutrinos}}},\ \bibfield{journal}{%
  \Doi{10.1103/PhysRevD.95.039901, 10.1103/PhysRevD.94.053001}{\bibinfo
  {journal} {Phys. Rev.}}\ }%
  \textbf{\bibinfo {volume} {D94}},\ \bibinfo {pages} {053001} (\bibinfo {year}
  {2016}),\ \bibinfo {note} {[Erratum: Phys. Rev.D95,no.3,039901(2017)]},\
  \Eprint{http://arxiv.org/abs/1606.04140}{arXiv:1606.04140 [hep-ph]}.%
  \bibAnnoteFile{Stop}{Cvetic:2016fbv}%
\bibitem{Cvetic:2015naa}%
  \BibitemOpen
  \bibfield{author}{%
  \bibinfo {author} {\bibfnamefont{G.}~\bibnamefont{Cvetic}}, \bibinfo {author}
  {\bibfnamefont{C.}~\bibnamefont{Dib}}, \bibinfo {author}
  {\bibfnamefont{C.~S.}\ \bibnamefont{Kim}}\ and\ \bibinfo {author}
  {\bibfnamefont{J.}~\bibnamefont{Zamora-Saa}},\ }%
  \emph{\bibinfo {title} {{Probing the Majorana neutrinos and their CP
  violation in decays of charged scalar mesons $\pi, K, D, D_s, B, B_c$}}},\
  \bibfield{journal}{%
  \Doi{10.3390/sym7020726}{\bibinfo {journal} {Symmetry}}\ }%
  \textbf{\bibinfo {volume} {7}},\ \bibinfo {pages} {726} (\bibinfo {year}
  {2015}),\ \Eprint{http://arxiv.org/abs/1503.01358}{arXiv:1503.01358
  [hep-ph]}.%
  \bibAnnoteFile{Stop}{Cvetic:2015naa}%
\bibitem{Wang:2014lda}%
  \BibitemOpen
  \bibfield{author}{%
  \bibinfo {author} {\bibfnamefont{Y.}~\bibnamefont{Wang}}, \bibinfo {author}
  {\bibfnamefont{S.-S.}\ \bibnamefont{Bao}}, \bibinfo {author}
  {\bibfnamefont{Z.-H.}\ \bibnamefont{Li}}, \bibinfo {author}
  {\bibfnamefont{N.}~\bibnamefont{Zhu}}\ and\ \bibinfo {author}
  {\bibfnamefont{Z.-G.}\ \bibnamefont{Si}},\ }%
  \emph{\bibinfo {title} {{Study Majorana Neutrino Contribution to B-meson
  Semi-leptonic Rare Decays}}},\ \bibfield{journal}{%
  \Doi{10.1016/j.physletb.2014.08.006}{\bibinfo {journal} {Phys. Lett.}}\ }%
  \textbf{\bibinfo {volume} {B736}},\ \bibinfo {pages} {428} (\bibinfo {year}
  {2014}),\ \Eprint{http://arxiv.org/abs/1407.2468}{arXiv:1407.2468 [hep-ph]}.%
  \bibAnnoteFile{Stop}{Wang:2014lda}%
\bibitem{Cvetic:2010rw}%
  \BibitemOpen
  \bibfield{author}{%
  \bibinfo {author} {\bibfnamefont{G.}~\bibnamefont{Cvetic}}, \bibinfo {author}
  {\bibfnamefont{C.}~\bibnamefont{Dib}}, \bibinfo {author}
  {\bibfnamefont{S.~K.}\ \bibnamefont{Kang}}\ and\ \bibinfo {author}
  {\bibfnamefont{C.~S.}\ \bibnamefont{Kim}},\ }%
  \emph{\bibinfo {title} {{Probing Majorana neutrinos in rare K and D, $~D_s$,
  B, $B_c$ meson decays}}},\ \bibfield{journal}{%
  \Doi{10.1103/PhysRevD.82.053010}{\bibinfo {journal} {Phys. Rev.}}\ }%
  \textbf{\bibinfo {volume} {D82}},\ \bibinfo {pages} {053010} (\bibinfo {year}
  {2010}),\ \Eprint{http://arxiv.org/abs/1005.4282}{arXiv:1005.4282 [hep-ph]}.%
  \bibAnnoteFile{Stop}{Cvetic:2010rw}%
\bibitem{Zhang:2010um}%
  \BibitemOpen
  \bibfield{author}{%
  \bibinfo {author} {\bibfnamefont{J.-M.}\ \bibnamefont{Zhang}}\ and\ \bibinfo
  {author} {\bibfnamefont{G.-L.}\ \bibnamefont{Wang}},\ }%
  \emph{\bibinfo {title} {{Lepton-Number Violating Decays of Heavy Mesons}}},\
  \bibfield{journal}{%
  \Doi{10.1140/epjc/s10052-011-1715-1}{\bibinfo {journal} {Eur. Phys. J.}}\ }%
  \textbf{\bibinfo {volume} {C71}},\ \bibinfo {pages} {1715} (\bibinfo {year}
  {2011}),\ \Eprint{http://arxiv.org/abs/1003.5570}{arXiv:1003.5570 [hep-ph]}.%
  \bibAnnoteFile{Stop}{Zhang:2010um}%
\bibitem{Helo:2010cw}%
  \BibitemOpen
  \bibfield{author}{%
  \bibinfo {author} {\bibfnamefont{J.~C.}\ \bibnamefont{Helo}}, \bibinfo
  {author} {\bibfnamefont{S.}~\bibnamefont{Kovalenko}}\ and\ \bibinfo {author}
  {\bibfnamefont{I.}~\bibnamefont{Schmidt}},\ }%
  \emph{\bibinfo {title} {{Sterile neutrinos in lepton number and lepton flavor
  violating decays}}},\ \bibfield{journal}{%
  \Doi{10.1016/j.nuclphysb.2011.07.020}{\bibinfo {journal} {Nucl. Phys.}}\ }%
  \textbf{\bibinfo {volume} {B853}},\ \bibinfo {pages} {80} (\bibinfo {year}
  {2011}),\ \Eprint{http://arxiv.org/abs/1005.1607}{arXiv:1005.1607 [hep-ph]}.%
  \bibAnnoteFile{Stop}{Helo:2010cw}%
\bibitem{Ali:2001gsa}%
  \BibitemOpen
  \bibfield{author}{%
  \bibinfo {author} {\bibfnamefont{A.}~\bibnamefont{Ali}}, \bibinfo {author}
  {\bibfnamefont{A.~V.}\ \bibnamefont{Borisov}}\ and\ \bibinfo {author}
  {\bibfnamefont{N.~B.}\ \bibnamefont{Zamorin}},\ }%
  \emph{\bibinfo {title} {{Majorana neutrinos and same sign dilepton production
  at LHC and in rare meson decays}}},\ \bibfield{journal}{%
  \Doi{10.1007/s100520100702}{\bibinfo {journal} {Eur. Phys. J.}}\ }%
  \textbf{\bibinfo {volume} {C21}},\ \bibinfo {pages} {123} (\bibinfo {year}
  {2001}),\ \Eprint{http://arxiv.org/abs/hep-ph/0104123}{arXiv:hep-ph/0104123
  [hep-ph]}.%
  \bibAnnoteFile{Stop}{Ali:2001gsa}%
\bibitem{Bifani:2018zmi}%
  \BibitemOpen
  \bibfield{author}{%
  \bibinfo {author} {\bibfnamefont{S.}~\bibnamefont{Bifani}}, \bibinfo {author}
  {\bibfnamefont{S.}~\bibnamefont{Descotes-Genon}}, \bibinfo {author}
  {\bibfnamefont{A.}~\bibnamefont{Romero~Vidal}}\ and\ \bibinfo {author}
  {\bibfnamefont{M.-H.}\ \bibnamefont{Schune}},\ }%
  \emph{\bibinfo {title} {{Review of Lepton Universality tests in $B$
  decays}}},\ \bibfield{journal}{%
  \Doi{10.1088/1361-6471/aaf5de}{\bibinfo {journal} {J. Phys. G}}\ }%
  \textbf{\bibinfo {volume} {46}},\ \bibinfo {pages} {023001} (\bibinfo {year}
  {2019}),\ \Eprint{http://arxiv.org/abs/1809.06229}{arXiv:1809.06229
  [hep-ex]}.%
  \bibAnnoteFile{Stop}{Bifani:2018zmi}%
\bibitem{Cvetic:2017gkt}%
  \BibitemOpen
  \bibfield{author}{%
  \bibinfo {author} {\bibfnamefont{G.}~\bibnamefont{Cvetič}}, \bibinfo
  {author} {\bibfnamefont{F.}~\bibnamefont{Halzen}}, \bibinfo {author}
  {\bibfnamefont{C.}~\bibnamefont{Kim}}\ and\ \bibinfo {author}
  {\bibfnamefont{S.}~\bibnamefont{Oh}},\ }%
  \emph{\bibinfo {title} {{Anomalies in (semi)-leptonic $B$ decays $B^{\pm} \to
  \tau^{\pm} \nu$, $B^{\pm} \to D \tau^{\pm} \nu$ and $B^{\pm} \to D^*
  \tau^{\pm} \nu$, and possible resolution with sterile neutrino}}},\
  \bibfield{journal}{%
  \Doi{10.1088/1674-1137/41/11/113102}{\bibinfo {journal} {Chin. Phys. C}}\ }%
  \textbf{\bibinfo {volume} {41}},\ \bibinfo {pages} {113102} (\bibinfo {year}
  {2017}),\ \Eprint{http://arxiv.org/abs/1702.04335}{arXiv:1702.04335
  [hep-ph]}.%
  \bibAnnoteFile{Stop}{Cvetic:2017gkt}%
\bibitem{Mandal:2020htr}%
  \BibitemOpen
  \bibfield{author}{%
  \bibinfo {author} {\bibfnamefont{R.}~\bibnamefont{Mandal}}, \bibinfo {author}
  {\bibfnamefont{C.}~\bibnamefont{Murgui}}, \bibinfo {author}
  {\bibfnamefont{A.}~\bibnamefont{Peñuelas}}\ and\ \bibinfo {author}
  {\bibfnamefont{A.}~\bibnamefont{Pich}},\ }%
  \emph{\bibinfo {title} {{The role of right-handed neutrinos in $b \to c \tau
  \bar{\nu}$ anomalies}}} (\bibinfo {month} {4}\ \bibinfo {year} {2020}),\
  \Eprint{http://arxiv.org/abs/2004.06726}{arXiv:2004.06726 [hep-ph]}.%
  \bibAnnoteFile{Stop}{Mandal:2020htr}%
\bibitem{Azatov:2018kzb}%
  \BibitemOpen
  \bibfield{author}{%
  \bibinfo {author} {\bibfnamefont{A.}~\bibnamefont{Azatov}}, \bibinfo {author}
  {\bibfnamefont{D.}~\bibnamefont{Barducci}}, \bibinfo {author}
  {\bibfnamefont{D.}~\bibnamefont{Ghosh}}, \bibinfo {author}
  {\bibfnamefont{D.}~\bibnamefont{Marzocca}}\ and\ \bibinfo {author}
  {\bibfnamefont{L.}~\bibnamefont{Ubaldi}},\ }%
  \emph{\bibinfo {title} {{Combined explanations of B-physics anomalies: the
  sterile neutrino solution}}},\ \bibfield{journal}{%
  \Doi{10.1007/JHEP10(2018)092}{\bibinfo {journal} {JHEP}}\ }%
  \textbf{\bibinfo {volume} {10}},\ \bibinfo {pages} {092} (\bibinfo {year}
  {2018}),\ \Eprint{http://arxiv.org/abs/1807.10745}{arXiv:1807.10745
  [hep-ph]}.%
  \bibAnnoteFile{Stop}{Azatov:2018kzb}%
\bibitem{Greljo:2018ogz}%
  \BibitemOpen
  \bibfield{author}{%
  \bibinfo {author} {\bibfnamefont{A.}~\bibnamefont{Greljo}}, \bibinfo {author}
  {\bibfnamefont{D.~J.}\ \bibnamefont{Robinson}}, \bibinfo {author}
  {\bibfnamefont{B.}~\bibnamefont{Shakya}}\ and\ \bibinfo {author}
  {\bibfnamefont{J.}~\bibnamefont{Zupan}},\ }%
  \emph{\bibinfo {title} {{R(D$^{(*)}$) from W$^{\prime}$ and right-handed
  neutrinos}}},\ \bibfield{journal}{%
  \Doi{10.1007/JHEP09(2018)169}{\bibinfo {journal} {JHEP}}\ }%
  \textbf{\bibinfo {volume} {09}},\ \bibinfo {pages} {169} (\bibinfo {year}
  {2018}),\ \Eprint{http://arxiv.org/abs/1804.04642}{arXiv:1804.04642
  [hep-ph]}.%
  \bibAnnoteFile{Stop}{Greljo:2018ogz}%
\bibitem{Robinson:2018gza}%
  \BibitemOpen
  \bibfield{author}{%
  \bibinfo {author} {\bibfnamefont{D.~J.}\ \bibnamefont{Robinson}}, \bibinfo
  {author} {\bibfnamefont{B.}~\bibnamefont{Shakya}}\ and\ \bibinfo {author}
  {\bibfnamefont{J.}~\bibnamefont{Zupan}},\ }%
  \emph{\bibinfo {title} {{Right-handed neutrinos and R(D$^{(*)}$)}}},\
  \bibfield{journal}{%
  \Doi{10.1007/JHEP02(2019)119}{\bibinfo {journal} {JHEP}}\ }%
  \textbf{\bibinfo {volume} {02}},\ \bibinfo {pages} {119} (\bibinfo {year}
  {2019}),\ \Eprint{http://arxiv.org/abs/1807.04753}{arXiv:1807.04753
  [hep-ph]}.%
  \bibAnnoteFile{Stop}{Robinson:2018gza}%
\bibitem{Duarte:2019rzs}%
  \BibitemOpen
  \bibfield{author}{%
  \bibinfo {author} {\bibfnamefont{L.}~\bibnamefont{Duarte}}, \bibinfo {author}
  {\bibfnamefont{J.}~\bibnamefont{Peressutti}}, \bibinfo {author}
  {\bibfnamefont{I.}~\bibnamefont{Romero}}\ and\ \bibinfo {author}
  {\bibfnamefont{O.~A.}\ \bibnamefont{Sampayo}},\ }%
  \emph{\bibinfo {title} {{Majorana neutrinos with effective interactions in B
  decays}}},\ \bibfield{journal}{%
  \Doi{10.1140/epjc/s10052-019-7104-x}{\bibinfo {journal} {Eur. Phys. J.}}\ }%
  \textbf{\bibinfo {volume} {C79}},\ \bibinfo {pages} {593} (\bibinfo {year}
  {2019}),\ \Eprint{http://arxiv.org/abs/1904.07175}{arXiv:1904.07175
  [hep-ph]}.%
  \bibAnnoteFile{Stop}{Duarte:2019rzs}%
\bibitem{Aaij:2014aba}%
  \BibitemOpen
  \bibfield{author}{%
  \bibinfo {author} {\bibfnamefont{R.}~\bibnamefont{Aaij}} \emph{et~al.}
  (\bibinfo {collaboration} {LHCb}),\ }%
  \emph{\bibinfo {title} {{Search for Majorana neutrinos in $B^- \to
  \pi^+\mu^-\mu^-$ decays}}},\ \bibfield{journal}{%
  \Doi{10.1103/PhysRevLett.112.131802}{\bibinfo {journal} {Phys. Rev. Lett.}}\
  }%
  \textbf{\bibinfo {volume} {112}},\ \bibinfo {pages} {131802} (\bibinfo {year}
  {2014}),\ \Eprint{http://arxiv.org/abs/1401.5361}{arXiv:1401.5361 [hep-ex]}.%
  \bibAnnoteFile{Stop}{Aaij:2014aba}%
\bibitem{Gelb:2018end}%
  \BibitemOpen
  \bibfield{author}{%
  \bibinfo {author} {\bibfnamefont{M.}~\bibnamefont{Gelb}} \emph{et~al.}
  (\bibinfo {collaboration} {Belle}),\ }%
  \emph{\bibinfo {title} {{Search for the rare decay of $B^+ \to \ell^{+}
  \nu_{\ell} \gamma$ with improved hadronic tagging}}},\ \bibfield{journal}{%
  \Doi{10.1103/PhysRevD.98.112016}{\bibinfo {journal} {Phys. Rev.}}\ }%
  \textbf{\bibinfo {volume} {D98}},\ \bibinfo {pages} {112016} (\bibinfo {year}
  {2018}),\ \Eprint{http://arxiv.org/abs/1810.12976}{arXiv:1810.12976
  [hep-ex]}.%
  \bibAnnoteFile{Stop}{Gelb:2018end}%
\bibitem{Wudka:1999ax}%
  \BibitemOpen
  \bibfield{author}{%
  \bibinfo {author} {\bibfnamefont{J.}~\bibnamefont{Wudka}},\ }%
  \emph{\bibinfo {title} {{A Short course in effective Lagrangians}}},\
  \bibfield{journal}{%
  \Doi{10.1063/1.1315034}{\bibinfo {journal} {AIP Conf.Proc.}}\ }%
  \textbf{\bibinfo {volume} {531}},\ \bibinfo {pages} {81} (\bibinfo {year}
  {2000}),\ \Eprint{http://arxiv.org/abs/hep-ph/0002180}{arXiv:hep-ph/0002180
  [hep-ph]}.%
  \bibAnnoteFile{Stop}{Wudka:1999ax}%
\bibitem{Weinberg:1979sa}%
  \BibitemOpen
  \bibfield{author}{%
  \bibinfo {author} {\bibfnamefont{S.}~\bibnamefont{Weinberg}},\ }%
  \emph{\bibinfo {title} {{Baryon and Lepton Nonconserving Processes}}},\
  \bibfield{journal}{%
  \Doi{10.1103/PhysRevLett.43.1566}{\bibinfo {journal} {Phys. Rev. Lett.}}\ }%
  \textbf{\bibinfo {volume} {43}},\ \bibinfo {pages} {1566} (\bibinfo {year}
  {1979}).%
  \bibAnnoteFile{Stop}{Weinberg:1979sa}%
\bibitem{Arzt:1994gp}%
  \BibitemOpen
  \bibfield{author}{%
  \bibinfo {author} {\bibfnamefont{C.}~\bibnamefont{Arzt}}, \bibinfo {author}
  {\bibfnamefont{M.}~\bibnamefont{Einhorn}}\ and\ \bibinfo {author}
  {\bibfnamefont{J.}~\bibnamefont{Wudka}},\ }%
  \emph{\bibinfo {title} {{Patterns of deviation from the standard model}}},\
  \bibfield{journal}{%
  \Doi{10.1016/0550-3213(94)00336-D}{\bibinfo {journal} {Nucl.Phys.}}\ }%
  \textbf{\bibinfo {volume} {B433}},\ \bibinfo {pages} {41} (\bibinfo {year}
  {1995}),\ \Eprint{http://arxiv.org/abs/hep-ph/9405214}{arXiv:hep-ph/9405214
  [hep-ph]}.%
  \bibAnnoteFile{Stop}{Arzt:1994gp}%
\bibitem{Bolton:2019pcu}%
  \BibitemOpen
  \bibfield{author}{%
  \bibinfo {author} {\bibfnamefont{P.~D.}\ \bibnamefont{Bolton}}, \bibinfo
  {author} {\bibfnamefont{F.~F.}\ \bibnamefont{Deppisch}}\ and\ \bibinfo
  {author} {\bibfnamefont{P.}~\bibnamefont{Bhupal~Dev}},\ }%
  \emph{\bibinfo {title} {{Neutrinoless double beta decay versus other probes
  of heavy sterile neutrinos}}},\ \bibfield{journal}{%
  \Doi{10.1007/JHEP03(2020)170}{\bibinfo {journal} {JHEP}}\ }%
  \textbf{\bibinfo {volume} {03}},\ \bibinfo {pages} {170} (\bibinfo {year}
  {2020}),\ \Eprint{http://arxiv.org/abs/1912.03058}{arXiv:1912.03058
  [hep-ph]}.%
  \bibAnnoteFile{Stop}{Bolton:2019pcu}%
\bibitem{KamLAND-Zen:2016pfg}%
  \BibitemOpen
  \bibfield{author}{%
  \bibinfo {author} {\bibfnamefont{A.}~\bibnamefont{Gando}} \emph{et~al.}
  (\bibinfo {collaboration} {KamLAND-Zen}),\ }%
  \emph{\bibinfo {title} {{Search for Majorana Neutrinos near the Inverted Mass
  Hierarchy Region with KamLAND-Zen}}},\ \bibfield{journal}{%
  \Doi{10.1103/PhysRevLett.117.109903, 10.1103/PhysRevLett.117.082503}{\bibinfo
  {journal} {Phys. Rev. Lett.}}\ }%
  \textbf{\bibinfo {volume} {117}},\ \bibinfo {pages} {082503} (\bibinfo {year}
  {2016}),\ \bibinfo {note} {[Addendum: Phys. Rev.
  Lett.117,no.10,109903(2016)]},\
  \Eprint{http://arxiv.org/abs/1605.02889}{arXiv:1605.02889 [hep-ex]}.%
  \bibAnnoteFile{Stop}{KamLAND-Zen:2016pfg}%
\bibitem{Tanabashi:2018oca}%
  \BibitemOpen
  \bibfield{author}{%
  \bibinfo {author} {\bibfnamefont{M.}~\bibnamefont{Tanabashi}} \emph{et~al.}
  (\bibinfo {collaboration} {Particle Data Group}),\ }%
  \emph{\bibinfo {title} {{Review of Particle Physics}}},\ \bibfield{journal}{%
  \Doi{10.1103/PhysRevD.98.030001}{\bibinfo {journal} {Phys. Rev.}}\ }%
  \textbf{\bibinfo {volume} {D98}},\ \bibinfo {pages} {030001} (\bibinfo {year}
  {2018}).%
  \bibAnnoteFile{Stop}{Tanabashi:2018oca}%
\bibitem{Kou:2018nap}%
  \BibitemOpen
  \bibfield{author}{%
  \bibinfo {author} {\bibfnamefont{W.}~\bibnamefont{Altmannshofer}}
  \emph{et~al.} (\bibinfo {collaboration} {Belle-II}),\ }%
  \emph{\bibinfo {title} {{The Belle II Physics Book}}},\ \bibfield{journal}{%
  \Doi{10.1093/ptep/ptz106}{\bibinfo {journal} {PTEP}}\ }%
  \textbf{\bibinfo {volume} {2019}},\ \bibinfo {pages} {123C01} (\bibinfo
  {year} {2019}),\ \bibinfo {note} {[Erratum: PTEP 2020, 029201 (2020)]},\
  \Eprint{http://arxiv.org/abs/1808.10567}{arXiv:1808.10567 [hep-ex]}.%
  \bibAnnoteFile{Stop}{Kou:2018nap}%
\bibitem{Hirose:2016wfn}%
  \BibitemOpen
  \bibfield{author}{%
  \bibinfo {author} {\bibfnamefont{S.}~\bibnamefont{Hirose}} \emph{et~al.}
  (\bibinfo {collaboration} {Belle}),\ }%
  \emph{\bibinfo {title} {{Measurement of the $\tau$ lepton polarization and
  $R(D^*)$ in the decay $\bar{B} \to D^* \tau^- \bar{\nu}_\tau$}}},\
  \bibfield{journal}{%
  \Doi{10.1103/PhysRevLett.118.211801}{\bibinfo {journal} {Phys. Rev. Lett.}}\
  }%
  \textbf{\bibinfo {volume} {118}},\ \bibinfo {pages} {211801} (\bibinfo {year}
  {2017}),\ \Eprint{http://arxiv.org/abs/1612.00529}{arXiv:1612.00529
  [hep-ex]}.%
  \bibAnnoteFile{Stop}{Hirose:2016wfn}%
\bibitem{Hirose:2017dxl}%
  \BibitemOpen
  \bibfield{author}{%
  \bibinfo {author} {\bibfnamefont{S.}~\bibnamefont{Hirose}} \emph{et~al.}
  (\bibinfo {collaboration} {Belle}),\ }%
  \emph{\bibinfo {title} {{Measurement of the $\tau$ lepton polarization and
  $R(D^*)$ in the decay $\bar{B} \rightarrow D^* \tau^- \bar{\nu}_\tau$ with
  one-prong hadronic $\tau$ decays at Belle}}},\ \bibfield{journal}{%
  \Doi{10.1103/PhysRevD.97.012004}{\bibinfo {journal} {Phys. Rev. D}}\ }%
  \textbf{\bibinfo {volume} {97}},\ \bibinfo {pages} {012004} (\bibinfo {year}
  {2018}),\ \Eprint{http://arxiv.org/abs/1709.00129}{arXiv:1709.00129
  [hep-ex]}.%
  \bibAnnoteFile{Stop}{Hirose:2017dxl}%
\bibitem{Kamenik:2017ghi}%
  \BibitemOpen
  \bibfield{author}{%
  \bibinfo {author} {\bibfnamefont{J.}~\bibnamefont{Kamenik}}, \bibinfo
  {author} {\bibfnamefont{S.}~\bibnamefont{Monteil}}, \bibinfo {author}
  {\bibfnamefont{A.}~\bibnamefont{Semkiv}}\ and\ \bibinfo {author}
  {\bibfnamefont{L.}~\bibnamefont{Silva}},\ }%
  \emph{\bibinfo {title} {{Lepton polarization asymmetries in rare semi-tauonic
  $ b \rightarrow s $ exclusive decays at FCC-$ee$}}},\ \bibfield{journal}{%
  \Doi{10.1140/epjc/s10052-017-5272-0}{\bibinfo {journal} {Eur. Phys. J. C}}\
  }%
  \textbf{\bibinfo {volume} {77}},\ \bibinfo {pages} {701} (\bibinfo {year}
  {2017}),\ \Eprint{http://arxiv.org/abs/1705.11106}{arXiv:1705.11106
  [hep-ph]}.%
  \bibAnnoteFile{Stop}{Kamenik:2017ghi}%
\bibitem{Abada:2019zxq}%
  \BibitemOpen
  \bibfield{author}{%
  \bibinfo {author} {\bibfnamefont{A.}~\bibnamefont{Abada}} \emph{et~al.}
  (\bibinfo {collaboration} {FCC}),\ }%
  \emph{\bibinfo {title} {{FCC-ee: The Lepton Collider}: {Future Circular
  Collider Conceptual Design Report Volume 2}}},\ \bibfield{journal}{%
  \Doi{10.1140/epjst/e2019-900045-4}{\bibinfo {journal} {Eur. Phys. J. ST}}\ }%
  \textbf{\bibinfo {volume} {228}},\ \bibinfo {pages} {261} (\bibinfo {year}
  {2019}).%
  \bibAnnoteFile{Stop}{Abada:2019zxq}%
\bibitem{Beneke:2011nf}%
  \BibitemOpen
  \bibfield{author}{%
  \bibinfo {author} {\bibfnamefont{M.}~\bibnamefont{Beneke}}\ and\ \bibinfo
  {author} {\bibfnamefont{J.}~\bibnamefont{Rohrwild}},\ }%
  \emph{\bibinfo {title} {{B meson distribution amplitude from $B \to \gamma l
  \nu$}}},\ \bibfield{journal}{%
  \Doi{10.1140/epjc/s10052-011-1818-8}{\bibinfo {journal} {Eur. Phys. J.}}\ }%
  \textbf{\bibinfo {volume} {C71}},\ \bibinfo {pages} {1818} (\bibinfo {year}
  {2011}),\ \Eprint{http://arxiv.org/abs/1110.3228}{arXiv:1110.3228 [hep-ph]}.%
  \bibAnnoteFile{Stop}{Beneke:2011nf}%
\bibitem{Wang:2016qii}%
  \BibitemOpen
  \bibfield{author}{%
  \bibinfo {author} {\bibfnamefont{Y.-M.}\ \bibnamefont{Wang}},\ }%
  \emph{\bibinfo {title} {{Factorization and dispersion relations for radiative
  leptonic $B$ decay}}},\ \bibfield{journal}{%
  \Doi{10.1007/JHEP09(2016)159}{\bibinfo {journal} {JHEP}}\ }%
  \textbf{\bibinfo {volume} {09}},\ \bibinfo {pages} {159} (\bibinfo {year}
  {2016}),\ \Eprint{http://arxiv.org/abs/1606.03080}{arXiv:1606.03080
  [hep-ph]}.%
  \bibAnnoteFile{Stop}{Wang:2016qii}%
\bibitem{DescotesGenon:2002mw}%
  \BibitemOpen
  \bibfield{author}{%
  \bibinfo {author} {\bibfnamefont{S.}~\bibnamefont{Descotes-Genon}}\ and\
  \bibinfo {author} {\bibfnamefont{C.~T.}\ \bibnamefont{Sachrajda}},\ }%
  \emph{\bibinfo {title} {{Factorization, the light cone distribution amplitude
  of the B meson and the radiative decay $B\to \gamma l \nu_{l}$}}},\
  \bibfield{journal}{%
  \Doi{10.1016/S0550-3213(02)01066-0}{\bibinfo {journal} {Nucl. Phys.}}\ }%
  \textbf{\bibinfo {volume} {B650}},\ \bibinfo {pages} {356} (\bibinfo {year}
  {2003}),\ \Eprint{http://arxiv.org/abs/hep-ph/0209216}{arXiv:hep-ph/0209216
  [hep-ph]}.%
  \bibAnnoteFile{Stop}{DescotesGenon:2002mw}%
\bibitem{Korchemsky:1999qb}%
  \BibitemOpen
  \bibfield{author}{%
  \bibinfo {author} {\bibfnamefont{G.~P.}\ \bibnamefont{Korchemsky}}, \bibinfo
  {author} {\bibfnamefont{D.}~\bibnamefont{Pirjol}}\ and\ \bibinfo {author}
  {\bibfnamefont{T.-M.}\ \bibnamefont{Yan}},\ }%
  \emph{\bibinfo {title} {{Radiative leptonic decays of B mesons in QCD}}},\
  \bibfield{journal}{%
  \Doi{10.1103/PhysRevD.61.114510}{\bibinfo {journal} {Phys. Rev.}}\ }%
  \textbf{\bibinfo {volume} {D61}},\ \bibinfo {pages} {114510} (\bibinfo {year}
  {2000}),\ \Eprint{http://arxiv.org/abs/hep-ph/9911427}{arXiv:hep-ph/9911427
  [hep-ph]}.%
  \bibAnnoteFile{Stop}{Korchemsky:1999qb}%
\bibitem{Wang:2018wfj}%
  \BibitemOpen
  \bibfield{author}{%
  \bibinfo {author} {\bibfnamefont{Y.-M.}\ \bibnamefont{Wang}}\ and\ \bibinfo
  {author} {\bibfnamefont{Y.-L.}\ \bibnamefont{Shen}},\ }%
  \emph{\bibinfo {title} {{Subleading-power corrections to the radiative
  leptonic $B \to \gamma \ell \nu$ decay in QCD}}},\ \bibfield{journal}{%
  \Doi{10.1007/JHEP05(2018)184}{\bibinfo {journal} {JHEP}}\ }%
  \textbf{\bibinfo {volume} {05}},\ \bibinfo {pages} {184} (\bibinfo {year}
  {2018}),\ \Eprint{http://arxiv.org/abs/1803.06667}{arXiv:1803.06667
  [hep-ph]}.%
  \bibAnnoteFile{Stop}{Wang:2018wfj}%
\bibitem{Beneke:2018wjp}%
  \BibitemOpen
  \bibfield{author}{%
  \bibinfo {author} {\bibfnamefont{M.}~\bibnamefont{Beneke}}, \bibinfo {author}
  {\bibfnamefont{V.~M.}\ \bibnamefont{Braun}}, \bibinfo {author}
  {\bibfnamefont{Y.}~\bibnamefont{Ji}}\ and\ \bibinfo {author}
  {\bibfnamefont{Y.-B.}\ \bibnamefont{Wei}},\ }%
  \emph{\bibinfo {title} {{Radiative leptonic decay $B\to \gamma \ell \nu_\ell$
  with subleading power corrections}}},\ \bibfield{journal}{%
  \Doi{10.1007/JHEP07(2018)154}{\bibinfo {journal} {JHEP}}\ }%
  \textbf{\bibinfo {volume} {07}},\ \bibinfo {pages} {154} (\bibinfo {year}
  {2018}),\ \Eprint{http://arxiv.org/abs/1804.04962}{arXiv:1804.04962
  [hep-ph]}.%
  \bibAnnoteFile{Stop}{Beneke:2018wjp}%
\bibitem{Prim:2019gtj}%
  \BibitemOpen
  \bibfield{author}{%
  \bibinfo {author} {\bibfnamefont{M.}~\bibnamefont{Prim}} \emph{et~al.}
  (\bibinfo {collaboration} {Belle}),\ }%
  \emph{\bibinfo {title} {{Search for $B^+ \to \mu^+\, \nu_\mu$ and $B^+ \to
  \mu^+\, N$ with inclusive tagging}}},\ \bibfield{journal}{%
  \Doi{10.1103/PhysRevD.101.032007}{\bibinfo {journal} {Phys. Rev. D}}\ }%
  \textbf{\bibinfo {volume} {101}},\ \bibinfo {pages} {032007} (\bibinfo {year}
  {2020}),\ \Eprint{http://arxiv.org/abs/1911.03186}{arXiv:1911.03186
  [hep-ex]}.%
  \bibAnnoteFile{Stop}{Prim:2019gtj}%
\bibitem{Adachi:2012mm}%
  \BibitemOpen
  \bibfield{author}{%
  \bibinfo {author} {\bibfnamefont{I.}~\bibnamefont{Adachi}} \emph{et~al.}
  (\bibinfo {collaboration} {Belle}),\ }%
  \emph{\bibinfo {title} {{Evidence for $B^- \to \tau^- \bar{\nu}_\tau$ with a
  Hadronic Tagging Method Using the Full Data Sample of Belle}}},\
  \bibfield{journal}{%
  \Doi{10.1103/PhysRevLett.110.131801}{\bibinfo {journal} {Phys. Rev. Lett.}}\
  }%
  \textbf{\bibinfo {volume} {110}},\ \bibinfo {pages} {131801} (\bibinfo {year}
  {2013}),\ \Eprint{http://arxiv.org/abs/1208.4678}{arXiv:1208.4678 [hep-ex]}.%
  \bibAnnoteFile{Stop}{Adachi:2012mm}%
\bibitem{Kronenbitter:2015kls}%
  \BibitemOpen
  \bibfield{author}{%
  \bibinfo {author} {\bibfnamefont{B.}~\bibnamefont{Kronenbitter}}
  \emph{et~al.} (\bibinfo {collaboration} {Belle}),\ }%
  \emph{\bibinfo {title} {{Measurement of the branching fraction of $B^{+} \to
  \tau^{+} \nu_{\tau}$ decays with the semileptonic tagging method}}},\
  \bibfield{journal}{%
  \Doi{10.1103/PhysRevD.92.051102}{\bibinfo {journal} {Phys. Rev.}}\ }%
  \textbf{\bibinfo {volume} {D92}},\ \bibinfo {pages} {051102} (\bibinfo {year}
  {2015}),\ \Eprint{http://arxiv.org/abs/1503.05613}{arXiv:1503.05613
  [hep-ex]}.%
  \bibAnnoteFile{Stop}{Kronenbitter:2015kls}%
\bibitem{Lees:2012ju}%
  \BibitemOpen
  \bibfield{author}{%
  \bibinfo {author} {\bibfnamefont{J.~P.}\ \bibnamefont{Lees}} \emph{et~al.}
  (\bibinfo {collaboration} {BaBar}),\ }%
  \emph{\bibinfo {title} {{Evidence of $B^+ \to \tau^+\nu$ decays with hadronic
  B tags}}},\ \bibfield{journal}{%
  \Doi{10.1103/PhysRevD.88.031102}{\bibinfo {journal} {Phys. Rev.}}\ }%
  \textbf{\bibinfo {volume} {D88}},\ \bibinfo {pages} {031102} (\bibinfo {year}
  {2013}),\ \Eprint{http://arxiv.org/abs/1207.0698}{arXiv:1207.0698 [hep-ex]}.%
  \bibAnnoteFile{Stop}{Lees:2012ju}%
\bibitem{Aubert:2009wt}%
  \BibitemOpen
  \bibfield{author}{%
  \bibinfo {author} {\bibfnamefont{B.}~\bibnamefont{Aubert}} \emph{et~al.}
  (\bibinfo {collaboration} {BaBar}),\ }%
  \emph{\bibinfo {title} {{A Search for $B^+ \to \ell^+ \nu_{\ell}$ Recoiling
  Against $B^{-}\to D^{0} \ell^{-}\bar{\nu} X$}}},\ \bibfield{journal}{%
  \Doi{10.1103/PhysRevD.81.051101}{\bibinfo {journal} {Phys. Rev.}}\ }%
  \textbf{\bibinfo {volume} {D81}},\ \bibinfo {pages} {051101} (\bibinfo {year}
  {2010}),\ \Eprint{http://arxiv.org/abs/0912.2453}{arXiv:0912.2453 [hep-ex]}.%
  \bibAnnoteFile{Stop}{Aubert:2009wt}%
\bibitem{Liventsev:2013zz}%
  \BibitemOpen
  \bibfield{author}{%
  \bibinfo {author} {\bibfnamefont{D.}~\bibnamefont{Liventsev}} \emph{et~al.}
  (\bibinfo {collaboration} {Belle}),\ }%
  \emph{\bibinfo {title} {{Search for heavy neutrinos at Belle}}},\
  \bibfield{journal}{%
  \Doi{10.1103/PhysRevD.95.099903, 10.1103/PhysRevD.87.071102}{\bibinfo
  {journal} {Phys. Rev.}}\ }%
  \textbf{\bibinfo {volume} {D87}},\ \bibinfo {pages} {071102} (\bibinfo {year}
  {2013}),\ \bibinfo {note} {[Erratum: Phys. Rev.D95,no.9,099903(2017)]},\
  \Eprint{http://arxiv.org/abs/1301.1105}{arXiv:1301.1105 [hep-ex]}.%
  \bibAnnoteFile{Stop}{Liventsev:2013zz}%
\end{thebibliography}%

\end{document}